\documentclass[useAMS,usenatbib]{mn2e}
\usepackage{graphicx}
\usepackage[draft]{hyperref}
\usepackage{amsfonts,amsmath}
\usepackage{subfigure}

\usepackage{times}
\newcommand{\twco}{$^{12}$CO}
\newcommand{\thco}{$^{13}$CO}

\newcommand{\virial}{$\alpha$}
\newcommand{\mach}{$\mathcal{M}$}
\newcommand{\drive}{$k$}
\newcommand{\plasbeta}{$\beta$}
\newcommand{\solfrac}{$\zeta$}
\newcommand\aj{{AJ}}%
\newcommand\araa{{ARA\&A}}%
\newcommand\apj{{ApJ}}%
\newcommand\apjl{{ApJ}}%
\newcommand\apjs{{ApJS}}%
\newcommand\aap{{A\&A}}%
\newcommand\mnras{{MNRAS}}%
\newcommand\prd{{Phys.~Rev.~D}}%
\title[Tools for Spectral-line Data Cube Comparison]{Identifying Tools for Comparing Simulations and Observations of Spectral-line Data Cubes}

\author[Koch et al.]{Eric W. Koch$^{1,2}$\thanks{E-mail:
ekoch@ualberta.ca (EWK); calebgward@gmail.com (CGW);
jason.loeppky@ubc.ca (JLL); erosolow@ualberta.ca (EWR)} Caleb
  G. Ward$^{1}$, Stella Offner$^{3}$,
\newauthor
Jason L. Loeppky$^{1}$ and Erik W. Rosolowsky$^{1,2}$\\
$^{1}$University of British Columbia, Okanagan
  Campus, Department of Physics, 3333 University
  Way, Kelowna BC V1V 1V7 Canada\\
$^{2}$University of Alberta, Department of Physics, 4-183 CCIS, Edmonton AB T6G 2E1, Canada \\
$^{3}$University of Massachusetts Amherst, Astronomy Department, 710 North Pleasant St., Amherst MA 01003, USA}

\begin{document}

\date{Draft date: \today}

\pagerange{\pageref{firstpage}--\pageref{lastpage}} \pubyear{2017}

\maketitle

\label{firstpage}

\begin{abstract}
We present a statistical framework to compare spectral-line data cubes of molecular clouds and use the framework to perform an analysis of various statistical tools developed from methods proposed in the literature.  We test whether our methods are sensitive to changes in the underlying physical properties of the clouds or whether their behaviour is governed by random fluctuations.  We perform a set of 32 self-gravitating magnetohydrodynamic simulations that test all combinations of five physical parameters -- Mach number, plasma parameter, virial parameter, driving scales, and solenoidal driving fraction -- each of which can be set to a low or high value.  We create mock observational data sets of \thco(1-0) emission from each simulation.  We compare these mock data to a those generated from a set of baseline simulations using pseudo-distance metrics based on 18 different statistical techniques that have previously been used to study molecular clouds.  We analyze these results using methods from the statistical field of experimental design and find that several of the statistics can reliably track changes in the underlying physics.  Our analysis shows that the interactions between parameters are often among the most significant effects.  A small fraction of statistics are also sensitive to changes in magnetic field properties.  We use this framework to compare the set of simulations to observations of three nearby star-forming regions: NGC 1333, Oph A, and IC 348.  We find that no one simulation agrees significantly better with the observations, although there is evidence that the high Mach number simulations are more consistent with the observations.
\end{abstract}


\section{Introduction}
\label{sec:intro}

Star formation is one of the fundamental agents of galaxy evolution.  The star formation rate and stellar initial mass function establish metal enrichment patterns as well as the mass, momentum, and energy injection rates into the interstellar and intergalactic medium.  While the stellar initial mass function appears robust \citep[e.g.,][]{kroupa-imf,bastian2010,offner-imf}, the star formation rate on a galactic and local scale shows significant variation \citep[e.g.,][]{k98,leroy-kslaw}.  We seek a broad theoretical understanding of star formation that can predict these broad behaviours as a function of initial conditions.  Many attempts to explain these statistical trends have been proposed \citep[e.g.,][]{kdm12} and have shown some promise for general predictions.

Despite successes in unravelling these general trends, the details of star formation remain difficult to predict.  For example, the mass distributions of binary stars \citep{duchene2013,offner-imf} or variations in star formation rate with galaxy type \citep{leroy-kslaw} are not described by modern general theories of star formation.  The major complication in this description arises from the wide variety of physics thought to be important in the details of star formation.  The emergent behaviour of a wide range of competing physical processes must give rise to the initial mass function or the star formation laws.  Observational studies of star forming regions throughout the Milky Way broadly suggest that star formation could be influenced by a wide range of effects including: gravity, (magnetized) turbulence, gas pressure, radiative heating and cooling, metallicity, plasma effects and chemistry \citep{mo07}.  This long list of important physical effects has been guided by observations of star forming regions and informed by an ongoing dialogue with theoretical analysis.

Capturing the details of the rich interplay between all of these physical effects requires numerical simulation.  We have not yet developed a complete analytic or statistical theory that describes the details of the star formation process, but we can identify the relevant physical effects and consequently their governing laws.  Given these laws, numerical simulations aim to reproduce the process of star formation by numerically solving the differential equations at the heart of the physics.  The outcome of these simulations is a model of a star forming region, where the values of all quantities tracked through the simulation are known through the observational domain.  The results of such a study can then be compared to our expected values for the underlying physical quantities derived from observations.

The importance of turbulence in star formation drives the pressing need for simulations.  While turbulence has a well developed statistical theory, even in the compressible case \citep{gs-mhd1,gs-mhd2,compressible-mhd}, the combination with other physical effects (in particular gravity) has stymied a generalized statistical theory to date.  Simulating turbulent flows is computationally expensive, and even so, no simulation captures the full range of a turbulent cascade from the driving scale ($1$ to $100$ pc) to the viscous dissipation scale \citep[$\sim$ km;][]{gs-mhd2}.  A further complication is that turbulent flows are chaotic in the sense that they show exponentially divergent final states given small differences in initial conditions.  For a given set of initial conditions, this sensitivity requires many simulations with small random fluctuations in order to capture the full range of conditions the physics produces.  Without such a probe of the fluctuations, it is impossible to discern the aspects of the output that are attributable to physics rather than happenstance.  The state of the art in star formation simulations include many effects given above, but a simulation of the full range of the physics remains computationally prohibitive.  The state of the art simulations of star formation only have computational resources to produce one or a few simulation runs.  While such studies provide excellent insight into the different physical effects at work, they cannot capture the statistical behaviour of the star formation process. Since the sensitivity to initial conditions makes it impossible to replicate the physics at work in any real star forming region, these simulations are generally compared to observational work qualitatively.  The studies that do provide a quantitative comparison rely on broadly matching global properties of star forming regions such as the line width, mass density, or reproducing the IMF \citep[e.g.,][]{padoan2001,offner2008,kirk2009,Krumholz2012ApJ...754...71K,Bate2014MNRAS.442..285B}.  However, the simulations frequently do not agree with observations of star forming regions, possibly due to approximations of boundary conditions (e.g., isolated clouds or periodic conditions) in the simulations. A richer suite of comparisons that emulate molecular gas in star forming regions is needed. This suite must be able to execute the comparison between simulations and observations using a broad range of measurements \citep{taste-testing, rosolowsky-scma}.

Fortunately, there is a rich literature of quantitative analyses that have been applied to the observed molecular gas structures.  These quantifications can be applied to mock observational data generated from simulations.  Simulations that correctly capture the relevant physical processes in star formation should produce mock observational data that are statistically indistinguishable from real observations under a wide variety of these measurements.  However, all of these results are based on observational data, which cannot be used to uniquely infer the full suite of physical conditions in the region.  For example, observations of a molecular cloud in CO emission alone cannot be used to infer the density of the gas nor its motion in the plane of the sky. A good agreement between simulations and observations is thus necessary but not sufficient to claim consistency with the physical conditions present in an environment.

In this work, we focus on the tools that compare position-position-velocity (PPV) data cubes of molecular line emission to each other.  Our goal is to identify good tools for making this comparison: they should be sensitive to the physical conditions governing star formation but show minimum sensitivity to ``happenstance,'' i.e., random fluctuations in the turbulent field.  In \citet{Yer14}, we advocated an approach based on the statistical field of experimental design.  By quantifying the differences between two different PPV data cubes in terms of a pseudo-distance, \citet{Yer14} showed that the design of parameter studies mattered in star formation and that effects of physical conditions, such as self-gravity or turbulence, had significant interaction effects.  Changes in individual physical parameters cannot be studied in isolation, and tracking the influence of physical effects requires non-trivial parameter studies.  With the \citet{Yer14} framework in place, we proceed to find good tools to compare different PPV data cubes.  In this work, we identify a suite of methods that have been developed in the literature and reformulate their approaches to work with the distance framework.  Then, using a designed set of simulations, we find those tools that respond to given physical effects.  We also test for significant responses to physical parameters.  This test identifies the physical effects to which a given method is sensitive.  \citet{Boyden2016ApJ...833..233B} have used this approach in analyzing the influence of stellar feedback on mock observational data from star forming regions.

This work focuses on tools that are related to the turbulent properties of the molecular gas or to the probability density functions of the emission.  In Section \ref{sec:overview}, we highlight how our approach has been used in other fields and the complementarity to other studies using astrophysical simulation.  We then present a set of simulations that we use to test the behaviour of different statistical tools (Sec.~\ref{sec:sims}).  We describe a set of tools that have been suggested in the literature as good descriptors of ISM physics in Sec.~\ref{sec:stats}.  Given these simulations and statistics, we describe the results of an analysis based on the experimental design in Sec.~\ref{sec:analysis}, discuss limitations in these results in Sec.~\ref{sec:limitations_of_this_study}, and summarize our findings in Sec.\ref{sec:summary}.

\section{Overview of Approach}
\label{sec:overview}

Owing to the complex interplay of multiple physical effects, numerical simulation has been an extremely fruitful method for understanding different astrophysical phenomena, including star formation.  This ample body of work uses physical insight gleaned from simulations to propose models for the star formation process.  In this work, we forward an approach that uses the tools from the statistical field of experimental design to analyze simulations of star formation.  This approach is fundamentally different than the typical simulation approach: instead of exploring the outputs for detailed tracking of the physics, we adopt a ``black-box'' method wherein we consider inputs to the simulations (the initial conditions) and the resulting outputs (spectral-line data cubes) with less emphasis on the physical processes being tracked in the simulation.  This approach necessarily loses some physical insight.  However, the approach gains the backing of an extensive body of statistical work on the {\it calibration} of computer models that allows us to make direct comparisons to observations and thereby infer behaviours of different regions in the context of the space of initial conditions.  This approach was introduced by \citet{sacks1989design} where the results of computationally expensive simulations (termed ``computer experiments'') could be emulated using simple predictors to understand the behaviour of the experiment in unexplored portions of parameter space.  The applications of these techniques span many fields including hydrodynamics, economics, and climate science \citep{santner2013design}.

The utility of these methods relies on defining some optimization criterion on the outputs of the simulations (e.g., minimizing the difference in temperatures predicted vs.~observed temperatures in a climate model).  The principal application of these approaches in astrophysics is in the field of cosmological modelling, where the approach measured the optimal cosmological parameters required to produce the observed matter power spectrum \citep{Heitmann06,Habib07,Schneider08,Heitmann09,Heitmann10,Lawrence10}.  In this case, the power spectrum can be easily measured and compared using a simple $L^2$ norm (i.e., a $\chi^2$-like statistic) describing the difference between simulations and observations.  Said differently, the reduction from a fully three-dimensional universe to a simple measure of its structure is well defined and theoretically relevant.  Such reductions, for the problem of star formation and molecular cloud structure are not as well defined.  Instead, there are substantial ambiguities in using observable quantities to infer the properties of star forming molecular clouds, it is not possible to map the observables into a three-dimensional distribution of matter, and even column density maps are subject to biases.  The 2D (images) and 3D (spectral line data cubes) cannot be uniquely simulated even if the average physical conditions are already known.  When faced with high-dimensional outputs, the statistical field of experimental design applies dimensionality reduction \citep{higdon2012computer} and we adopt a similar approach here.  In this work, we recast many of the tools developed previously to quantify the structure and kinematics of molecular clouds as dimensionality reduction tools.  Our principal goal is to identify those tools that would be suitable for this reduction, namely those that show significant response when the initial conditions of the simulations are changed.

Our approach makes the fundamental exchange of quantity over quality: we run a large number of low-resolution simulations with a range of physical conditions to understand the effect.  The large number of simulations (37 in our case) are needed to validate the use the statistical tools.

\section{Simulations}
\label{sec:sims}

In this work, we are interested in determining how the metrics drawn from the literature respond to linear changes in parameter settings.  For example, does a given analysis tool respond to increases in the magnetic field strength? It is thus sufficient to consider each of the factors at only two levels \citep{Yer14}.  To account for interactions between parameter changes that we expect to be important, we consider two levels for five different physical properties (``factors'') and construct a full factorial design that consists of all $2^5=32$ different combinations of the factors.  This design allows us to estimate all main effects and interactions among the factors.

To create a set simulated observations on which we can test different distance metrics, we conduct simulations using the {\sc Enzo} adaptive mesh refinement (AMR) code \citep{enzo}.  We use {\sc Enzo}'s constrained transport magnetohydrodynamics solver \citep{collins10} and assume an isothermal ($T=10~\mbox{K}$) gas.  
We set the initial conditions using a set of dimensionless parameters that define magnetohydrodynamic systems \citep{mckee10}.  These include the 3D sonic Mach number $\mathcal{M}\equiv \sqrt{3}\sigma_v/c_s$, the virial parameter $\alpha_{\mathrm{vir}} = 5\sigma_v^2 L/(2 G M)$, and the plasma parameter $\beta \equiv 8\pi M c_s^2 / (L^3 B^2)$.  Here, $\sigma_v$ is the one-dimensional velocity dispersion of the gas, $c_s$ is the isothermal sound speed, $L$ is the linear size of the simulation domain, $M$ is the total mass in the domain, and $B$ is the initial field strength.  For all simulations, we take $L=10\mbox{ pc}$ and assume an initially uniform field in the $z$ direction. We set the turbulent velocity dispersion by applying a fixed random field to the gas velocity, which is normalized to give the desired initial velocity dispersion.  We further characterize the system by adopting different parameters that describe the input random field.  We vary the wavenumbers \drive\ over which energy is injected, where the input power spectrum over this wavenumber range is $P(k)\propto k^{-2}$ in all cases.  We also decompose the vector field into its compressive ($\nabla \times \mathbf{F}=0$) and solenoidal ($\nabla \cdot \mathbf{F}=0$) components and create fields with varying mixtures of these modes.  Following \citet{federrath10}, we parameterize the relative fraction of power found in these two modes using the parameter \solfrac such that $\zeta = 0$ corresponds to purely compressive modes and $\zeta = 1$ corresponds to purely solenoidal driving.  We adopt the same random seed unless otherwise specified.

We set these simulation parameters to different values corresponding to low, high, and fiducial (middle) values as described in Table \ref{tab:design}.  Given  these dimensionless parameters and the fixed physical scale of the simulation box, we define the initial density, the total mass in the domain, the energy injection required to maintain turbulence, the initial magnetic field strength, and the vector field for turbulent driving.

\begin{table}
\begin{center}
\caption{\label{tab:design} Experimental Design}
\begin{tabular}{l|ccc}
Parameter & Low Value & High Value & Fiducial \\
\hline
Virial Parameter (\virial) & 2 & 10 & 6 \\
Plasma Parameter (\plasbeta) & 0.5 & 2.0 & 1.0 \\
Mach Number (\mach) & 5 & 12 & 8.5\\
Driving Scale & $k\in[2,4]$ & $k\in[4,8]$ & $k\in[2,8]$\\
Solenoidal Fraction (\solfrac) & 0.33 & 0.66 & 0.5 \\
\end{tabular}
\end{center}
\end{table}

\begin{figure*}
\includegraphics[width=0.9\textwidth]{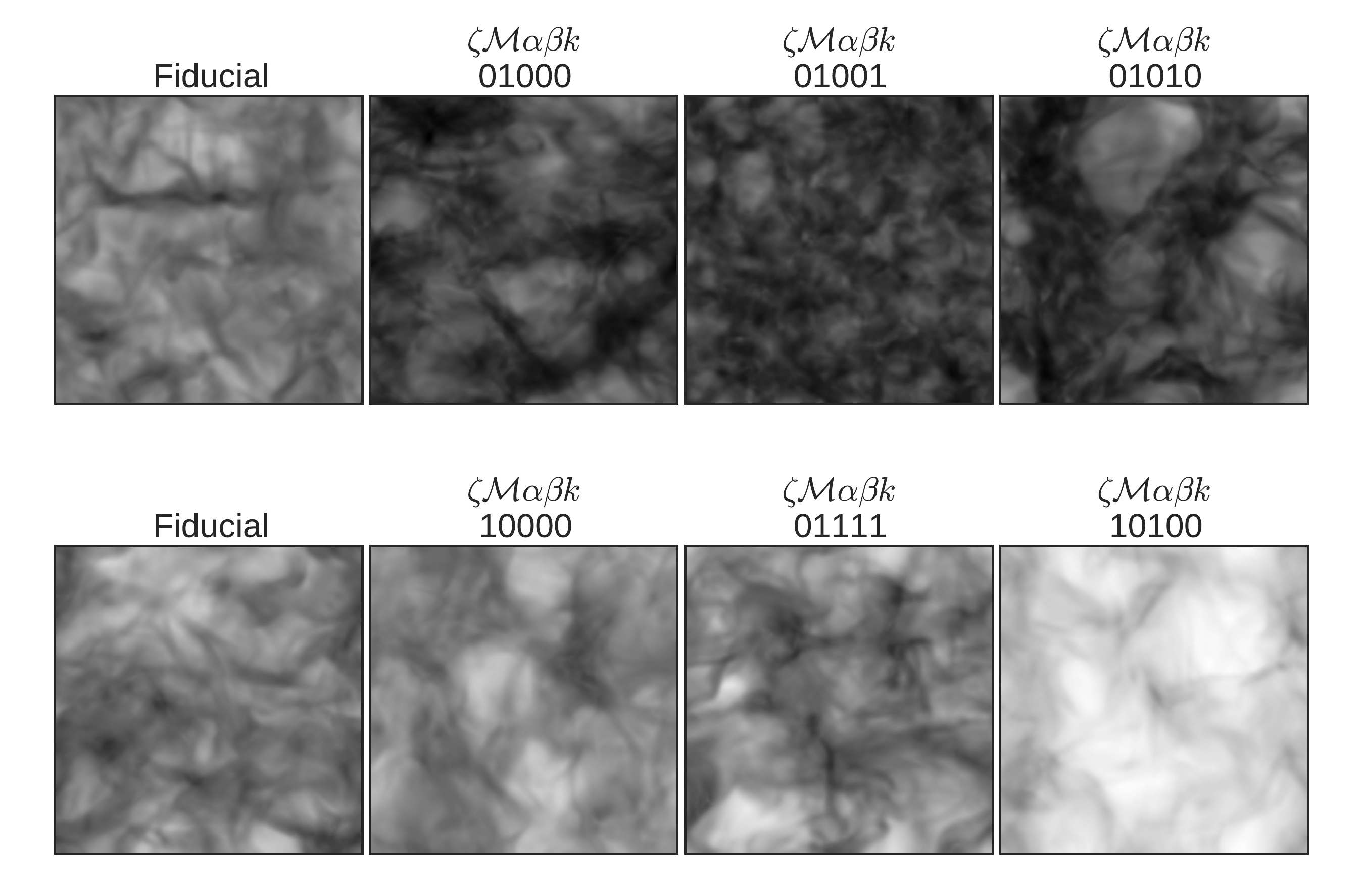}
\caption{\label{fig:moment0s} Examples of the velocity integrated intensities for eight of the simulations in the set are shown on a common square-root scale. Two fiducial simulations are shown on the far left, which have the same physical inputs. The parameter settings are shown for each of the six design simulations. The coding corresponds to the low (0) and high (1) settings, which are indicated in Table \ref{tab:design}.}
\end{figure*}

For all $2^5$ combinations of low and high parameters, we conduct a magnetohydrodynamic {\sc Enzo} simulation.  All simulations have a $128^3$ root grid representing a region (10 pc)$^3$ with periodic boundary conditions. Each volume is intended to represent a sub-region of a turbulent molecular cloud having different physical conditions. Once the size scale is set, we derive all other simulation inputs given the parameters in Table \ref{tab:design}.   The initial density field is uniform, but it is quickly perturbed by the driven turbulence.   Similar to the work of \citet{Yer14}, we find that the driving formalism in {\sc Enzo} produces Mach numbers consistent with the design requirements, but there can be significant deviations for small values of the plasma parameter (i.e., strong fields).  For each simulation, we drive the turbulence without gravity for two crossing times [$t_c \equiv t/(\mathcal{M}\sigma_v)$] without AMR to develop a turbulent cascade and reach a statistical steady state \citep[e.g.,][]{maclow99}.  After this period, we turn on gravity and AMR.  Refinement is added following the \citet{truelove97} criterion for a Jeans number of $N_J=0.125$. We insert sink particles \citep{wang10} wherever the Jeans criterion is exceeded on the finest level ($l=4$) .     The simulations are evolved for another crossing time with both gravity and turbulent driving, at which point $t/t_c=3.0$ has elapsed since the start of the simulation.  We recorded snapshots every $0.1 t_c$.  A few strongly self-gravitating simulations generated too many sink particles to accurately track on the computational resources available. These simulations were terminated before $t/t_c=3.0$, though they still generate more than five snapshots during the self-gravitating phase.

We supplement our simulation design with five more simulations at the fiducial values for the parameters given in Table \ref{tab:design}.  These simulations have identical physical conditions except for a different random seed that is used to generate the turbulent driving field.  The role of these fiducial simulations is to create many realizations of identical physical conditions that quantify the importance of random effects.

Because of limited computational resources, we rely on a root grid that is coarse with respect to the current state of the art.  Though not without precedent in recent studies \citep{padoan2012}, the $128^3$ root grid leads to a limited inertial range in our simulations: only a factor of 2 beyond the driving range (to $k\sim 16$; see Appendix \ref{app:resn}).  For $k>16$, the velocity power spectrum is artificially damped.  We minimize the influence of this damping by restricting fitted spectra to the inertial range. However, we note that results based on the analysis of turbulence may not be directly comparable to higher resolution simulations \citep{Kitsionas2009A&A...508..541K}.

We post-process the simulation checkpoints using the {\sc radmc-3d}\footnote{\url{http://www.ita.uni-heidelberg.de/~dullemond/software/radmc-3d/}} code, creating mock position-position-velocity data cubes of $^{13}\mathrm{CO}(J=1\to 0)$ from the simulation data sets. The cubes are constructed using the 128$^3$ root grid data. The level populations are solved using a large velocity gradient approximation \citep{shetty-xfac} with collisional and radiative transition probability adapted from the Leiden Atomic and Molecular Database \citep[LAMDA;][]{lamda}.  The {\sc radmc-3d} processing includes a microturbulence with a velocity dispersion of $0.1\mbox{ km s}^{-1}$ to avoid calculating radiative transfer at velocity resolutions smaller than the thermal line width and the channel width of observations.  We adopt a uniform abundance for $^{13}\mbox{CO}$ of $1.5\times 10^{-6}$ per H nucleus, appropriate for an active star forming region like Perseus \citep{pineda08}.  The simulated data sets have a channel width of $\delta v = 0.2\mbox{ km s}^{-1}$.

These observational data sets are designed to mimic the observations in the COMPLETE survey \citep{complete-data}.  We project the data onto sky coordinates based on a distance of 260 pc, the distance adopted in COMPLETE for the Perseus molecular cloud.   In addition to noiseless data, we also create a set of ``noisy'' data sets with Gaussian white noise data added to emulate thermal noise in real observational data.  For each data set, we set the noise level to 5\% of the peak intensity, corresponding to a signal-to-noise ratio of 20.  For our simulations, this translates into a typical noise level of 0.1 K on the $T_A^*$ scale, again consistent with the COMPLETE survey.

This analysis averages over the checkpoints in each simulation for comparisons using the distance metric.  In total there are $(2^5+5)$ simulations, with at most 10 checkpoints. Each of these is generated with and without observational noise, leading to an analysis suite of 740 simulated data sets.

\section{Statistical Comparison of Data Cubes}
\label{sec:stats}

Drawing from several statistical techniques presented in the literature,  we propose pseudo-distance metrics that measure the similarities between data sets.  These statistics will ideally measure differences in the physics (i.e., temperature, magnetic field, Mach number) in the data sets.  The ideal properties of a distance metric are described in \citet{Yer14}; we provide a summary of those properties here. Consider two datasets $I_1(\mathbf{x},v)$ and $I_2(\mathbf{x},v)$. In general, a pseudo-distance metric parameterizes the differences between two data sets with a scalar value.  The metric must satisfy two properties: it must be non-negative ($d(I_1,I_2)\in \mathbb{R}^{0+}$) and symmetric ($d(I_1,I_2)=d(I_2,I_1)$).   Synthetic observations created from the same physical processes should have distances near zero, $d(I_1,I_2)\approx 0$.  This should also hold true for observations of real ISM structures in similar evolutionary states.  Distance metrics should not be sensitive to spatial relationships between pixels, i.e., comparing the same dataset with a spatial offset should produce a distance of zero.  Similarly, the distance should not be sensitive to the velocity frame from which the dataset is viewed. However, the metrics should be sensitive to the spatial structure of their brightness and differences in physical scale.

A valid distance metric will be insensitive to the noise levels present in the dataset.
Ideally, $d(I_1,I_2)\sim 0$ if $I_1$ and $I_2$ map the same object and only differ in realizations of their noise. Practically, this requirement means that statistics should rely on high intensity levels within the data sets that are well above the noise levels.

A similar approach to this formalism is presented by \citet{adams94} and \citet{wiseman94}.  Where that work focuses on establishing the behaviour of the pseudo-distance metrics under transformation, here we focus on using these metrics to examine their sensitivity to the underlying changes in physics.  This approach is enabled by the significant improvements in numerical algorithms and computing speed in the past two decades.

Below, we describe several distance metrics developed based on existing statistical techniques presented in the literature. We also include the three distance metrics proposed by \citet{Yer14} as the sensitivity analysis (Sec.~\ref{sec:analysis}) is performed on a new simulation suite that explores a different region of parameter space with a different design. \citet{Yer14} tested different physical parameters (\drive, \mach, magnetic field strength, and temperature), and we have refined the values used for the parameters that overlap in the studies. Since the basis for this work comes from a variety of sources, we present a standardized notation in Table \ref{tab:notation}.  Note that many of the statistics rely on velocity moments of the data.  We use the notation that
\begin{eqnarray}
M_0(\mathbf{x}) &=& \sum_v I(\mathbf{x},v) \delta v, \\
M_1(\mathbf{x}) &=& M_0^{-1}\sum_v I(\mathbf{x},v) v \delta v, \mbox{ and} \\
M_2(\mathbf{x}) &=& M_0^{-1}\sum_v I(\mathbf{x},v) [v-M_1(\mathbf{x})]^2 \delta v,
\end{eqnarray}
for the zeroth, first, and second moments, respectfully. We use $\delta v$ to denote the width of a velocity resolution element (channel).

We have implemented these statistics in a {\sc Python} package called {\sc TurbuStat}, which we make freely available to the community\footnote{\url{http://turbustat.readthedocs.io}}.  The methodology and features of {\sc Turbustat} will be presented in a future paper (Koch et al.~in prep.). In addition to the distance metrics outlined below, the package also returns typical results for many of the statistics in the literature.  The comparison methods implemented in {\sc TurbuStat}  fall into three broad categories: the analysis of identified structures and objects, the analysis of the properties of turbulence, and the analysis of distributions.

\begin{table*}
\begin{center}
\caption{\label{tab:notation} Summary of Notation}
\begin{tabular}{c|l}
Symbol & Meaning \\
\hline
$\mathbf{x},v$ & Position and velocity in an observed data cube\\
$\mathcal{F}(\cdot)$ & Fourier transform operator \\
$f_i$ & The $i$th element in a set of basis functions \\
$I(\mathbf{x},v)$ & Intensity of observed radiation field in a data cube\\
$d_{\mathrm{Statistic}}$ & Distance metric for a given statistic \\
$N$ & Number of data in sums \\
$\mathcal{I}(\mathbf{k},k_v)$ & Three-dimensional Fourier transform of
intensity\\
$\mathbf{k}$ & Wavenumber corresponding to the spatial dimensions\\
$k_v$ & Wavenumber corresponding to the velocity dimension\\
$M_0(\mathbf{x})$ & Velocity-integrated zeroth moment of the data cube (integrated intensity) \\
$M_1(\mathbf{x})$ & Velocity-integrated first moment of the data cube \\
$M_2(\mathbf{x})$ & Velocity-integrated second moment of the data cube \\
$\mathcal{M}_0(\mathbf{k})$ & Fourier transform of the integrated intensity map, $\mathcal{F}[M_0(\mathbf{x})]$\\
$\mathcal{M}_1(\mathbf{k})$ & Fourier transform of the first moment map, $\mathcal{F}[M_1(\mathbf{x})]$\\
$\beta_i$& Linear model coefficients\\
$\sigma_i$ & Linear model uncertainties\\
$p(S)$ & Empirical probability density function (histogram) for a data $S$\\
$\tilde{S}$ & Standardized value of data $S$ with mean of 0 and standard deviation of 1\\
$H(p_i,p_j)$ & Hellinger distance between two empirical probability density functions \\
$(a_i, b_j)^{\mathrm{T}}$ & Vector formed by concatenating the elements of vectors $\mathbf{a}$ and $\mathbf{b}$ \\
$P_{1\mathrm{D}}$ & One-dimensional power spectrum \\
$P_{2\mathrm{D}}$ & Two-dimensional power spectrum \\
$\boldsymbol{\ell}$ & Scale or correlation function lag \\
$K(\mathbf{x},\ell)$ & Convolution kernel defined with scale $\ell$ \\
$\left|\left|y_1 - y_2\right|\right|$ & $L^2$ norm defined as $\left(\sum_i \left( y_{i,1} - y_{i, 2} \right)^2 \right)^{1/2}$
\end{tabular}
\end{center}
\end{table*}

\subsection{Structure Analysis} 
\label{sub:structure_identification}

These methods analyze the properties of emission in PPV data sets and include both object cataloging methods and generalized structural analysis.  Cataloging methods can be thought of as a ``parametric'' view of the molecular ISM.  Cataloging focuses on dividing the ISM into objects and using the distributions of objects to make inferences about the star formation process.  Objects can include clumps \citep{clumpfind}, cores \citep{enoch06}, and the stellar initial mass function \citep{offner-imf}.  Other approaches can be thought of as ``non-parametric'' and do not attempt to divide the ISM into a population, relying instead on characterizing the emission with image processing approaches.  We do not directly address object cataloging in this paper, focusing instead on statistical descriptions of the data. We note that dendrograms (\S\ref{sub:dendrograms}) are a cataloging method, but the statistics we use are not based on the properties of a single set of objects or regions.

\subsubsection{Genus Statistics} 
\label{sub:genus_statistics}

The use of the genus statistic on column density maps was introduced qualitatively in \citet{Kowal2007} and quantitatively in \citet{Chepurnov2008}. In this work, we compute the statistic using the integrated intensity (zeroth moment), $M_0(\mathbf{x})$. The genus statistic provides a measure of a region's topology. At a given intensity value $I$, the genus value is defined as,
\begin{eqnarray}
  \label{eqn:genus}
  G(I_{0}) \equiv N_{>I_{0}} - N_{<I_{0}},
\end{eqnarray}
where $N_{>I_{0}}$ is the number of isolated regions in the data cube above a threshold intensity $I_0$ and $N_{<I_{0}}$ is the number of regions below the threshold. The genus curve is constructed by varying the threshold $I_0$ over the range of intensities in the datacube.

To compare the shapes of genus curves from two $M_0(\mathbf{x})$ maps, we standardize the curves such that they have a mean of 0 and a standard deviation of 1. This step is necessary to place the two curves on the same scale for arbitrary data sets. We then define the distance between the two curves as the $L^2$ norm,
\begin{equation}
  \label{eq:genus_distance}
  d_{\mathrm{genus}} = \left|\left|\frac{G_{1}\left(I_{0,i}\right)}{A_1} - \frac{G_{2}\left(I_{0,i}\right)}{A_2}\right|\right|
\end{equation}
where the $I_{0,i}$ are the standardized intensity values compared over a set of $N$ intensity levels, and $A_1$ and $A_2$ are the areas over which the genus values were computed.  This comparison method differs from the analysis presented in \citet{Chepurnov2008}, where a fifth-order polynomial was fit to the curves to estimate the $x$-intercept. The location of the $x$-intercept relative to a genus curve of a Gaussian noise distribution yields how {\it clumpy} the topology is.  We note that our simulations do not suffer from large-scale velocity gradients nor a lack of compact features, both of which can bias the genus curve \citep{Chepurnov2008}.

\subsubsection{Dendrograms} 
\label{sub:dendrograms}

\citet{dendrograms} and \citet{dendrograms-nature} introduced the use of dendrograms for describing hierarchical structure in molecular clouds. Dendrograms are able to characterize the fractal structure of the ISM and thereby estimate key physical properties.  A dendrogram can be visualized as a tree representing a Reeb graph \citep[see][]{dendrograms}. We refer to the peaks of the trees as {\it leaves} and the connected components as {\it branches}.  Each leaf is a local maximum of the intensity distribution $I(\mathbf{x},v)$ and defined to be brighter than the highest level isosurface containing two local maxima by an interval $\delta_I$. Following the analysis developed in \citet{burkhart-dendrograms}, we propose two distance metrics based on the dendrogram method.

The first proposed statistic is based on the relation between $n$, the total number of structures including leaves and branches, and the value of $\delta_I$. \citet{burkhart-dendrograms} found that increasing $\delta_I$ past the mean value in the data created a power-law relationship between $n$ and $\delta_I$. We extract this power-law by considering the number of structures at $\delta_I$ values past the mean.  We limit $\delta_I$  to avoid removing the fundamental tree structure; increasing $\delta_I$ too much leads to no structures in the tree.  Performing a dendrogram analysis for two data cubes, we measure $n_1(\delta_I)$ and $n_2(\delta_I)$ for a range of $\delta_I$.  To  characterize the similarity between these relationships, we fit a linear model to the log-transformed data of $N_1$ and $N_2$:
\begin{equation}
    \label{eqn:linmodel}
    \log N = \beta_0 + \beta_1 z + \beta_2 \delta_I + \beta_3 \delta_I z,
\end{equation}
where $N=n=(n_1,n_2)^\mathrm{T}$ is a vector formed by concatenating the values for the two data sets and $z\in\{0,1\}$ is a dummy variable that tracks which data set is being fit (i.e., $z=0$ for a datum in $n_1$ and $z=1$ for a datum in $n_2$). The last term describes the interaction between the two fits.   If the transforms have identical slopes, $\beta_3=0$.  If $\beta_3$ is small with respect to its uncertainty $\sigma_3$, the slopes of the fits are indistinguishable and the $n(\delta_I)$ curves have similar behaviour.  However, if $\beta_3$ becomes large with respect $\sigma_3$, the two curves are different. From this, we define a distance metric based on the $t$-statistic of the interaction term:
\begin{equation}
  \label{eq:dendro_num_dist}
  d_{\mathrm{DendNum}} = |\beta_3| / \sigma_3.
\end{equation}
This formulation as a linear model provides a measure of the significance of this interaction and thus should serve as a good distance metric.

The second proposed dendrogram statistic in \citet{burkhart-dendrograms} translates the tree structure of the dendrogram into an empirical distribution of structures (leaves and branches), $p(I)$.
 The intensity of a structure is defined as the maximum intensity contained within the contour that defines that structure. This method provides an alternate view of analyzing the dendrogram's features.  Given two dendrograms with structures (branches) and different intensity levels $I$, we normalize these distributions of intensity to have mean 0 and standard deviation 1, $p_1(\tilde{I}),p_2(\tilde{I})$.   We then generate a distance metric between the two normalized histograms based on a Hellinger distance \citep{Huber-RobustStatistics}:
\begin{equation}
  \label{eq:hellinger_dist}
  H(p_1,p_2) = \frac{1}{\sqrt{2}}\left\{\sum_{\tilde{I}} \left[ \sqrt{p_1(\tilde{I})} - \sqrt{p_{2}(\tilde{I})} \right]^2\right\}^{1/2}.
\end{equation}
Because $p_1$ and $p_2$ are empirical PDFs and are thus normalized to have a sum of 1, the Hellinger distance is also normalized.  Given two data sets, we compute dendrograms for a range of  $\delta_I$ values used for the previous statistic.  For each pair of dendrograms we calculate the Hellinger distances between their respective $p(\tilde{I}_{\mathrm{struct}})$.  Dendrograms that have few branches ($<50$), caused by choosing large $\delta$ values, are discarded from the comparison.  The distance between two data sets is then given as the average Hellinger distance over the values of $\delta_I$:

\begin{equation}
  \label{eq:dendro_hist_dist}
  d_{\mathrm{Dend Hist}} = \left[\sum H(p_{1,\delta_I},p_{2,\delta_I})\right]/N_\delta
\end{equation}
where $N_\delta$ is the number of $\delta_I$ levels compared. The appropriate value of $\delta I(\mathbf{x},v)$ for a given dataset is difficult to determine a priori. Thus, averaging over a range of values minimizes bias for the choice of this parameter.

To compute the dendrograms, we use the {\sc Python} package {\sc Astrodendro}\footnote{\url{http://dendrograms.org/}}. This implementation has three free parameters: the minimum intensity value to be used to construct the dendrogram, the minimum number of pixels a leaf must contain, and the minimum height of a branch (i.e., the minimum difference in intensity between the hierarchical levels, $\delta I(\mathbf{x},v)$). For the set of simulated noiseless data cubes used, we use a minimum intensity value of 0.01 K, which avoids numerical artifacts, and a minimum pixel limit of 80 pixels corresponding to the highest intensity leaves clearly discernible by-eye. When running comparisons with added noise in the data cubes (see \S\ref{sub:calculating_distances}), the minimum intensity value was set to twice the estimated noise level.

For each data cube, we compute the dendrogram and prune the corresponding tree by varying the $\delta I(\mathbf{x},v)$ parameter. We vary this parameter using 100 logarithmic steps between $10^{-2.5}$~K to $10^{0.5}$~K. We verified by visual inspection that this range captures the entirety of the variations for all data cubes in the simulation set (i.e., the tree is reduced to a single leaf). It should be noted that this range is dependent on the specific data cubes in use and does not describe a universal range for all comparisons.

Since dendrograms can be extended to multiple dimensions, we note that these statistics are valid for 2D and 3D data sets. While in our analysis, we compute the dendrograms of full data cubes, these comparisons apply equally to the integrated intensity images making these useful observational diagnostics.


\subsection{Properties of Turbulence} 
\label{sub:properties_of_turbulence}

The properties of interstellar turbulence offer several avenues for comparing observations to simulations.  Turbulence is ubiquitous in star forming clouds \citep[e.g.,][]{larson81,heyer04} and the physics that characterizes the turbulent flow has close connections to the star formation process \citep{padoan02,Federrath08,kdm12}.  The theory of astrophysical turbulence \citep[e.g.,][]{gs-mhd1,gs-mhd2,compressible-mhd}, while complex, can be related to observational quantities \citep[e.g.,][]{vca,lp04,lp06}, focusing on properties such as the turbulent power spectrum, intermittency, and anisotropy.  The distance measures developed here are based on methods that have been shown to track the signatures of turbulent flow in the observational domain.

For all cases in which a power-spectrum is fit, we limit the fitting to the inertial range in our simulations. In Appendix \ref{app:resn}, we show that the inertial range for our $128^3$ is between $k\sim 5$ and $k\sim 15$.

\subsubsection{Modified Velocity Centroids} 
\label{sub:modified_velocity_centroids}

The Modified Velocity Centroid (MVC) method was proposed by \citet{Lazarian2003} to provide a better statistical measure of velocity-density correlations primarily using velocity centroids. For a spectral line data cube with $N_v$ velocity indices, the MVC may be calculated in the Fourier-domain, as described in Section 4 of \citet{Lazarian2003}, from which a power spectrum can be extracted:

\begin{equation}
    \label{eq:mvc_fourier}
    P_{\mathrm{MVC}}(\mathbf{k}) = |\mathcal{M}_0(\mathbf{k}) \mathcal{M}_1(\mathbf{k})]|^2 - \langle M_2(\mathbf{x}) \rangle_{\mathbf{x}}|\mathcal{M}_0(\mathbf{k})|^2.
\end{equation}
The first term is the Fourier transform of the unnormalized velocity centroid. $P_{\mathrm{MVC}}(\mathbf{k})$ is a two-dimensional power-spectrum from which a one-dimensional power-spectrum is attained by azimuthal averaging.

The one-dimensional power spectra follow a power-law relation, as is expected for a measure of turbulent motion. As noted by \citet{Lazarian2003}, this method can be used when the turbulence does not follow a power-law relation, such as when self-gravity dominates over small scales. Such deviations are difficult to robustly compare, particularly when comparing low-resolution data cubes. Again, we fit a linear model, using log quantities of the MVC transform for two data sets but only over the scales that follow a power-law:
\begin{equation}
    \label{eqn:mvc_plaw}
    \log P_{1\mathrm{D}} = \beta_0 + \beta_1 z + \beta_2 \log k + \beta_3 z \log k
\end{equation}
The $t$-statistic of the interaction term $\beta_3$ is the measure of distance:
\begin{equation}
    \label{eqn:mvc_distance}
    d_{\mathrm{MVC}} = \frac{|\beta_3|}{\sigma_3}
\end{equation}
This distance should be sensitive to differences in the turbulence, based on the slope of the power spectra. The results in Figure 1 of \citet{Lazarian2003} show that MVC extends the power-law to smaller scales than methods using the standard centroid (i.e., spectral analysis of $\mathcal{M}_1$ alone), indicating its sensitivity to velocity-density correlations at small scales. The low resolution of the data cubes used here may hinder our ability to probe such small physical scales.

\subsubsection{Spatial Power Spectrum} 
\label{sub:power_spectrum}

The Spatial Power Spectrum (SPS) is computed from the Fourier transform of the two-point autocorrelation function using the velocity-integrated intensity:
\begin{equation}
  P(k) = \sum_{|\mathbf{k}|=k} \left|\mathcal{M}_0(\mathbf{k})\right|^2.
   \label{eq:pow_spec}
\end{equation}
Previous studies by \citet{burkhart-bispectrum} and \citet{stan01-sps} have shown the slope of the SPS to be sensitive to properties of turbulence. A one-dimensional power spectrum is constructed from Equation \ref{eq:pow_spec} by radially averaging over the surface. To compare two power spectra, we fit a similar linear model as was introduced in \S\ref{sub:dendrograms} between the logs of the power-spectra and the scales. This gives the same form as Equation \ref{eqn:mvc_plaw}. The distance is again defined as the $t$-statistic of the interaction term:
\begin{equation}
    \label{eqn:sps_distance}
    d_{\mathrm{SPS}} = \frac{|\beta_3|}{\sigma_3}
\end{equation}


\subsubsection{Bispectrum and Bicoherence} 
\label{sub:bicoherence}

The bispectrum is the Fourier transform of the three-point correlation function and represents a complex extension of the power spectrum:
\begin{equation}
  \label{eq:bispectrum}
  B(k_1, k_2) = \sum_{|\mathbf{k}_1|=k_1} \sum_{|\mathbf{k}_2|=k_2} \mathcal{M}_0(\mathbf{k}_1) \cdot \mathcal{M}_0(\mathbf{k}_2) \cdot \mathcal{M}_0^{\star}(\mathbf{k}_1+\mathbf{k}_2)
\end{equation}
\citet{burkhart-bispectrum} applied the bispectrum to the {\sc Hi} column density data of  the Small Magellanic Cloud (SMC).  They use a visual comparison between the bispectrum results for SMC data and turbulence simulations, supporting the applicability of their simulation set.  Quantifying this similarity is more difficult than in the case of the power spectrum, which can be reduced to a one-dimensional representation. As the bispectrum is complex-valued, it is necessary to retain the two-dimensional plane to analyze the statistic.  Since measuring distances between complex numbers can be poorly defined, we use a real-valued and normalized form of the bispectrum called the bicoherence \citep{Bicoherence-Hagihira}:
\begin{equation}
  \label{eq:bicoh}
  b(k_1, k_2) = \frac{|B(k_1, k_2)|}{\sum_{\mathbf{k}_1, \mathbf{k}_2} |\mathcal{M}_0(\mathbf{k}_1) \cdot \mathcal{M}_0(\mathbf{k}_2) \cdot \mathcal{M}_0^{\star}(\mathbf{k}_1+\mathbf{k}_2)|}.
\end{equation}
The denominator represents the bispectrum with phases set to zero. A bicoherence of 0 indicates completely random phases, while a value of 1 shows complete phase coupling. Since the expression in Equation \ref{eq:bicoh} is both normalized and real, the bicoherence is significantly easier to use when defining a measure of distance. To calculate the bicoherence, we randomly sample the direction of the vectors $\mathbf{k}_1$ and $\mathbf{k}_2$ at all possible magnitudes (up to half of the image size in each direction). The samples are normalized by the number of samples taken at each pixel. We then define the distance metric as the absolute difference between the average of the bicoherence surfaces calculated from the integrated intensity images:
\begin{equation}
  \label{eq:bispec_distance}
    d_{\mathrm{Bispec}} = \left|\overline{b_{1}} - \overline{b_{2}}\right|
\end{equation}
Since the metric is computed only between two values, this form is equivalent to the $L^2$ norm. This form effectively parameterizes the difference of the overall phase correlation within the integrated intensity images; similar distance measures have been used for image comparisons \citep{farid_bispectrum}.  Since the slope of the SPS has been shown to reflect the power spectrum of the underlying fluid flow, it is natural to expect that those properties and more are somehow captured within the bispectrum. What remains unclear is how it relates the turbulence, since the statistic does not have a clear slope-like property to compare.


\subsubsection{Velocity Coordinate Spectrum and Velocity Channel Analysis} 
\label{sub:velocity_coordinate_spectrum_and_analysis}

The Velocity Coordinate Spectrum (VCS) and Velocity Channel Analysis (VCA) are related methods that rely on different manipulations of the full three-dimensional Fourier transform of a PPV data cube. The works of \citet{vca,lp04,lp06} introduced these methods and quantitatively related their values to the properties of the turbulent flow.  We largely follow the approach of \citet{chepurnov09} and \citet{chepurnov10}. For a PPV data cube $I(\mathbf{x},v)$, we calculate the three-dimensional power spectrum from its Fourier transform:
\begin{equation}
  \label{eq:vca_vcs_pspec3D}
  P_{\mathrm{3D}}(\mathbf{k}, k_v) = \sum \mathcal{I}(\mathbf{k},k_v) \cdot \mathcal{I}^{\star}(\mathbf{k},k_v)
\end{equation}
To calculate the VCS, we average Equation \ref{eq:vca_vcs_pspec3D} over the spatial wavenumbers:
\begin{equation}
  \label{eq:vcs_func}
  P_1(k_v) = N^{-1}_{\mathbf{k}} \sum_{\mathbf{k}} P_{\mathrm{3D}}(\mathbf{k}, k_v).
\end{equation}
This gives a one-dimensional power-spectrum. When the data are noiseless, the VCS shows two distinct power-law relations, one at larger scales where variations in the velocity field are likely to dominate, and the second at smaller scales where both the density and velocity may dominate, depending on the properties of the fields \citep{chepurnov09}. To account for the possibility of two regimes, we fit a linear model to the power spectrum that also fits the transition between the regions. This technique is known as segmented linear regression, and we follow the method presented by \citet{muggeo2003estimating} where the breakpoint is estimated by iteratively minimizing a term in the model corresponding to the difference between the line segments at the break point. The method relies on the likelihood being well-approximated by a first order Taylor expansion in the neighbourhood around the break point, which should hold in our application. Unlike the linear models for the statistics presented above, we fit separate linear models to each VCS curve.  We found that attempting to include the interaction terms for the differences in the power-laws caused the iterative minimization on the breakpoint to become unstable.  Although this method is introduced to account for two different physical regimes, the limited resolution of our simulation sets will likely lead to the second power law fitting to the damped region rather than the physical transition.

Each fit to the VCS curve returns the slope for each region, along with their standard respective errors. These can be combined to give the $t$-statistic of the interaction terms, equivalent to those discussed above, in both of the regions. The total distance is the sum of the $t$-statistics in each region:
\begin{equation}
  \label{eq:vcs_distance}
  d_{\mathrm{VCS}} = t_{\mathrm{large-scale}} + t_{\mathrm{small-scale}}
\end{equation}
Since one term may dominate over the other, we also consider the individual terms separately in our analysis (\S\ref{sec:analysis}).

Information about the turbulence at small scales is destroyed even with the modest addition of Gaussian noise and cannot be readily recovered from real observational data, given current noise limits.  In this case, we follow the same fitting procedure, however, the power-law on the smallest scales is flat, simply representing the noise. Thus, only the power-law at larger scales is relevant to the turbulent properties in the region, and we use this $t$-statistic by itself for a measure of distance:
\begin{equation}
  \label{eq:vcs_distance_noise}
  d_{\mathrm{VCS}} = t_{\mathrm{large-scale}}
\end{equation}
In all cases, this measure is dominated by the large-scale portion, but is likely influenced by whatever remaining portion of the small-scale region is above the noise level.  The fitted slope in the presence of noise for the large-scale region will then be steeper than in the noiseless case. The extent of this effect is dependent on the noise level in the data cube and the range of scales the small-scale region affects.

VCA is calculated by integrating over the velocity channels to yield a two-dimensional power-spectrum:
\begin{equation}
  \label{eq:vca_func}
  P_2(\mathbf{k}) = \sum_{k_v} P_{\mathrm{3D}}(\mathbf{k}, k_v)
\end{equation}
Like the SPS (\S\ref{sub:power_spectrum}), we radially average over the two-dimensional surface to yield a one-dimensional power spectrum. We then characterize the slope of the power-law. Unlike the VCS, the VCA can be fit to a single power-law across a wide range of scales. We note that in our simulations, we did not encounter significant distortion from finite resolution, as has been shown to occur in \citet{chepurnov09}. As such, we did not change the velocity slice-thickness, keeping the original velocity resolution in the data (\S\ref{sec:sims}).  With an improved simulation set that tracks turbulent flow better, including changing slice thickness may improve the discriminatory powers of the VCA.

To model VCA, we use the a linear model like that in Equation \ref{eqn:mvc_plaw} and define the distance metric to be the $t$-statistic of the interaction term:
\begin{equation}
  \label{eq:vca_distance}
  d_{\mathrm{VCA}} = |\beta_3|/ \sigma_3
\end{equation}

We found that, for data cubes containing noise, the scales dominated by noise (i.e., the smallest scales, as in VCS) varied between data cubes. In this case, we limited the scales fit to those that followed the power-law and discarded any smaller scales. We found that a reliable measure of the transition to noisy scales could be found by adopting the same segmented linear regression as was used for VCS.  A potential improvement for this method would be to include the noise model directly into the formulation, provided the noise characteristics are well known.

Our simulation set is of limited resolution and thus is equally affected by shot noise, which  \citet{chepurnov09} have shown will cause positive deviations at large $\mathbf{k}$ and $k_v$.


\subsubsection{Delta Variance} 
\label{sub:delta_variance}

The delta-variance method was introduced by \citet{stutzki98} to quantify the fractal nature within molecular clouds.  Structure resulting from fractional Brownian motion follows a power law power spectrum with a random distribution of phases, which was shown to be consistent with the Polaris Flare \citep{stutzki98,bensch01-delvar}. Qualitatively, the delta-variance shares many features with the wavelet transform (\S\ref{sub:wavelet_transform}).  This method can be computed in both the spatial and Fourier domains on integrated intensity or velocity centroid images \citep{bensch01-delvar}.  We calculate the delta-variance in the Fourier domain using the ``improved'' method introduced in \citet{oss08I-delvar,oss08II-delvar}.   This requires defining a weight map $W(\mathbf{x})$ at each position.  We define weights as the inverse variance of the map; in the noiseless case, this is equivalent to the weighting by the number of channels integrated, as was used in \citet{oss08II-delvar}.

Following \citet{oss08II-delvar}, we convolve the integrated intensity image and its weight array with a Mexican hat wavelet: $K(\mathbf{x},\ell)$ with a diameter ratio of 1.5 over a range of scales $\ell$. To account for finite sized data and the presence of noise in the data, the convolution kernel is broken into its constituent Gaussian components, $K_{\mathrm{core}}$ and $K_{\mathrm{ann}}$, and convolved separately.  These are aggregated into a single, edge-corrected, convolved map:
\begin{eqnarray}
  \label{eq:del_var_convimg}
  F(\mathbf{x},\ell) & \equiv &
  \frac{K_{\mathrm{core}}(\mathbf{x},\ell) * M(\mathbf{x})}
  {{K_{\mathrm{core}}(\mathbf{x},\ell) * W(\mathbf{x})}} -\nonumber\\
  &&\frac{K_{\mathrm{ann}}(\mathbf{x},\ell) * M(\mathbf{x})}{{K_{\mathrm{ann}}(\mathbf{x},\ell) * W(\mathbf{x})}},
\end{eqnarray}
where $M(\mathbf{x})$ is either the zeroth or first moments maps, and $W(\mathbf{x})$ is the inverse-square of its respective uncertainty.

The delta-variance is then:
\begin{equation}
  \label{eq:del_var}
  \sigma_{\Delta}^2 (\ell) = \frac{\sum_{\mathbf{x}} \left( F_{\ell}(\mathbf{x}) - \left< F_{\ell} \right> \right)^2 W_{\ell}(\mathbf{x})}{\sum_{\mathbf{x}} W_{\ell}(\mathbf{x})}.
\end{equation}

The expected delta-variance spectra will not follow a single power-law over all scales due to beam smoothing (on small scales) and edge effects (on large scales). We define two metrics to capture the delta-variance behaviour: the power-law slope within set spatial limits, and the $L^2$ norm between the two normalized curves.

We use a linear model as defined in Equation \ref{eqn:mvc_plaw} and fit on scales between $1/10$ and $1/5$ of the resolution of the image, similar to the range used in previous work \citep{Bertram2015MNRAS.451..196B}\footnote{\citet{Bertram2015MNRAS.451..196B} use $1/10$ to $1/4$ of the box size. We find large deviations at scales of $1/4$ the box due to changes in the driving scale.}. We find that this fitting region correctly captures the power-law behaviour for all data cubes used here, though the extent of the power-law component extends to smaller and larger regions depending on the driving scale. The distance between the slopes is then:
\begin{equation}
  \label{eq:delvar_slopedistance}
  d_{\Delta-{\rm slope}} = |\beta_3| / \sigma_3.
\end{equation}

To attempt to capture changes outside of this defined fitting range, we introduce a non-parametric distance measure based on the $L^2$ norm between the two delta-variance curves:
\begin{equation}
  \label{eq:delvar_distance}
  d_{\Delta} = \left|\left|\frac{\sigma_{\Delta,1}^2 (\ell)}{\sum_i \sigma_{\Delta,1}^2 (\ell)} - \frac{\sigma_{\Delta,2}^2 (\ell)}{\sum_i \sigma_{\Delta,2}^2 (\ell)}\right|\right|
\end{equation}
Each curve is normalized such that the fraction of variance is the quantity compared at each scale, which would otherwise only parameterize the differences in the total variance in each data set. This form is able to account for general shape differences between the curves that may arise due to differing physical parameters. For example, the delta-variance spectrum curve changes slope on scales where large-scale velocity gradients dominate \citep{oss08II-delvar}.  Due to the effect of beam smearing on small scales \citep{bensch01-delvar}, this comparison metric is valid only if the angular resolution is the same between the data being compared.


\subsubsection{Wavelet Transform} 
\label{sub:wavelet_transform}

\citet{gill-wavelet} proposed the first use of a wavelet transform to spectral line data of molecular clouds. Their method involves convolving a column density or integrated intensity image with a kernel $K(\ell)$, while changing the scale $\ell$ of the kernel over a chosen range. At each scale $\ell$, the transform value is the average value of the positive regions of the convolved image, $T(\ell) = \langle \mbox{max} \{ M_0 * K(\ell),0\}\rangle_\mathbf{x}$. This yields an estimate of the amount of structure at each scale tested. \citet{gill-wavelet} showed that a significant portion of the transform followed a single power-law form, the slope of which they interpreted as a measure of dimensionality of the space they were testing. This dimensionality is in turn related to the turbulent properties of the region.

We follow the method of \citet{gill-wavelet} by convolving the integrated intensity image of the data cubes with a normalized Mexican Hat kernel using scales up to half of the data size. The deviations from a power-law on small and large scales are similar to those in the delta-variance (\S\ref{sub:delta_variance}) and we use the same fitting range of $1/10$ to $1/4$ of the spatial dimensions. We fit a linear model in this range between the log-values of the transform ($\log T$) and the scale $a$ as in \S\ref{sub:dendrograms}:
\begin{equation}
    \label{eqn:wavelet_plaw}
    \log T_g = \beta_0 + \beta_1 z + \beta_2 \log a + \beta_3 z \log a.
\end{equation}
\noindent
As in \S\ref{sub:dendrograms}, $\beta_3$ represents the interaction term between the slope of the power-laws, so we define the distance between the transforms to be its $t$-statistic:
\begin{equation}
    \label{eqn:wavelet_distance}
    d_{\mathrm{wavelet}} = |\beta_3| / \sigma_3
\end{equation}

There is significant similarity between the delta-variance (\S\ref{sub:delta_variance}) and the wavelet transform. Apart from the weighting scheme used in the delta-variance \citep{oss08I-delvar}, these techniques differ only in the statistic used to construct the 1D transform from the convolved maps: variance for delta-variance and the mean of the positive response here.


\subsubsection{Principal Component Analysis Eigenvalues} 
\label{sub:pca}

Principal Component Analysis (PCA) is a general technique which decomposes a covariance matrix into linear orthogonal components with vectors that maximize the variance. In this way, the dimensionality of a given data set can be reduced to a minimal set of components that capture the majority of the structure in the data (i.e., variance). The variance contained in each of these components is the eigenvalue of the decomposition.

The application of Principal Component Analysis to spectral-line data cubes was first proposed by \citet{heyer-pca} and has further been applied to describe the turbulent structure function \citep{brunt-pca1,brunt-pca2}. This method reconstructs the structure function by using the eigenvectors to construct a set of eigen-images (spatial structure) and eigen-spectra (spectral structure). The structure function is recovered by finding the spatial and spectra sizes contained from these sets. Given the limited resolution in our set of data cubes, we refrain from calculating the structure function and instead adopt a simplified comparison using the covariance matrix alone.

We follow the method presented in \citet{Yer14}, which describes differences in the proportion of variance in each principal component, and present it again here for completeness.  We note that while this differs from the Brunt \& Heyer work described above, the approach of analyzing the proportion of variance is used universally [e.g., in economics \citep{pca-econ} and genetics \citep{pca-genetics}] and we adopt it because it readily integrates into the distance metric framework under which we operate.

We construct a covariance matrix for each data cube by comparing the velocity channels:
\begin{equation}
  \label{eqn:pca_cov}
  C\left( v,v^\prime\right) =\sum _{\mathbf{x}}\left[I\left(\mathbf{x},v\right) - \overline{I\left(\mathbf{x},v\right)}\right] \left[I\left( \mathbf{x},v^\prime\right) - \overline{I\left( \mathbf{x},v^\prime\right)} \right],
\end{equation}
where each channel is centered by subtracting the mean \citep{Ivezic2014sdmm.book.....I}. A PCA decomposition is performed on this covariance matrix, yielding eigenvalues $\lambda$ and eigenvectors $\mathbf{u}$ which satisfy the matrix equation $C\mathbf{u}=\lambda\mathbf{u}$. The sum of all of the eigenvalues is the total variance contained in the data cube. We define a normalized set of eigenvalues, $\lambda_i'=\lambda_i / \sum_i \lambda_i$, such that each $\lambda_i'$ is the proportion of the total variance described in the $i$th principal component.

After performing the decomposition on the respective covariance matrices for two different data sets, we define a distance metric as the $L^2$ norm between the sets of normalized eigenvalues:
\begin{equation}
  \label{eq:pca_dist}
  d_{\mathrm{PCA}} = \left|\left|\lambda_{1}' - \lambda_{2}'\right|\right|.
\end{equation}
The sums are performed over the vector of eigenvalues. The distance metric measures the difference in the proportion of variance contained in the $i$th PC between the two data sets. As in \citet{Yer14}, we truncate the sum to first 50 terms. This is a safe approximation since the vast majority of the variance is accounted for by the first few principal components.

\subsubsection{Spectral Correlation Function} 
\label{sub:scf}

The Spectral Correlation Function (SCF) was proposed by \citet{scf} as a statistic to describe similarities in the spatial and velocity dimensions of a spectral-line data cube. Further studies have shown the SCF has the ability to discriminate properties of turbulence when comparing spectral-line data cubes \citep{padoan-scf,Yer14}. We follow the method introduced by \citet{Yer14}, where we compute the SCF as a normalized root-mean-square difference between spectra separated by a spatial offset $\boldsymbol{\ell}$:
\begin{equation}
  \label{eq:scf}
  S(\boldsymbol{\ell}) = 1 - \left\langle \sqrt{\frac{\sum_v
    |I(\mathbf{x},v)-I(\mathbf{x}+\boldsymbol{\ell},v)|^2}{\sum_v
    |I(\mathbf{x},v)|^2+\sum_v |I(\mathbf{x}+\boldsymbol{\ell},v)|^2}}\right\rangle_{\mathbf{x}}.
\end{equation}
The vector $\mathbf{x}$ corresponds to the two spatial dimensions in the data cube. For each data cube, a two-dimensional surface is returned by the statistic. We define the distance metric as the $L^2$ norm between the two surfaces, weighted by the distance from the center of the surfaces:
\begin{equation}
  \label{eq:scf_distance}
  d_{\mathrm{SCF}} = \left( \frac{\sum_{\boldsymbol{\ell}}[S_1(\boldsymbol{\ell})-S_2(\boldsymbol{\ell})]^2/|\boldsymbol{\ell}|}
{\sum_{\boldsymbol{\ell}} 1/|\boldsymbol{\ell}|}\right)^{1/2}.
\end{equation}
This choice of weighting decreases the importance of correlations over larger offsets, and we found that this improved the sensitivity of the statistic in our results (\S\ref{sec:analysis}). As in \citet{Yer14} we calculate the SCF for a 23-pixel squared patch.


\subsection{Analysis of Distributions} 
\label{sub:analysis_of_distributions}

The final category of statistical analyses study the properties of the distribution of values within observed data sets.  These can include just the distributions of the values in the map, like the commonly used column density PDF, or the distribution shape of statistics computed with some spatial information included.

\subsubsection{Intensity Probability Density Function} 
\label{sub:pdf}

The probability density function (PDF) of column density maps (observations) and density cubes (simulations) has been extensively studied by \citet{Kowal2007}. The PDF is an attractive tool due to the ease with which it can be calculated and its flexibility to apply to data of any dimension.  We construct PDFs from the integrated intensity images: $p(M_0)$. In the absence of noise, and with periodic boundary conditions, the structure of these PDFs is well-set. We fit the PDFs with the commonly adopted log-normal form \citep[e.g.,][]{federrath10}:
\begin{equation}
  \label{eq:pdf_lognormal}
  p(M_0) = (2\pi w^2)^{-1} \exp{\left[-\frac{(\log I / \bar{I} - w^2/2)^2}{2\pi w^2}\right]},
\end{equation}
where $I$ are values of the integrated intensity and $w$ is the distribution width. Since we stop our simulations before significant star formation occurs, high-intensity power-law tails are not produced in significance over a log-normal. We fit the distributions using a maximum likelihood estimate (MLE), and use the estimated standard errors from this procedure. An alternate MCMC fitting procedure was also explored, however the MLE converged to the same parameter values and the standard errors closely matched the equivalent 1-$\sigma$ range from the MCMC posterior distributions.

We define the distance between the log-normal fits to be the $t$-statistic of the distribution widths:
\begin{equation}
  \label{eq:pdf_distance}
  d_{\rm PDF} = \frac{|w_1 - w_2|}{\sqrt{\sigma_{w_1}^2 + \sigma_{w_2}^2}}.
\end{equation}
Unlike the other methods described here, which have clear applications to observational data, adopting this log-normal form will not allow for a simple extension to observational comparisons. \citet{Lombardi2015} show that the log-normal component of observed column-density PDFs is susceptible to both noise and boundary conditions. However, since our formalism is not model dependent, the PDF distance may be re-defined as the difference between the power-law slopes or a related parameter. Since power-law tails are not readily seen in our mock observational data, we defer this comparison to future work.


\subsubsection{Higher Order Statistical Moments} 
\label{sub:higher_order_statistical_moments}

Higher order statistical moments (namely skewness and kurtosis) have been previously used as a comparison tool between simulated and observed data sets \citep{Kowal2007,burkhart-bispectrum}.   Following the analysis of \citet{burkhart-bispectrum}, we create maps of the higher order moments by calculating the value of each moment within a circular aperture around each position in the integrated intensity image.  The empirical PDF of the moment values is then sensitive to the physical conditions underlying the emission.
For each position, the moments are calculated considering only those data within a radius $r$, i.e., $|\mathbf{x}'-\mathbf{x}|\le r$.  Within each circular aperture, we calculate the mean $\mu_r(\mathbf{x})$ and standard deviation $\sigma_r(\mathbf{x})$ around each position, along with the skewness
\begin{equation}
\gamma_{3,r}(\mathbf{x}) \equiv \frac{\sum_{|\mathbf{x}'-\mathbf{x}|\le r} w(\mathbf{x}')\left[\frac{M_0(\mathbf{x}')-\mu_r(\mathbf{x})}{\sigma_r(\mathbf{x})}\right]^3}{\sum_{|\mathbf{x}'-\mathbf{x}|\le r} w(\mathbf{x}')} ,
\end{equation}
and the kurtosis
\begin{equation}
\gamma_{4,r}(\mathbf{x}) \equiv \frac{\sum_{|\mathbf{x}'-\mathbf{x}|\le r} w(\mathbf{x}') \left[\frac{M_0(\mathbf{x}')-\mu_r(\mathbf{x})}{\sigma_r(\mathbf{x})}\right]^4}{\sum_{|\mathbf{x}'-\mathbf{x}|\le r} w(\mathbf{x}')} - 3.
\end{equation}
These are weighted quantities, where we define weights using the inverse squared integrated intensity uncertainty, $w(\mathbf{x}) = \left[\sigma_{M_0}(\mathbf{x})\right]^{-2}$. As is noted by \citet{burkhart-bispectrum}, the mean  $\mu_r(\mathbf{x})$ is not a physically useful value for comparing turbulent properties and the variance depends on the scaling of the data sets. Thus, we limit our comparisons to the skewness and kurtosis. By using the weighted forms of the skewness and kurtosis, we significantly down-weight regions dominated by noise.

We choose a radius of $r=5$ pixels for our kernel. While \citet{burkhart-bispectrum} makes a statistical argument for using a radius of 35 pixels based on the standard errors, such a size is too large for the spatial size of our simulations and acts to blur out important features in the maps. We find our chosen radius is well-matched to the relevant scales in our simulations and the inclusion of noise does not change our results.

We define the distance metrics for both the skewness and kurtosis using the Hellinger Distance (Equation \ref{eq:hellinger_dist}) applied to the normalized probability density functions of the maps: $p(\tilde{\gamma}_3)$ and $p(\tilde{\gamma}_4)$.  As before, a common set of bins for the empirical PDFs are constructed based on the extrema of the two quantities being compared.


\subsubsection{Cramer Statistic} 
\label{sub:cramer}

The Cramer statistic is a general statistical measure developed by \citet{cramer-test} for multivariate two-sample testing. This method was first applied to spectral-line data cubes by \citet{Yer14}. We follow the same method and provide a brief description.
Given two data sets, $I_1(\mathbf{x},v)$ and $I_2(\mathbf{x},v)$, we construct two representations of the data with reduced dimensionality by creating an array where each row consists of the top 20\% of the intensity values in a given velocity channel, sorted in order of decreasing brightness.  Selecting the top 20\% ensures that the response of the statistic is defined by the signal in the data. Each data cube is thus transformed into a two-dimensional data set with size $N_\mathrm{channels} \times N_D$ where $N_D$ is the number of the intensity values drawn from a velocity channel.  To remove scaling with the absolute intensity in this two-dimensional data set, we normalize each by its spectral norm\footnote{$||P|| \equiv \sqrt{\lambda_0}$ where $\lambda_0$ is the largest eigenvalue of $P^TP$.}, as is used in \citet{Yer14}.  The statistic is calculated by considering each row in the data set to be a point in an $N_D$ space and calculating the Euclidean distance between those data.  For notational compactness, we denote $I_1$ as $P$ and the set of $I_2$ as $Q$,
\begin{eqnarray}
d_{\mathrm{C}}(P,Q) & = &\frac{N_P N_Q}{N_P+N_Q} \left( \frac{1}{N_P N_Q}\sum_{p}^{N_P}\sum_{q}^{N_Q} || P_{p} - Q_{q} || \right. \nonumber \\
&&-\frac{1}{2N_P^{2}}\sum_{p_1,p_2=1}^{N_P} || P_{p_1}  - P_{p_2} || \nonumber \\
&&\left. -\frac{1}{2N_Q^{2}}\sum_{q_1,q_2=1}^{N_Q} || Q_{q_1} - Q_{q_2} ||\right). \label{eq:cramer_met} \end{eqnarray}
This metric compares the typical distance between the points comprising $P$ and $Q$ to the distance between the points within an individual data set.  When $P$ and $Q$ are significantly different from each other, the inter-set distance (first term) will be much larger than the intra-set distances (second and third terms). In this manner, the Cramer statistic is a model-free approach to estimate the difference in variance between two data sets.




\section{Analysis}
\label{sec:analysis}

Using the set of simulations (Sec.~\ref{sec:sims}), we evaluate each of the proposed measures of distance presented in Sec.~\ref{sec:stats}. Our evaluation and testing process is composed of three steps: calculating the distances between each of the design runs to the five fiducial runs, measuring the `quality' of a statistic using two forms of a permutation test, and performing the sensitivity analysis with respect to the experimental design. As an example, we show the results of our analysis for the SCF statistic (\S\ref{sub:scf}) in Figures \ref{fig:scf_clean_distances}, \ref{fig:scf_noise_distances}, \ref{fig:scf_clean_model}, \ref{fig:scf_noisy_model}, \ref{fig:scf_obs_to_fid}, and \ref{fig:scf_des_to_obs}. The figures for the remainder of the statistical tools are available as online content.

\subsection{Calculating distances} 
\label{sub:calculating_distances}

For each of the statistics presented in Section \ref{sec:sims}, we calculate the distances between the 32 design runs and each of the five fiducial runs, which have the same physical parameters and differ only in the random driving field.  A given simulation run produces up to ten data cubes, representing simulation outputs equally spaced at one-tenth of a crossing time.   We calculate distances for each output, matching the time between the runs being compared.  The distance between two runs is then taken to be the average over the distances at each output time.  Performing these comparisons using multiple fiducial runs allows for `pseudo-replicates' of the distances (i.e., there are five distance measurements for each of the 32 designs), with only the structure of the turbulent driving field in the simulation being different.  These replicates measure how a given distance measure responds to random fluctuations with no change in the underlying physics, which allows for the characterization of errors in the sensitivity analysis (Sec.~\ref{sub:determining_quality_of_statistics}).

We also inter-compare each of the five fiducial runs, again averaging over the distances from the 10 data cubes at the output times.   We expect that, since each fiducial run has the same physical parameters, distance measures which are tracing physical properties show distances from the fiducial runs that are significantly larger than the comparisons between the fiducial runs.

For the original data, we perform the sensitivity analysis using 160 datapoints (32 designs $\times$ 5 fiducial runs).  We also perform a parallel analysis, which tests for the influence of Gaussian noise by adding noise to each of the simulated data sets.  The level of the noise was scaled to 5\% of the maximum intensity in each of the cubes, i.e.~a peak signal-to-noise ratio of 20.   This signal-to-noise is comparable to the typical noise level in three observational \thco\ data cubes extracted around NGC 1333, Oph A and IC 348 from the COMPLETE survey \citep{complete-data}. A masking procedure is applied to the noise-added simulated data cubes, as would be applied to observational data. The procedure is described in \S\ref{sub:comparing_to_observational_data}.

\begin{figure*}
\includegraphics[width=0.9\textwidth]{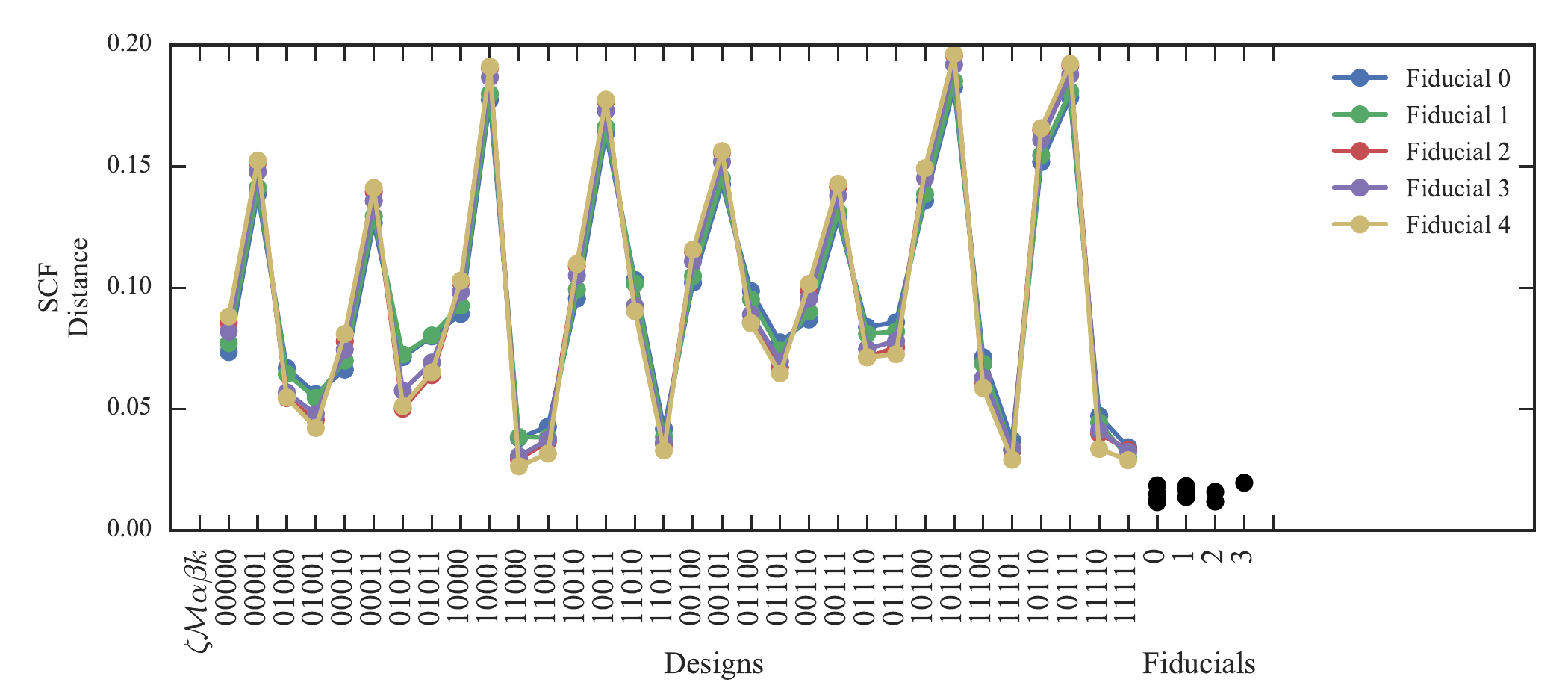}
\caption{\label{fig:scf_clean_distances} SCF distances (Sec.~\ref{sub:scf}) between fiducial and design simulations (coloured curves) and among fiducial runs (black points).  The $x$-axis is labelled with a code corresponding to the parameter changes from the experimental design.  Each digit in the code refers respectively to the solenoidal fraction (\solfrac), the  Mach number (\mach), the virial parameter (\virial), the plasma parameter (\plasbeta), and the driving scale (\drive), with 0 representing the low setting and 1 representing the high setting (Table \ref{tab:design}).
The SCF shows good behaviour for a distance measure: a small scatter between fiducial to design comparisons, and distances close to 0 when fiducial runs are compared to other fiducial runs.}
\end{figure*}

\begin{figure*}
\includegraphics[width=0.9\textwidth]{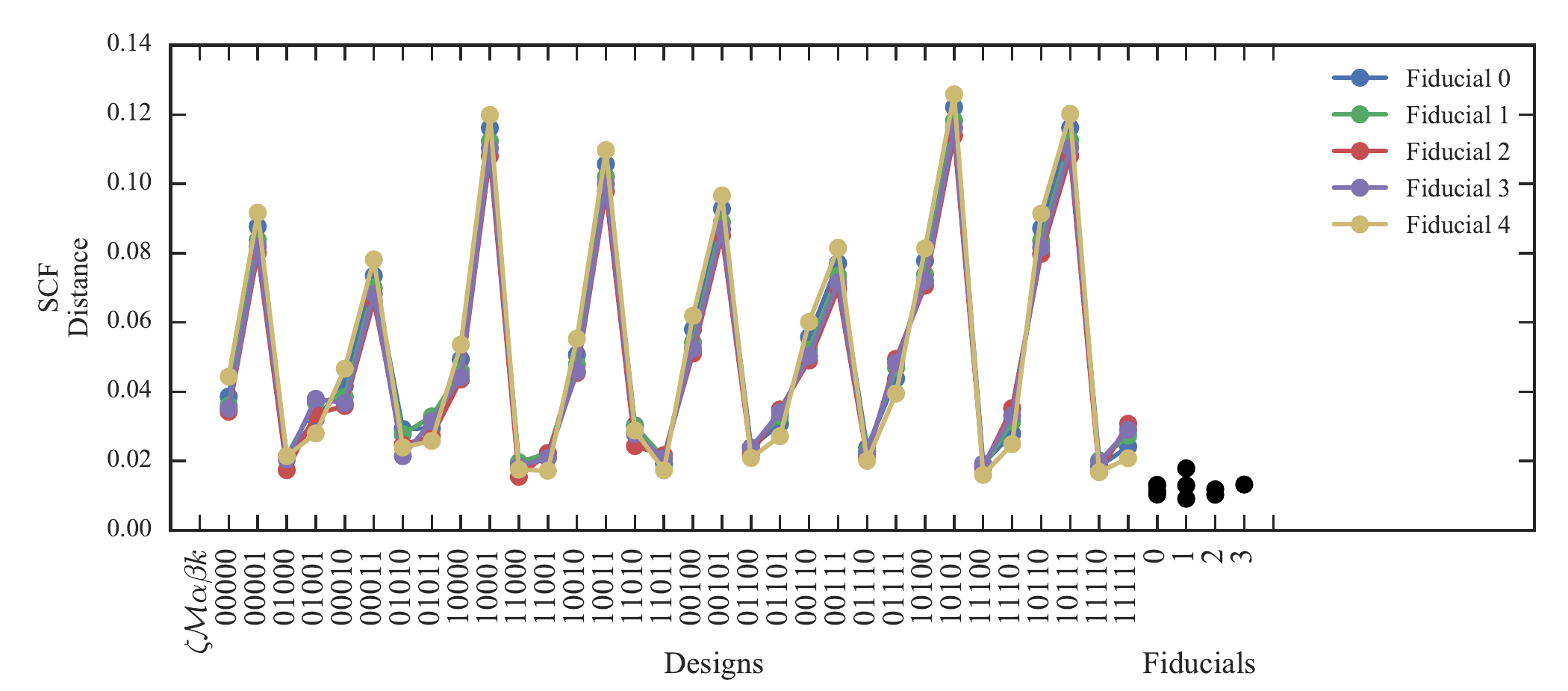}
\caption{\label{fig:scf_noise_distances} SCF distances as per Figure \ref{fig:scf_clean_distances} but with Gaussian noise added to the mock observational data sets.  The SCF still shows ideal behaviour for a measure of distance, however there are small differences from the distances without added noise. This corresponds to a loss of sensitivity to changes in \drive.}
\end{figure*}

We illustrate these distances for the SCF for noiseless data in Figure \ref{fig:scf_clean_distances} and with added noise in Figure \ref{fig:scf_noise_distances}.  These curves show the distances of the different design runs from each of the five fiducial runs.  The distances of the fiducial runs with respect to each other are shown on the right side of the Figure. The SCF is an example of a well-behaved measure of distance since the distances between the fiducial runs are small compared to the distances of the design simulations, and the scatter is small compared to the variations seen in the design runs.  In contrast, we also present the same results for our formulation of metrics based on the bicoherence and PCA eigenvalues in Figure \ref{fig:bispec_pca_distances}.  Both of these distances show a larger scatter in the fiducial comparisons and the design parameters than the SCF. The bispectrum, in particular, shows clear signal to a few design parameters, but the response of the others is unclear and likely a measure of ``happenstance.'' This effect is more pronounced in the PCA eigenvalue responses. This may suggest that the changes in the physical parameters we explore are not large enough for the method to reliably detect changes in our proposed formulations.

\subsection{Distances at $t/t_{\rm ff}=1$} 
\label{sub:distances_at_freefalltime}

Our analysis averages over all available checkpoints between $t/t_{c}=2.0$ and $3.0$.  However, self-gravitating systems will evolve on the free-fall timescale, $t_{\mathrm{ff}}$, in particular with the density probability density function and power spectra changing significantly during collapse \citep{collins-selfgrav}.  We have carried out a parallel analysis of our simulated data cubes, comparing only the time-step closest to $t/t_{\mathrm{ff}}=1$ after the onset of self-gravitation in each of the simulations. We find that only two methods -- bispectrum and PCA eigenvalues -- change appreciably compared to the time-averaged responses (see \S\ref{sub:determining_quality_of_statistics}).  Due to the limitations of this simulation set (\S\ref{sec:limitations_of_this_study}), we defer modeling time-related effects to future work.



\subsection{Determining quality of statistics} 
\label{sub:determining_quality_of_statistics}

The quality of a statistic is determined in two steps, where we test the {\it sensitivity} and the {\it scatter} for the different statistics.  The results of our quality tests are shown in Table \ref{tab:pvals}.  A reliable statistic will show large changes in the distance metrics for the design runs when compared to the fluctuations between the fiducial runs, which all have the same underlying physics.  In addition, the scatter among the five different design-to-fiducial comparisons should be small with respect to the differences between those statistics.  First, we test whether there is a significant difference between the distances of the design-to-fiducial comparisons versus the fiducial-to-fiducial comparisons, which is denoted ``Noise p-value'' in Table \ref{tab:pvals}. We fit a linear model between the two groups such that the slope of the line corresponds to the difference in the means of the groups. We then perform a permutation test, randomly exchanging values between the two sets and calculating the slope from the linear model applied to the permuted data.  A good statistic should have a significantly larger slope when the data are not mixed between fiducial runs and simulations compared to when they are.  For a large number of permutations ($N=10^4$), we report the fraction of permutations which give a slope larger than the un-permuted data.  The ideal result of this test is a p-value of 0.  Statistics which show a significant fraction of permutations with slopes larger than the un-permuted data are not {\it sensitive} and do not show significant response to real physical changes.

In the second step of quality testing, we develop a test for scatter within the design measurements.  We fit another linear model, where we define a parameter that distinguishes between each design. Since we test for 5 factors, the parameter has 32 levels.  Each simulation in the design has five distance measures, one for each fiducial, so we can test how important the particular set of fiducial-to-design connections actually is.  We do this by re-sampling (with replacement) the distance measures for a given simulation in the design from the five distance values.  We then fit a linear model and measure the $R^2$ value of the correlation.  We repeat this re-sampling $10^4$ times and measure the fraction of fits with $R^2>0.9$.  A statistic passes this scatter test if a large fraction of the fits show high correlation, and thus the ideal result if a p-value of 1.  This test shows that any measured response to physical parameters is well defined, regardless of the set of fiducial runs used to define it.  Statistics without a large fraction of ``good'' fits show significant scatter among their design-to-fiducial comparisons and thus do not have a clear response to physical signal.

The SCF, whose results we show throughout the paper as an example, returns the ideal p-values in all cases: noiseless, free-fall time only, and with added noise. Other statistics which show either the ideal result, or close to it, in all cases include the skewness, VCS, and the large-scale VCS component alone.

\begin{figure*}
\includegraphics[width=0.9\textwidth]{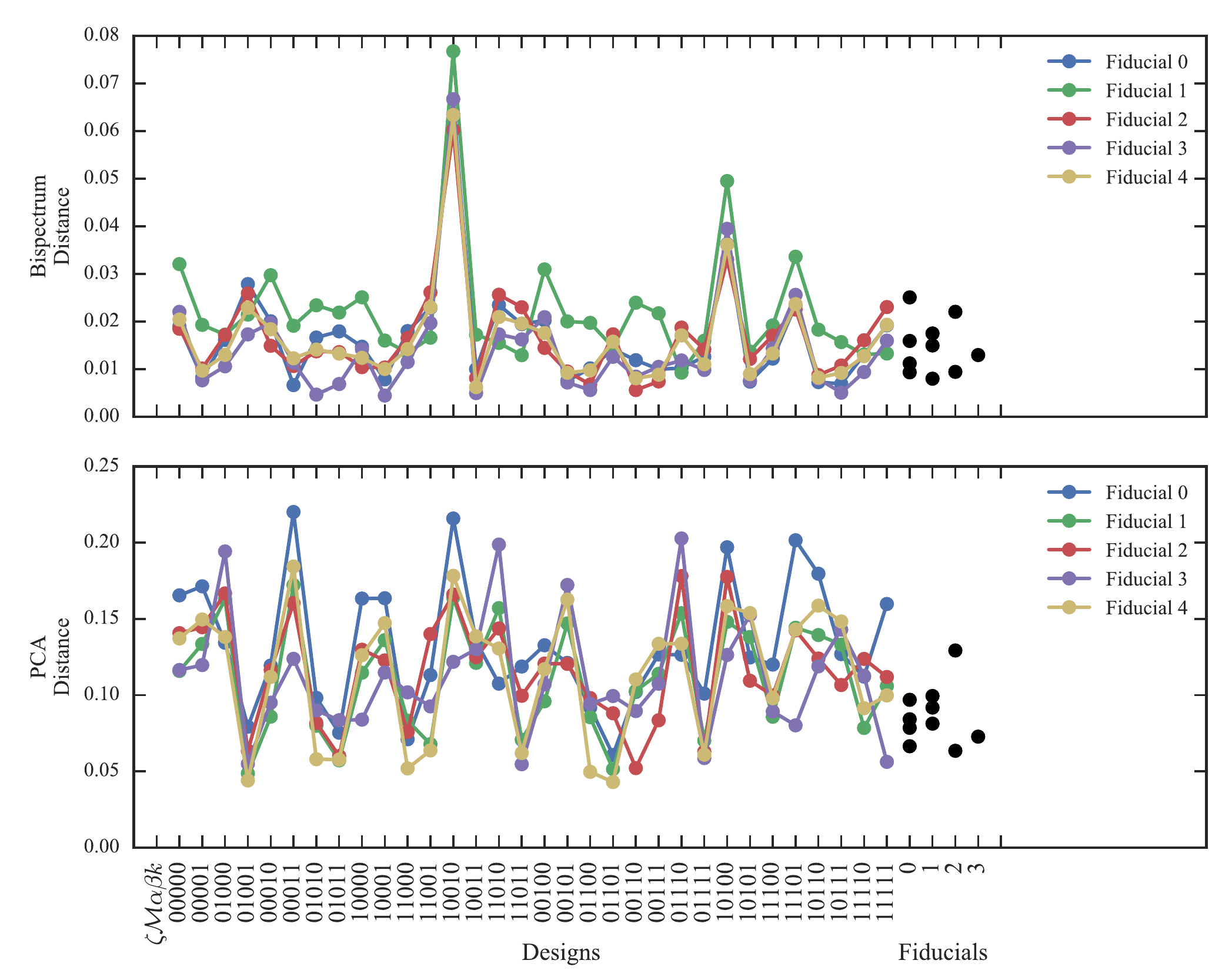}
\caption{\label{fig:bispec_pca_distances} Same as shown in Figure \ref{fig:scf_clean_distances} for the Bispectrum statistic (top, \S\ref{sub:bicoherence}) and PCA eigenvalue statistics (bottom, \S\ref{sub:pca}) in the noiseless case. These demonstrate cases where additional scatter in the fiducial-fiducial or fiducial-design distances makes it unclear whether the response is driven by physical differences. This uncertainty is indicated by quality testing scores in Table \ref{tab:pvals}.}
\end{figure*}

Many other statistics give an ideal response for at least one case or only one of the tests. We show two examples of statistics where this is the case in Figure \ref{fig:bispec_pca_distances}. The bispectrum shows a strong response to a small number of design parameters, however there is significant scatter between the fiducial distances. This leads to the large Noise p-values. However, this scatter is significantly reduced when only considering distances at the free-fall time, and accordingly, the associated p-values indicate a much higher quality response. The PCA eigenvalues exhibits a similar behaviour to the bispectrum, albeit with significantly more fiducial-distance scatter as illustrated by Figure \ref{fig:bispec_pca_distances}. This drives its poor quality scores in the noiseless and added-noise cases and suggests that the responses can only be reliably be modeled for the free-fall distances.

Only a small number of statistics have poor quality in all three cases, and we stress that these quality results do not necessarily condemn any of the presented methods. We discuss possible reasons for a lack of a response in a statistic in \S\ref{sec:limitations_of_this_study}.

We complete the sensitivity analysis only for the statistics that have a high quality response in our testing since we are confident the responses are driven by physical changes in the parameters.

\begin{table*}
\begin{minipage}{\textwidth}
\centering
\caption{\label{tab:pvals} The results of our quality testing for the proposed statistics. As described in \S\ref{sub:determining_quality_of_statistics}, we use two tests to determine the quality of the statistic. The ``Noise $p$-value'' tests whether there is a significant difference between the fiducial-to-design comparisons and the design-to-design comparisons. The $p$-value is based on a permutation test between these two groups; a good statistic should return a $p$-value of 0. The ``Signal $p$-value'' tests the significance of the distances between different design comparisons. The $p$-value is calculated by permuting the design-to-fiducial distances amongst the designs and recording the number of fits which are `good' (based on $R^2>0.9$). We show the quality scores for three analyses: noiseless, free-fall, and added noise. The noiseless and added noise cases are based on averaging over all time steps, and the free-fall scores are the distances at the free-fall time in each of the simulations. Note that the small scale component of the VCS is dominated by noise, and accordingly, its $p$-values are excluded from the noise-added case.}
\begin{tabular}{lllllll}
 & \multicolumn{2}{c}{Noiseless} & \multicolumn{2}{c|}{Free-fall} & \multicolumn{2}{c|}{Added Noise} \\ \hline
Statistic & \multicolumn{1}{l|}{\begin{tabular}[c]{@{}l@{}}Noise \\ $p$-value\end{tabular}} & \begin{tabular}[c]{@{}l@{}}Signal \\ $p$-value\end{tabular} & \begin{tabular}[c]{@{}l@{}}Noise\\ $p$-value\end{tabular} & \begin{tabular}[c]{@{}l@{}}Signal\\ $p$-value\end{tabular} & \begin{tabular}[c]{@{}l@{}}Noise\\ $p$-value\end{tabular} & \begin{tabular}[c]{@{}l@{}}Signal\\ $p$-value\end{tabular} \\ \hline
VCS                      &                 0 &                  1 &                    0 &                     1 &                 0.0751 &                  0 \\
VCS Large Scale          &                 0 &                  1 &                    0 &                     1 &                 0.0689 &                  0 \\
SCF                      &                 0 &                  1 &                    0 &                     1 &                 0 &                  1 \\
Skewness                 &                 0 &                  1 &                    0 &                     1 &                 0 &                  1 \\
VCS Small Scale          &                 0.0153 &                  1 &                    0.1534 &                     0.9968 &                 -- &                  -- \\
Del. Var. Centroid Curve &                 0.0001 &                  1 &                    0.0002 &                     0.0147 &                 0 &                  0.9997 \\
Kurtosis                 &                 0 &                  1 &                    0.0002 &                     0.0001 &                 0 &                  0.2387 \\
VCA                      &                 0 &                  0.5308 &                    0 &                     0.2985 &                 0 &                  0.4956 \\
Bispectrum               &                 0.2442 &                  0.3581 &                    0 &                     1 &                 0.0993 &                  0.3777 \\
PDF Lognormal            &                 0.0229 &                  0.1963 &                    0.0224 &                     0.1306 &                 0.0047 &                  0.0012 \\
Cramer                   &                 0 &                  0.0299 &                    0 &                     0.0186 &                 0 &                  0.0272 \\
Del. Var. Centroid Slope &                 0.0009 &                  0.0412 &                    0.0002 &                     0.0056 &                 0.0002 &                  0.0961 \\
MVC                      &                 0.0001 &                  0.0008 &                    0.0382 &                     0 &                 0.0003 &                  0.0074 \\
Spatial Power Spec.      &                 0.0001 &                  0.0008 &                    0.0384 &                     0 &                 0.0003 &                  0.0090 \\
Genus                    &                 0.0051 &                  0.0001 &                    0.0109 &                     0 &                 0.0183 &                  0.0004 \\
Dendro. Num.             &                 0.0992 &                  0 &                    0.0001 &                     0.1830 &                 0.0418 &                  0.0007 \\
PCA Eigenvalues                      &                 0.0060 &                  0 &                    0 &                     0.6199 &                 0.0404 &                  0 \\
Wavelet                  &                 0.0097 &                  0 &                    0.0870 &                     0 &                 0.0328 &                  0 \\
Del. Var. Curve          &                 0.0315 &                  0 &                    0.0076 &                     0 &                 0.0848 &                  0 \\
Del. Var. Slope          &                 0.0451 &                  0 &                    0.0044 &                     0 &                 0.0847 &                  0 \\
Dendro. Hist.            &                 0.0122 &                  0 &                    0.0233 &                     0 &                 0.0083 &                  0 \\
\end{tabular}
\end{minipage}
\end{table*}



\subsection{Sensitivity Analysis} 
\label{sub:sensitivity_analysis}


Following \cite{Yer14}, we quantify the sensitivities of the distance measures to each of the physical parameters by fitting a linear model to the data.  Practically, this model has a form:
\begin{equation}
\mathbf{d} = \mathbf{X}\boldsymbol{\beta} + \boldsymbol{\epsilon}.
\end{equation}
Here, $\mathbf{d}$ is the set of the 160 distance measurements for a given statistic (32 designs $\times$ 5 fiducials).  The matrix $\mathbf{X}$ encodes the design, consisting of values $\pm 1$ depending on whether a given parameter is low or high. Each row in $\mathbf{X}$ is orthogonal to the other rows, ensuring the effects are uncorrelated. We optimize the model to find the vector of coefficients $\boldsymbol{\beta}$, which represents the magnitude of the response of the distance metric to changing physical parameters.  This vector has 32 elements, representing all $2^5$ combinations of effects including main effects and interaction terms.  Variation among the fiducial runs is treated as part of the noise, $\boldsymbol{\epsilon}$, so the resulting model is a generalization of a weighted least squares using a non-diagonal covariance matrix.\footnote{The response to replicated, fiducial runs is sometimes accomplished by adding a categorical variable with 5 levels \citep{R_Faraway_2006}.  However, this treats the specific seeds we chose as having a meaningful context, which is not the case.}  We use the {\sc lme4} package \citep{lme4} in the software package R to measure the coefficients and their uncertainties.

\begin{figure}
\includegraphics[width=0.45\textwidth]{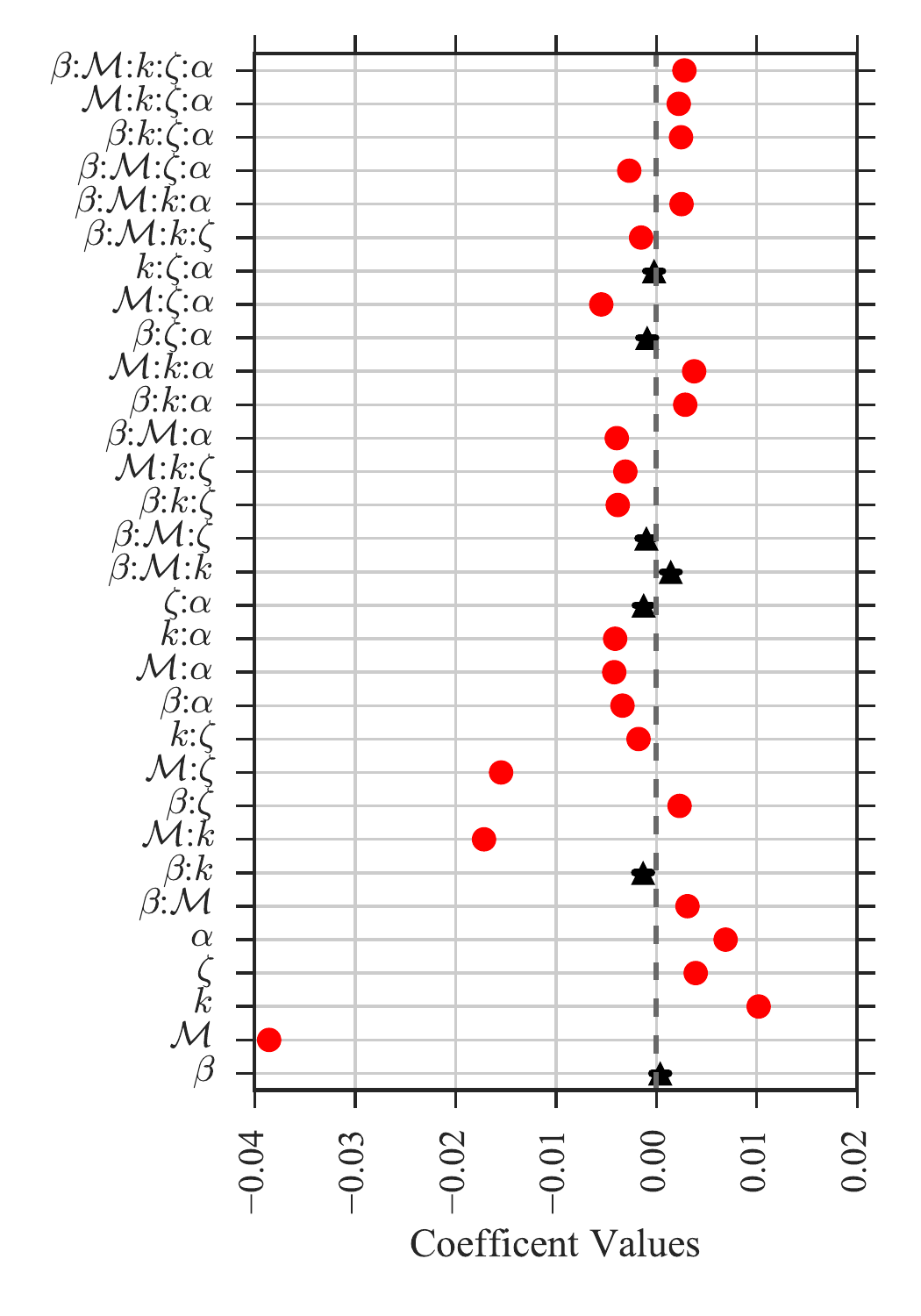}
\caption{\label{fig:scf_clean_model}  The coefficient values, with error-bars in the x-axis indicating the standard error for each term,  from fitting to the SCF distances in Figure \ref{fig:scf_clean_distances}.  Red circles indicate significant terms in the fit, where we define significance as a $t$-statistic above 3.46, the 99.9\% confidence level. Black triangles are insignificant terms in the model.  The standard errors are roughly equal for each term, and so a larger coefficient is essentially equivalent to a larger $t$-statistic. However, since higher-order terms have fewer simulations to fit to, the highest order terms have an effectively lower significance. For the SCF, we find a significant response to most of the physical parameters, though the largest by far is \mach\ and its second order terms with \drive\ and \solfrac.}
\end{figure}

\begin{figure}
\includegraphics[width=0.45\textwidth]{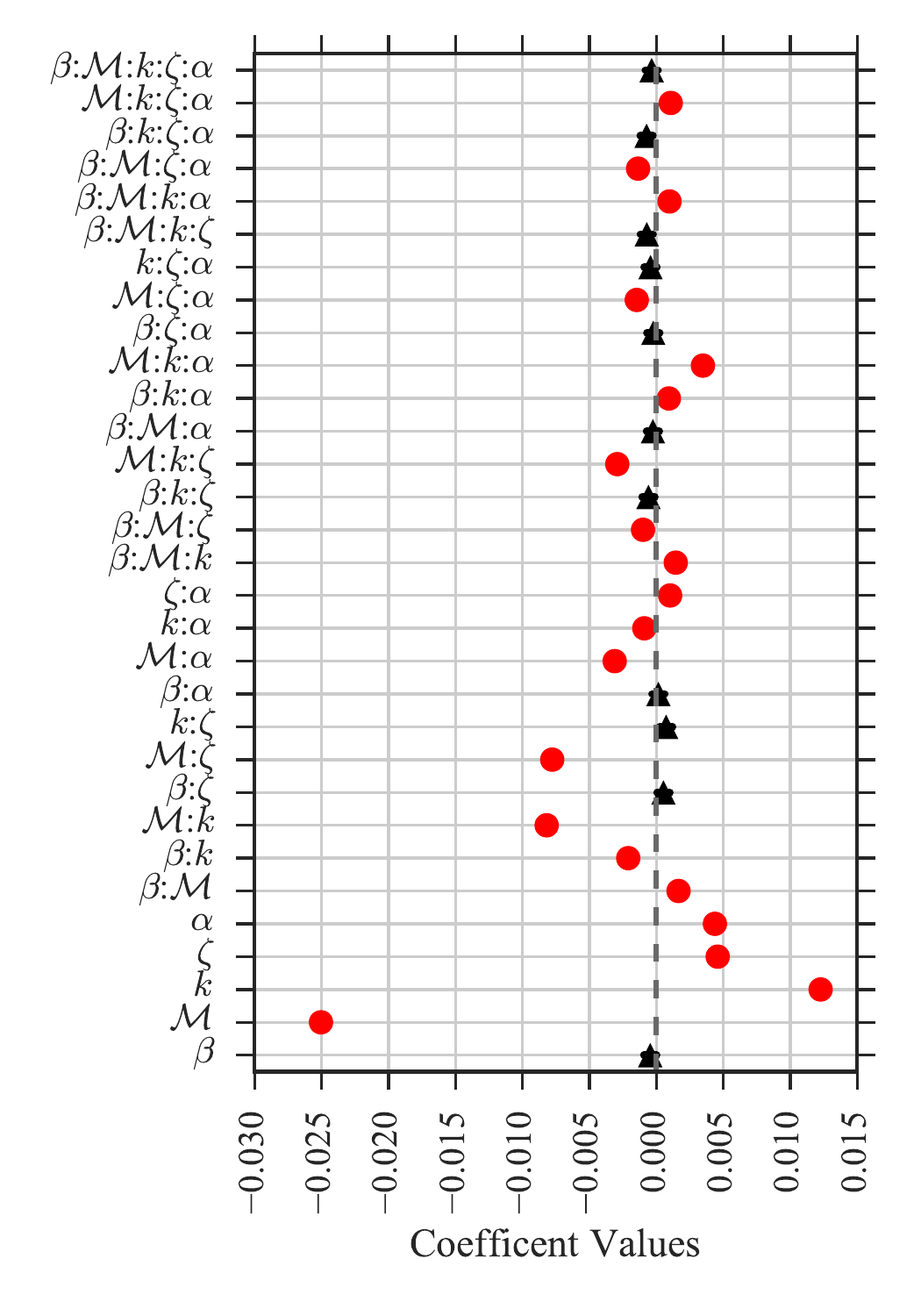}
\caption{\label{fig:scf_noisy_model} Same as Figure \ref{fig:scf_clean_model} but with noise added. The SCF response is qualitatively unchanged in the presence of noise. Quantitatively, the significance of several terms has been reduced compared to the noiseless case due to increased scatter between the distances. For example, the \mach\ coefficient value is lower and the standard error has increased, though it remains a highly significant term. We see additional scatter in the higher-order terms since each category has fewe samples and therefore has a lower significance compared to the lower-order terms.}
\end{figure}

After running the regression, we compute the standard error for each coefficient, which we use to judge the significance of each factor using a $t$-statistic.  Highly significant factors that demonstrate a lot of sensitivity have large $t$-statistics.  The easiest coefficients to interpret are the {\it main effects} which represent how a given distance metric changes by partitioning the simulation set into two parts: those runs with ``high'' parameters (e.g., high Mach number) and those runs with ``low'' parameters (e.g., low Mach number).  A significant main effect shows that a given statistic has a difference between these two sets, marginalizing over all the other effects in the design.  We show the results of the sensitivity analysis for the SCF statistics in both the noiseless case (Figure \ref{fig:scf_clean_model}) and the case of added noise (Figure \ref{fig:scf_noisy_model}). The colour and shape of the points in these figures indicate whether a term is significant or not. The red circles are significant, where we consider significant terms to have a $t$-value greater than 3.46, the 99.9\% single-tail confidence level for a Student-$t$ distribution.

Figures \ref{fig:scf_clean_model} and \ref{fig:scf_noisy_model} show the SCF strongly depends on \mach\ but has a significant response to all other first-order terms except \plasbeta. The significant advantage in adopting a full-factorial design is its ability to test the interactions between these parameters. The SCF indicates that its next most significant terms are the second order interactions of \mach\ with \drive\ and \solfrac. \citet{padoan-scf} find a similar response to changes in \mach\ but point out that it is unclear whether variations in the SCF result from changes in the Mach number or the line width.  In Appendix \ref{app:re_scaling_to_a_common_linewidth_and_intensity}, we investigate the effect of normalizing to a common line width to determine if this ambiguity remains.

The importance of interactions between the main effects is demonstrated for many of the statistics.  Many of the statistics show a strong response to changes in \mach, \drive, and \virial. Often one of the most significant terms is a combination of two of these three parameters. Thus it is critical to test for these interactions when interpreting the response of a statistic. In the experimental design presented here, some of these interactions are expected, since the parameters themselves must be degenerate (see \S\ref{sub:sensitivity_to_basic_observables}). For example, the $\alpha:\mathcal{M}$ interaction suggests that the effects of simultaneously increasing both the virial parameter and the Mach number can cancel out.  This interaction effect manifests in self-gravity affecting the turbulent motions.  Using the original simulation volumes rather than the derived PPV cubes, we fit the kinetic energy spectrum over the limited range where the turbulent cascade is apparent ($k=2$ to $k=16$, see Appendix \ref{app:resn}).  One quarter of the simulations with low virial parameter and high Mach number show a much steeper kinetic energy spectrum ($E(k)\propto k^{-2.3}$ instead of $E(k) \propto k^{-1.7}$) which arises from gravity driving large-scale motions.  The statistics that are sensitive to the energy spectrum detect these deviations and report a significant interaction effect.  The VCS and skewness show significant sensitivity to this effect.  However, our resolution of turbulent motions is limited and these results should be revisited with higher root grid resolution simulations.

Testing the sensitivity to interaction effects emphasizes the importance of adopting an experimental design that thoroughly, but efficiently, explores the parameter space \citep{Yer14}.  The commonly used one-factor-at-a-time style of designs are not sensitive to these interaction effects.  They would measure modest main effects in, for example, the virial parameter and Mach number, and without knowledge of interactions. From such studies they would infer that increasing both virial parameter and Mach number would lead to more differences from the fiducial sample than is actually found.  Figures \ref{fig:scf_clean_model} and \ref{fig:scf_noisy_model} also show higher-order interaction terms, though these become progressively less sensitive since the number of simulations in each category decreases with increasing order.  Thus, we primarily focus our analysis on the main effects and their second-order interactions.

\begin{figure*}
\mbox{
\subfigure{
\includegraphics[width=0.5\textwidth]{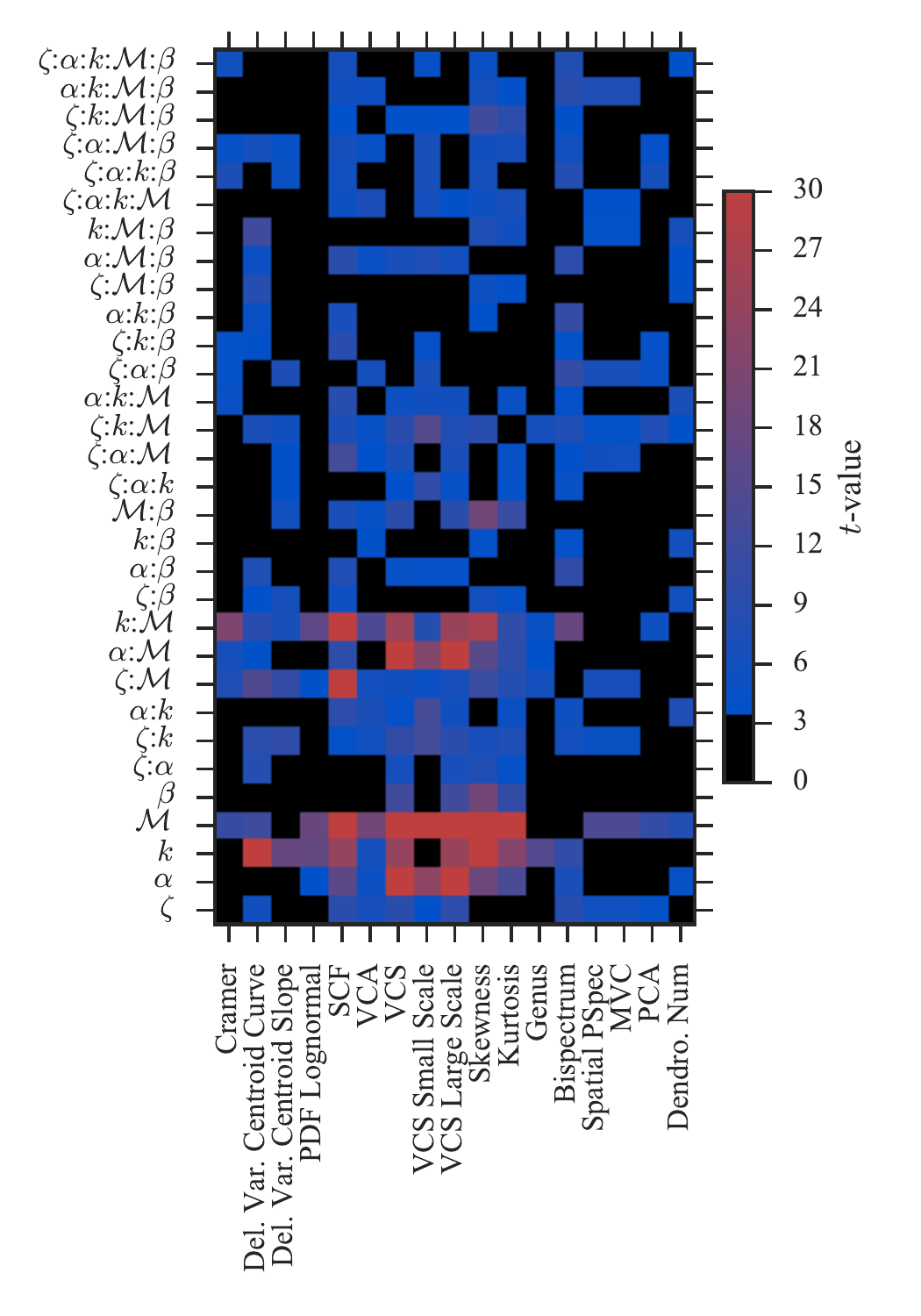}
\label{fig:all_results}
}\quad
\subfigure{
\includegraphics[width=0.5\textwidth]{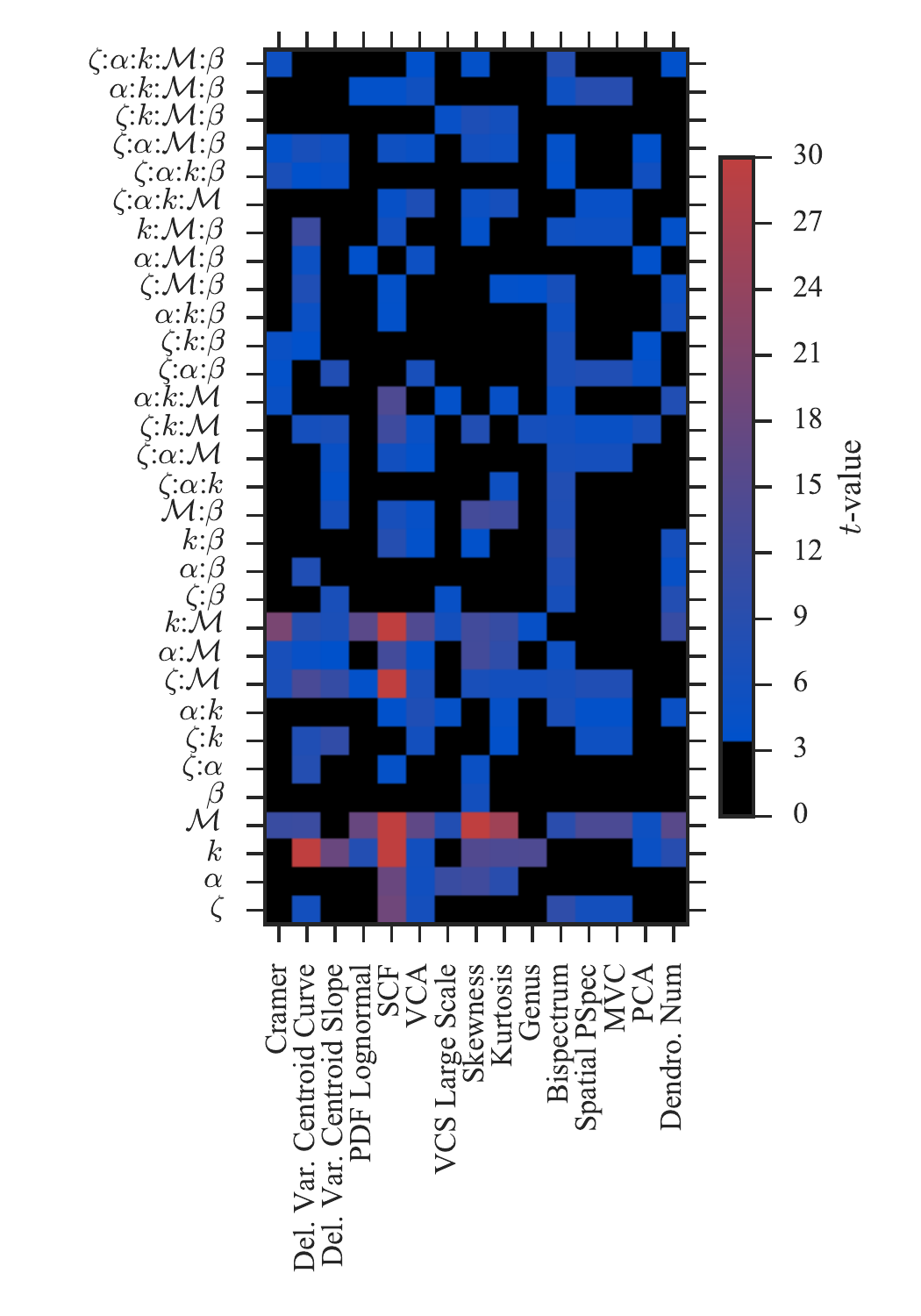}
\label{fig:noisy_all_results}
}
}
\caption{Model $t$-statistics for statistics with discernible signal for the noiseless (left) and noise-added (right) datasets.  Values below 3.46 (99th percentile) are considered unimportant to the fit and are shown in black.  There are two overall trends: the significance of the effects tends to be attenuated with noise; and the higher-order terms are more sensitive to the presence of noise.}
\end{figure*}

\begin{figure}
\includegraphics[width=0.5\textwidth]{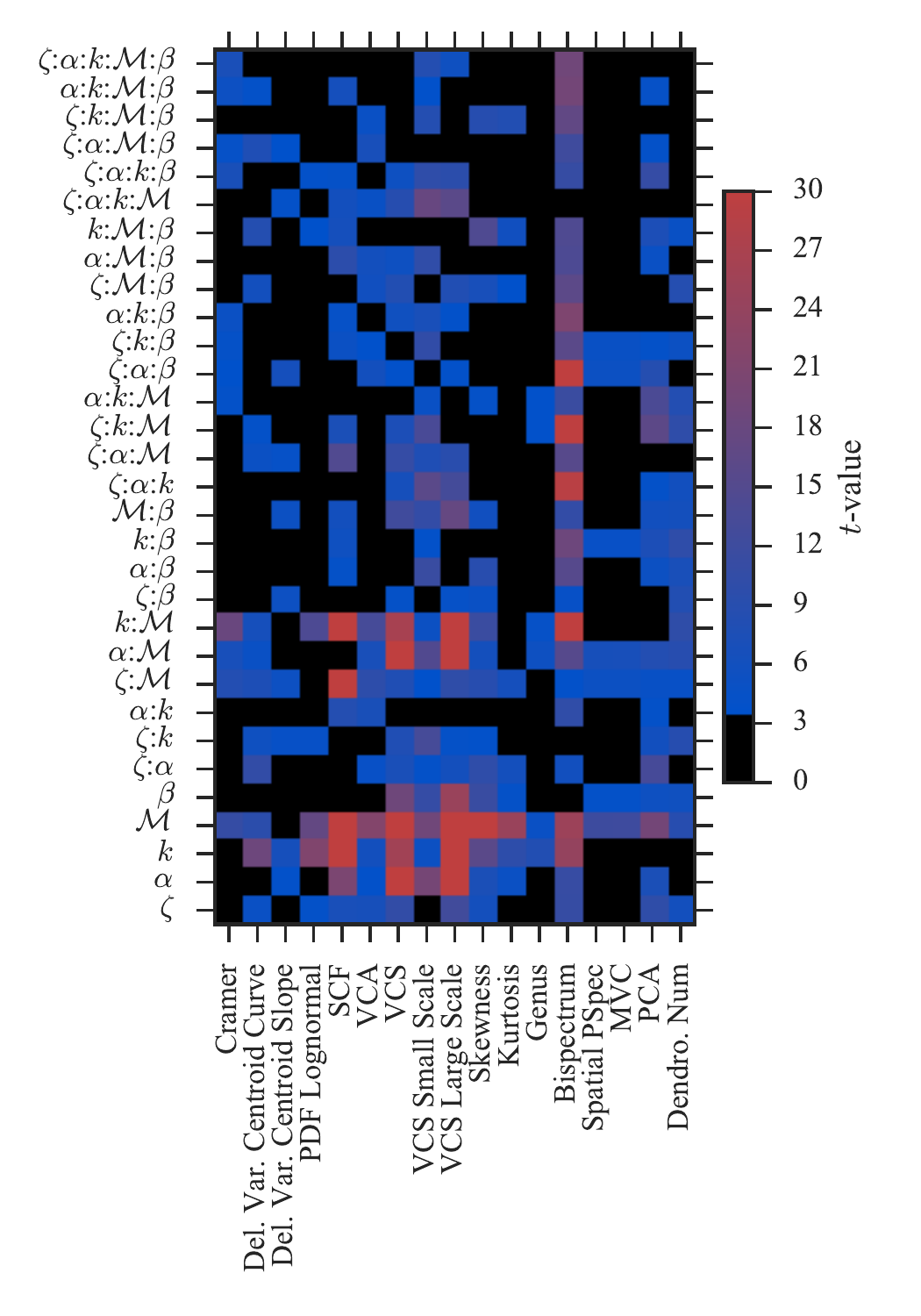}
\caption{\label{fig:freefall_all_results} Same as Figure \ref{fig:all_results} for the constant free-fall time analysis. For most of the statistics, time-averaging reduces scatter in the distances, resulting in more significant responses. However, the scatter in the bispectrum and PCA eigenvalues is {\it reduced} in the free-fall analysis (see also Table \ref{tab:pvals}). Compared to Figure \ref{fig:all_results}, its coefficients are far more significant. These metrics may be sensitive to time-evolution in the molecular cloud.}
\end{figure}

In Figures \ref{fig:all_results}, \ref{fig:noisy_all_results} and \ref{fig:freefall_all_results}, we present the $t$-statistic values for all methods where the quality testing has suggested a physical response is being reliably measured. We impose the same significance cut of $t=3.46$ to determine which terms in the model are significant. The most common strong response of these methods is to changes in \mach. Typically this response is coupled with strong second-order effects of \mach\ with \drive, \virial, or \solfrac. The relation between \mach\ and \virial\ to the turbulent energy spectrum is described above. Sensitivities to \drive\ and \solfrac, and their coupling to \mach\ are an indication that the method is sensitive to changes in spatial structure of the cloud. This is illustrated in Figure \ref{fig:moment0s}, where the effects on the sizes of the spatial structure from changing these parameters is evident. This is most prominent in the genus statistics (\S\ref{sub:genus_statistics}), since it measures the size distribution of regions in the data.

The physical parameters that the fewest methods are sensitive to are \plasbeta\ and \solfrac.  Those that are sensitive --- the VCS, VCA, skewness and kurtosis, and the SCF --- provide some measure of the anisotropy in the turbulent field.  Additionally, the curve distance of the delta-variance on the centroid field is sensitive to \solfrac, suggesting that the response of the symmetric delta-variance kernel to asymmetric structure is detectable through the shape of the delta-variance curve, but does not have a measurable effect on the slope within the fitting range.  This is at least partially due to our distance metric definitions, which in some cases average over azimuthal structure. For example, we convert the two-dimensional power spectra from the spatial power spectrum, MVC, and VCA into azimuthally-averaged one-dimensional spectra, from which the slopes are compared. Additional information may be accessed by comparing azimuthal structure \citep{kandel-vca}.

Given these results, we examine the behaviour of the modelled metrics in more detail below in Appendix \ref{app:method_sensitivities}.

\subsection{Sensitivity to Basic Observables} 
\label{sub:sensitivity_to_basic_observables}

The tools we use in the above analysis are, to varying degrees, a more complex approach for interpreting an observational dataset. The underlying assumption is that each method provides a particular advantage for measuring a property that cannot be attained through more standard analysis of the data. Here we perform a sensitivity analysis on some properties that are easily measured from a data cube: the total intensity (sum over the cube), the peak line temperature (maximum in the cube), and the average line width (averaged over the 2D line width map).  Given the known degeneracies in our experimental design, we expect to find significant responses for a few parameters.

\begin{figure}
\includegraphics[width=\columnwidth]{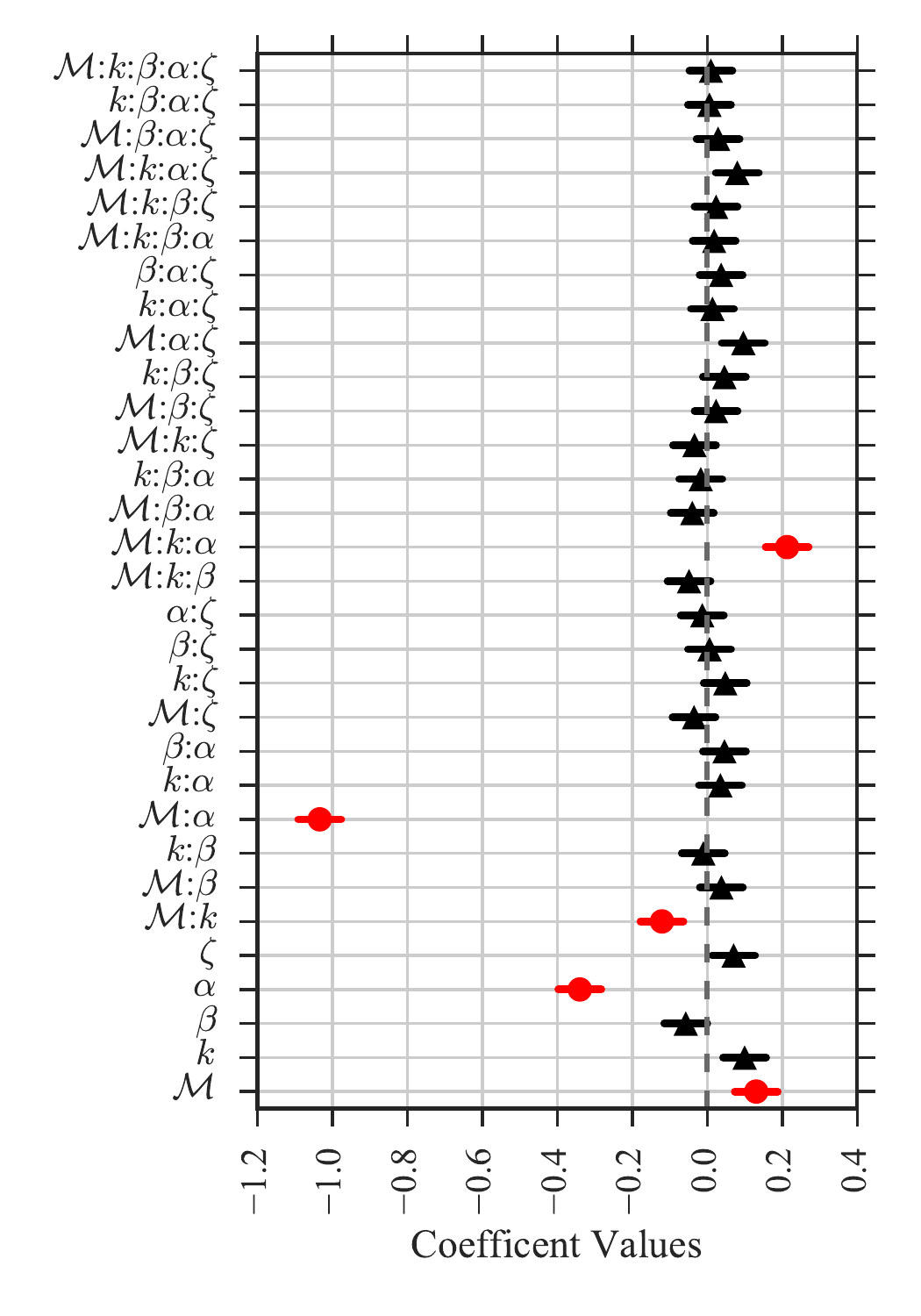}
\caption{\label{fig:tpeak_sensitivity} The parameter sensitivities from comparing the peak temperatures fit as described in \S\ref{sub:sensitivity_analysis}. The peak line temperature behaves as expected, with the most significant parameters being \virial\ and \mach.}
\end{figure}

For each of the quantities we compare, we define a measure of distance as the absolute difference between the quantities. As before, we average over the time steps.\footnote{Using the values at a common free-fall time did not change the qualitative results of this analysis.} In Figure \ref{fig:tpeak_sensitivity}, we show the parameter sensitivities of the peak line temperature. Unsurprisingly, we find a strong \mach:\virial\ interaction, with the first order terms of opposite sign. We also find a less significant \mach:\drive, and a third order interaction between these three variables.  The total intensity shows a nearly identical response.  The difference in average line width is driven again by two first order variables and their interaction: \mach\ and \drive.

None of the methods we test in \S\ref{sub:sensitivity_analysis} show an identical response to these quantities. In particular, none of the methods are sensitive solely to scalings between the mean or peak intensities. The greatest similarity we see here is between the SCF and the line width. As mentioned in \S\ref{sub:sensitivity_analysis}, there is ambiguity in whether the strongest response in the SCF is due only to the line width, and our comparison here suggests this is likely the case. The SCF has additional sensitivities, however, and so is useful beyond measuring a line width difference.

\subsection{Comparing to Observational Data} 
\label{sub:comparing_to_observational_data}

This analysis has shown that there is a suite of well-behaved statistics that are sensitive to changes in physical parameters.  While some statistics are more sensitive than others, Figure \ref{fig:all_results} demonstrates that most of the statistical approaches are sensitive to many different physical effects.  In particular, because of interaction effects, prior work may have conflated the response of the methods to changes in certain parameters.  For example, VCS has been shown to cleanly measure turbulence \citep[e.g.,][]{lp06}, but VCS will also register a change if the solenoidal driving fraction of turbulence changes.  This sensitivity argues for caution in interpreting the connection of any given method to specific physical parameters without explicitly controlling for other changes.  However, the sensitivity of most statistics to a wide variety of physical effects can be useful when comparing simulations to observations.  Obtaining a small distance with reliable distance methods would indicate that the simulations show some consistency with observations.

\begin{figure*}
\includegraphics[width=0.9\textwidth]{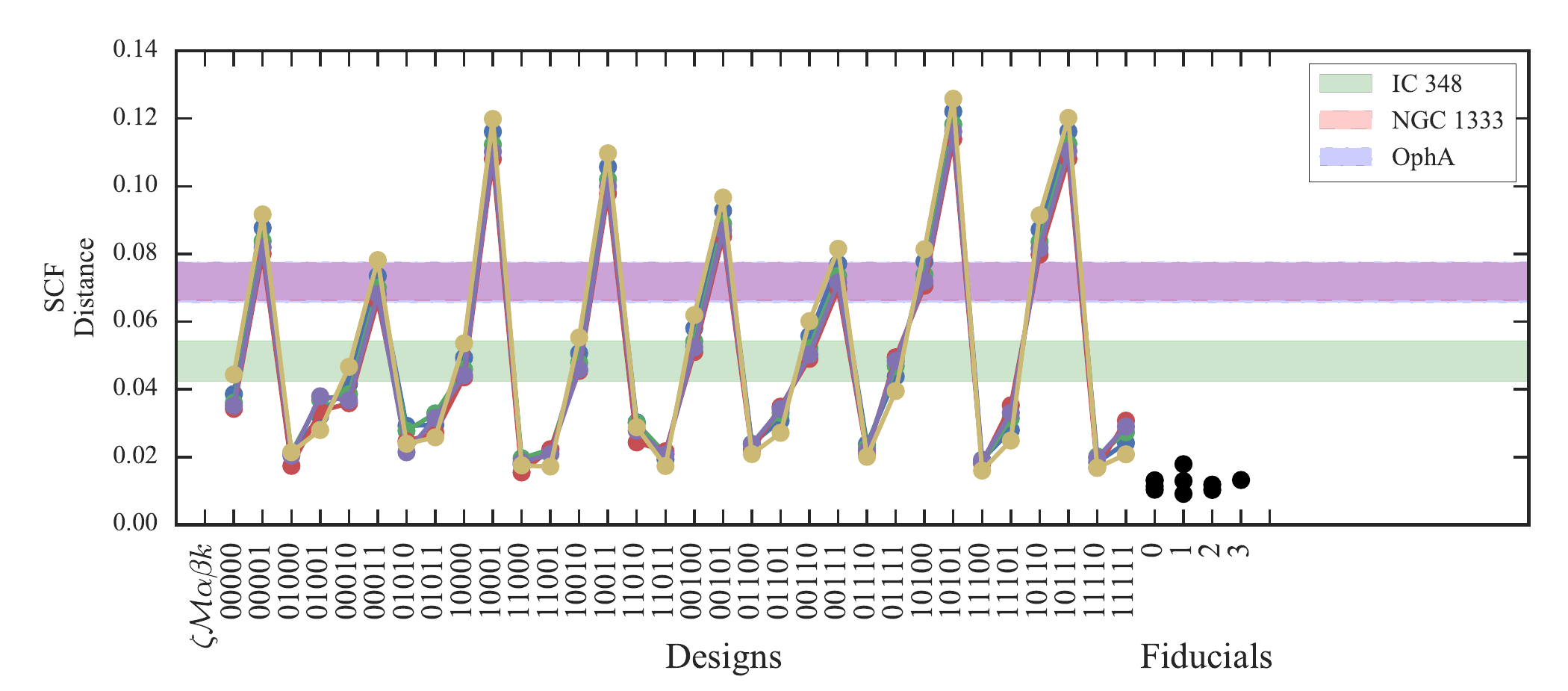}
\caption{\label{fig:scf_obs_to_fid} SCF distances between the design and the fiducial simulations (from Figure \ref{fig:scf_noise_distances}) where the distances between the fiducial simulations and the observations  are indicated by the colored bands.  The observation distances all fall within the range of design-fiducial distances but are larger than the fiducial-fiducial distances.}
\end{figure*}

As an example of this potential, we also compare our simulation suite to observational data of three nearby star-forming regions mapped in $^{13}$CO (1-0) emission as part of the COMPLETE survey \citep{complete-data}: NGC 1333 and IC 348 in Perseus ($d=260$~pc) and Oph A ($d=120$~pc).  Our simulation domain was chosen to have similar physical scales to these observations to facilitate this comparison.  We conduct two comparisons to the observational data: one of the distances between the fiducial and observations, and the other between the distances of the observations and the design run. Each of the comparisons uses the noise-added simulated data cubes to provide a more realistic comparison.

Before running the analysis, we use a common masking procedure for the data such that the methods are run only where appreciable signal is detected. This procedure is also used to mask out noise in the noise-added simulated data cubes. We define a signal mask based on separate searches in the spectral and spatial dimensions. First, we define a mask for each spectrum such that there is a connected component across three channels with a peak $>5\sigma$. We then extend the edges of this mask down to the $1.5\sigma$ level on each side. A region in the mask is considered valid if it spans seven consecutive channels, which given the thermal line width of $~200\, {\rm m\, s^{-1}}$ ($~3$ channels at 10 K) and reasonable estimates of the Mach number, is at least the expected width of a real line feature. After this, we filter the mask spatially by requiring a real spatial component be at least the twice the beam size. This is chosen since the FWHM of the beam major axis ($\sim46\arcsec$) in the COMPLETE data is just two pixels across in the map. At twice the beam size, a region need only cover $\sim16$ pixels. We remove regions smaller than this by performing the morphological opening and closing operators with a tophat kernel equivalent to the beam FWHM \citep{mathematical-morphology}. The opening operator removes all regions smaller than the kernel, while the closing operator restores the shapes of the regions that are larger than the kernel. Using the resulting mask applied to both observational data, we run the methods described in \S\ref{sec:stats}.

In Figure \ref{fig:scf_obs_to_fid}, we show the SCF measured for the observational data when compared to the fiducial runs.  The shaded bands indicate the range of distances obtained between the three observational data sets and the fiducial data sets.  This figure shows that the distances between the simulations and observations are comparable to the distances between the design and fiducials runs (\S\ref{sub:calculating_distances}).  Based on the range of distances to the observations, the IC 348 distance is consistent with six design runs, while NGC 1333 and Oph A, which have nearly identical distance ranges, are consistent with just three. Since the SCF response is quite complicated, inferring which parameter change results in a consistent distance is difficult. For the designs consistent with the IC 348 distance range, each of the five parameters is in both the high and low state. If these designs are truly consistent with the observations, the distance between them should be much smaller, ideally near the fiducial distances.

\begin{figure*}
\includegraphics[width=0.9\textwidth]{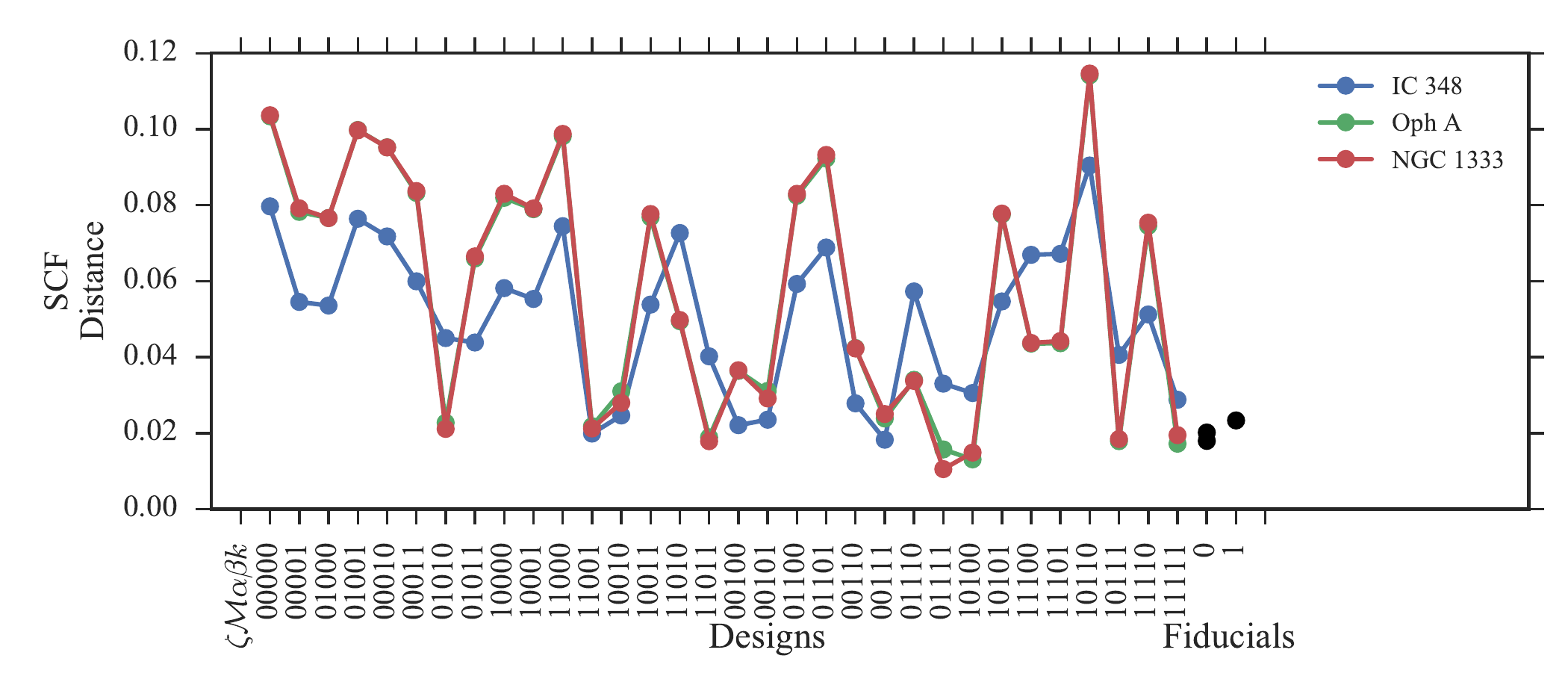}
\caption{\label{fig:scf_des_to_obs} Distances between the design runs and the observational data sets for the SCF.  Essentially, we treat the observational data as fiducial runs. While sensitivities cannot be derived in this way, these comparisons highlight whether the design simulations match the conditions in the observed star-forming regions. Significant non-zero distances that are of order the fiducial-design distances indicate that there is not a good match to those in Figure \ref{fig:scf_obs_to_fid}.}
\end{figure*}

To address this latter point, we also plot the distances obtained by comparing the design simulations to the observations, where the observed data sets play the role of the fiducials.  The results for the SCF are shown in Figure \ref{fig:scf_des_to_obs}.  Comparing to Figure \ref{fig:scf_obs_to_fid} shows that the design runs with similar distances to the observations do not imply a small distance when compared directly to that design. We do, however, see that that IC 348 is an outlier compared to NGC 1333 and Oph A.   A number of the distances are similar to the distances between the observations themselves, shown as the black points, and are therefore consistent with each other.  However, there is little overlap with the simulation we find to be consistent in Figure \ref{fig:scf_obs_to_fid}. For IC 348, Figure \ref{fig:scf_des_to_obs} shows five design distances consistent with the observation distances, but just one (10010) is consistent with the observation-fiducial distances in Figure \ref{fig:scf_obs_to_fid}. This is the same for the Oph A and NGC 1333 distances: only one (10100) is consistent in both. From this, it is difficult to make any conclusions connecting the physical parameters in the simulation set to the observations.

We find similar results for all of the methods modelled in \S\ref{sub:sensitivity_analysis}: there are no clear agreements, based on the observation-to-fiducial and observation-to-design distances, between the observational data and the simulation set presented here. Thus, we conclude that a full comparison to observations should adopt a broader range of parameters coupled with a more efficient design to explore the parameter space. The full-factorial design we have presented here is useful for establishing the behaviour of the statistics, but it does not efficiently sample the parameter space.  Other designs are better suited to that goal \citep{Yer14}.  In future work, we will compare these and other observational data to a more comprehensive set of simulations in pursuit of obtaining good agreement between the simulated and observed star forming regions.


\section{Limitations of this Study} 
\label{sec:limitations_of_this_study}

We note that the presented analysis has several limitations that should be considered when comparing with the presented results. We use this section to discuss the major limitations and their effects on the results presented in \S\ref{sec:analysis}.

\subsection{Simulation Resolution} 
\label{sub:simulation_resolution}

In using a large-set of low-resolution simulations, we have traded resolution in order to efficiently cover the parameter space without a monumental use of computational resources. The effects of changing the resolution on the methods we explore is shown in Appendix \ref{app:resn}. There are two important consequences to using low-resolution simulations: (1) the inertial range is small (see Figure \ref{fig:powerspec}), and (2) the fluctuations between different fiducial runs produce enhanced scatter for most power-spectrum based methods. For the former, we ensure that the fitting range of power spectra are limited to within the inertial range, and we find good fits despite the small available range. The latter effect decreases the sensitivity of most power-spectrum based methods, as well as the delta-variance. For those methods utilizing the entire PPV cube, such as the SCF, VCA, and VCS, this scatter is less prominent due to averaging over a larger amount of data points.


\subsection{Choice of Tracer} 
\label{sub:choice_of_tracer}

We only use one spectral line, \thco(1-0), in our analysis.  This choice is likely the reason for some methods' poor results in our analysis.  In particular, both distance metrics for the delta-variance when used on the integrated intensity show no discernable signal in the responses. However, the methods in its original formulation \citep{stutzki98} and subsequent work focussed on its application to column density maps. \citet{Burkhart2013ApJ...771..123B} test the prediction of \citet{lp04} that the spatial power-spectrum saturates to $-3$ in the optically thick limit. They find that a mix of optically thin and thick gas, traced by \thco, produces mixed behaviour with respect to the optically thick limit. This may explain the lack of signal we find applying the delta-variance on the integrated intensity maps, as well as the larger scatter in the SPS comparison. A similar result is found by \citet{Boyden2016ApJ...833..233B}, who use \twco(1-0) to target the influence of stellar winds on the surrounding molecular gas. They find that radiative transfer effects have a drastic effect on the shape of the integrated intensity delta-variance curves \citep[see Figure 20 in][]{Boyden2016ApJ...833..233B}.  These results demonstrate that an additional and important parameter to compare is the choice of tracer, and the component of gas that it traces.  One such study by \citet{Gaches2015ApJ...799..235G} focussed on the SCF response to a set of chemical tracers. Similar studies with the methods presented here will highlight the regimes where optical depth effects render a method less useful or insensitive to the underlying physical properties.


\subsection{Parameter Space} 
\label{sub:parameter_space}

Our experimental design is chosen to explore a parameter space with reasonable values based on observational evidence, as well as to refine the parameter space used in \citet{Yer14}. It is not, however, exhaustive.
There are clear cases where the parameter space could be improved, notably in resolution of turbulence, in relaxing some of the physical constraints such as isothermality, and in exploring a wider range of conditions ($\alpha < 2$).  A wider range of observational tracers may also highlight different effects.  Despite the limitations of the simulation set, the analysis still reveals several good tools for comparing the simulations and observations.  The construction of the design with several fiducial simulations used to evaluate the noise in these measurements means that these tools are likely to make stable comparisons, even in spite of a systematic bias in the resolution of turbulent effects.  Better formulations and a broader simulation set will likely lead to the identification of more useful tools.  Furthermore, our sensitivities are based on linear responses between the low and high settings. Thus, we are not sensitive to a non-linear response in any variable. To test this, an alternate experimental design will need to be explored. Finally, the parameters tested are a sub-set of those that may affect star formation and the evolution of molecular clouds.  For example, \citet{Boyden2016ApJ...833..233B} explore whether the methods presented here are sensitive to stellar feedback. They find a large fraction of the methods -- particularly those sensitive to structure (\S\ref{sub:structure_identification}) -- show a clear response to the level of feedback.  Expanding the parameter space, however, will begin to make parameter explorations computationally prohibitive due to the number of simulations that need to be run. Instead, a thorough exploration of one portion of the parameter space may be used to guide future efforts on where interesting variations are detected.  With well-understood sensitivities in a region of parameter space, a smaller set of simulations may be used to investigate a sub-set of the parameter variations \citep{Yer14,Boyden2016ApJ...833..233B}.  Regardless of the outcome of improved simulations, the basic concepts of establishing a framework for comparisons of observations to simulations remain of high utility.


\subsection{Formulation of Distance Metrics} 
\label{sub:formulation_of_distance_metrics}

Our results depend on how we have defined measures of distances. In many cases, we have sought to use similar measures, such as the power-spectrum slope, typically used in the literature. In others, namely the PCA, we have used a related, but simplified comparison approach.  A PCA metric based on the scale of the position and velocity components (i.e., closer to that present in the astronomical literature) will likely lead to behaviour similar to the other turbulent statistics.  Similarly, the bicoherence implementation of the bispectrum statistic may be recast to better capture the variations in the phase coherence.  There are several possible reductions of the bispectrum to a single, scalar metric value which can then be tested through our approach here.  The wavelet formulations and intensity power spectra have clear relations to the literature formulations of these methods but may simply be too sensitive to the particulars of a given physical scenario to be useful for our goals.  We also neglected some of the established features of successful tools such as changing the velocity width thickness in VCA \citep{lp04} or using the size vs. velocity width analysis of the PCA implementation of \citep{heyer-pca}.

Furthermore, there are cases where using the literature standard may reduce the sensitivity of a method to different physical parameters. For example, adopting the two-dimensional power-spectrum for the VCA, SPS, and MVC will produce distances sensitive to anisotropy \citep[e.g.,][]{kandel-vca}. Slight changes in the distance metric also lead to different sensitivities. We show this with the two formulations of the delta-variance distance, where the $L^2$ norm between the curves shows additional sensitivities not detected by the difference of the slopes.



\section{Summary}
\label{sec:summary}

In this work, we propose a comparison framework for evaluating several statistical techniques previously identified as promising tools for analyzing spectral line data cubes from both observations and numerical simulations.  Drawing from methods in the literature, we develop several measurements that quantify the differences between PPV data sets.  We then validate these tools using a common framework to evaluate their effectiveness for characterizing underlying physical changes and assessing whether they reflect the stochastic sampling of the physics associated with one instance of star formation. Our goal in this paper was to identify and characterize tools that are sensitive to physical processes and thus move beyond `by-eye' analysis.

To accomplish our objective, we first create a suite of numerical simulations designed to test the behaviour of statistics under all combinations of five different physical parameters set to low and high values (i.e., binary settings).  This represents an alternative approach to one factor at a time studies by running a suite of smaller simulations rather than a few higher-resolution studies.  An exhaustive exploration of parameter space, such as we do here, is required to understand the importance of interactions (similar to covariances) obtained from statistical analysis.   With our alternative approach, we run five different fiducial simulations, which we use to determine how susceptible these statistical tools are to random fluctuations.

The 18 statistical tools we develop based on the literature fall into three categories: statistics for (1) the identification of structure, (2) analysis of the properties of turbulence, and (3) analysis of the distributions of data values.  Given these approaches, we define a pseudo-distance metric for each of the different methods, creating a measure of distance between any two data cubes. This parameterizes the differences between the data with a single number.  We measure the distances between the design runs, which have varying physical parameters set by the experimental design and the fiducial data sets using all 18 approaches.  We summarize the resulting distances using a linear model, since this approach cleanly identifies the significance of the physical effects in the presence of other confounding variables.

Our analysis shows some of the proposed methods have a high degree of susceptibility to random fluctuations.  However, we find fourteen statistics that show a significant and reliable response to the underlying physical conditions: delta-variance applied to the centroid field, the Cramer statistic, VCS, VCA, SCF, Dendrograms, Higher Order Statistical Moments, and PDF log-normal width.  The majority of these well-behaved statistics show significant response to changes in the Mach number, the virial parameter, and driving scale. Frequently, the most significant terms in the modelling are the interactions between these parameters.

In exploring these well-behaved statistics further, we find that most work well with the addition of noise mimicking that found in observational data.  Using the distances and sensitivities calculated at a fixed free-fall time improved the reliability of the PCA and bispectrum. All other methods showed either additional scatter in the distances or no change in the sensitivities.

We made an initial comparison of the simulated data to real observations of three nearby molecular clouds using the tools determined to be reliable and sensitive.  We find that the observational data are at least as different from the fiducial data as some of the design simulations are.  However, our analysis shows that none of the simulations in the experimental design display a good agreement with the observations.

This work has identified several tools that have good potential to compare simulated observational data with real observations, broadening the dialogue between theoretical studies and observations.  Our experimental design framework provides a means of evaluating these tools using a well designed set of simulations. Future work to rigorously compare simulations and observations will require a new experimental design that efficiently and more broadly samples the parameter space using a higher resolution set of simulations.

The tools developed in this work are available as part of the {\sc TurbuStat} package (version 0.2), written in {\sc Python}\footnote{\url{http://turbustat.readthedocs.org/}}. The scripts to reproduce these results\footnote{\url{https://github.com/Astroua/AstroStat_Results}} and the simulated data cubes\footnote{\url{http://doi.org/10.11570/17.0006}} have also been made available.

\section*{Acknowledgments}

We are grateful for a series of fruitful discussions with Adam Ginsburg, Simon Glover, Paul Clark and Doug Johnstone in developing this paper.  Blakesley Burkhart provided good insight into the behaviour of several software tools.  The deep and insightful reading by an anonymous referee refined the conclusions and interpretation of the paper.  EWK and EWR are supported by a Discovery Grant from the Natural Sciences and Engineering Research Council Canada (NSERC; RGPIN-2012-355247). EWK is supported by a Postgraduate Scholarship from NSERC.  SSRO acknowledges support from NASA through Hubble Fellowship grant \# 51311.01 awarded by the Space Telescope Science Institute, which is operated by the Association of Universities for Research in Astronomy, Inc., for NASA, under contract NAS 5-26555. SSRO also acknowledges support from NSF grant AST-1510021.  JLL is supported by a Natural Sciences and Engineering Research Council of Canada Discovery Grant  (RGPIN-2015-03895).  This research was enabled in part by support provided by WestGrid (\url{www.westgrid.ca}), Compute Canada - Calcul Canada (\url{www.computecanada.ca}), and CANFAR (\url{www.canfar.net}).
\\
\textbf{Code Bibliography: } {\sc astropy} \citep{astropy}; {\sc radio-astro-tools} (\url{http://radio-astro-tools.github.io/}); {\sc matplotlib} \citep{matplotlib}; {\sc seaborn} (\url{https://seaborn.pydata.org/}); {\sc statsmodels} \citep{seabold2010statsmodels}; {\sc pandas} \url{http://pandas.pydata.org/}; {\sc astrodendro} \url{www.dendrograms.org}; {\sc TurbuStat}-v0.2 (Koch et al., in prep); {\sc lme4} \citep{lme4};  \\

\bibliographystyle{mn2e}
\bibliography{astrostat}

\begin{thebibliography}{}
\makeatletter
\def\mn@urlcharsother{%
\let\do\@makeother
\do\$\do\&\do\#\do\^\do\_\do\%\do\~}
\def\mn@doi{\begingroup
\mn@urlcharsother
\@ifnextchar[%
{\mn@doi@}
{\mn@doi@[]}}
\def\mn@doi@[#1]#2{%
\def\@tempa{#1}%
\ifx\@tempa\@empty
\href{http://dx.doi.org/#2}{doiXX:#2}%
\else
\href{http://dx.doi.org/#2}{#1}%
\fi
\endgroup
}
\def\mn@eprint#1#2{%
\mn@eprint@#1:#2::\@nil}
\def\mn@eprint@arXiv#1{\href{http://arxiv.org/abs/#1}{{\tt arXiv:#1}}}
\def\mn@eprint@dblp#1{\href{http://dblp.uni-trier.de/rec/bibtex/#1.xml}{dblp:#1}}
\def\mn@eprint@#1:#2:#3:#4\@nil{%
\def\@tempa{#1}%
\def\@tempb{#2}%
\def\@tempc{#3}%
\ifx\@tempc\@empty
\let\@tempc\@tempb
\let\@tempb\@tempa
\fi
\ifx\@tempb\@empty
\def\@tempb{arXiv}%
\fi
\@ifundefined{mn@eprint@\@tempb}
{\@tempb:\@tempc}
{\expandafter\expandafter\csname
  mn@eprint@\@tempb\endcsname\expandafter{\@tempc}}%
}

\bibitem[\protect\citeauthoryear{{Adams} \& {Wiseman}}{{Adams} \&
  {Wiseman}}{1994}]{adams94}
{Adams} F.~C.,  {Wiseman} J.~J.,  1994, \mn@doi [\apj] {10.1086/174847}, \href
  {http://adsabs.harvard.edu/abs/1994ApJ...435..693A} {435, 693}

\bibitem[\protect\citeauthoryear{{Astropy Collaboration} et~al.,}{{Astropy
  Collaboration} et~al.}{2013}]{astropy}
{Astropy Collaboration} et~al., 2013, \mn@doi [\aap]
  {10.1051/0004-6361/201322068}, \href
  {http://adsabs.harvard.edu/abs/2013A%26A...558A..33A} {558, A33}

\bibitem[\protect\citeauthoryear{Baringhaus \& Franz}{Baringhaus \&
  Franz}{2004}]{cramer-test}
Baringhaus L.,  Franz C.,  2004, Journal of multivariate analysis, 88, 190

\bibitem[\protect\citeauthoryear{Bastian, Covey \& Meyer}{Bastian
  et~al.}{2010}]{bastian2010}
Bastian N.,  Covey K.~R.,    Meyer M.~R.,  2010, \araa, 48, 339

\bibitem[\protect\citeauthoryear{{Bate}}{{Bate}}{2014}]{Bate2014MNRAS.442..285B}
{Bate} M.~R.,  2014, \mn@doi [\mnras] {10.1093/mnras/stu795}, \href
  {http://adsabs.harvard.edu/abs/2014MNRAS.442..285B} {442, 285}

\bibitem[\protect\citeauthoryear{Bates, M{\"a}chler, Bolker \& Walker}{Bates
  et~al.}{2015}]{lme4}
Bates D.,  M{\"a}chler M.,  Bolker B.,    Walker S.,  2015, \mn@doi [Journal of
  Statistical Software] {10.18637/jss.v067.i01}, 67, 1

\bibitem[\protect\citeauthoryear{Bensch, Stutzki \& Ossenkopf}{Bensch
  et~al.}{2001}]{bensch01-delvar}
Bensch F.,  Stutzki J.,    Ossenkopf V.,  2001, \aap, 366, 636

\bibitem[\protect\citeauthoryear{{Bertram}, {Klessen} \& {Glover}}{{Bertram}
  et~al.}{2015}]{Bertram2015MNRAS.451..196B}
{Bertram} E.,  {Klessen} R.~S.,    {Glover} S.~C.~O.,  2015, \mn@doi [\mnras]
  {10.1093/mnras/stv948}, \href
  {http://adsabs.harvard.edu/abs/2015MNRAS.451..196B} {451, 196}

\bibitem[\protect\citeauthoryear{{Boyden}, {Koch}, {Rosolowsky} \&
  {Offner}}{{Boyden} et~al.}{2016}]{Boyden2016ApJ...833..233B}
{Boyden} R.~D.,  {Koch} E.~W.,  {Rosolowsky} E.~W.,    {Offner} S.~S.~R.,
  2016, \mn@doi [\apj] {10.3847/1538-4357/833/2/233}, \href
  {http://adsabs.harvard.edu/abs/2016ApJ...833..233B} {833, 233}

\bibitem[\protect\citeauthoryear{Brunt \& Heyer}{Brunt \&
  Heyer}{2002a}]{brunt-pca1}
Brunt C.~M.,  Heyer M.~H.,  2002a, \apj, 566, 276

\bibitem[\protect\citeauthoryear{Brunt \& Heyer}{Brunt \&
  Heyer}{2002b}]{brunt-pca2}
Brunt C.~M.,  Heyer M.~H.,  2002b, \apj, 566, 289

\bibitem[\protect\citeauthoryear{{Burkhart} \& {Lazarian}}{{Burkhart} \&
  {Lazarian}}{2016}]{Burkhart2016ApJ...827...26B}
{Burkhart} B.,  {Lazarian} A.,  2016, \mn@doi [\apj]
  {10.3847/0004-637X/827/1/26}, \href
  {http://adsabs.harvard.edu/abs/2016ApJ...827...26B} {827, 26}

\bibitem[\protect\citeauthoryear{Burkhart, Falceta-Gon{\c c}alves, Kowal \&
  Lazarian}{Burkhart et~al.}{2009}]{burkhart-bispectrum}
Burkhart B.,  Falceta-Gon{\c c}alves D.,  Kowal G.,    Lazarian A.,  2009,
  \apj, 693, 250

\bibitem[\protect\citeauthoryear{Burkhart, Lazarian, Goodman \&
  Rosolowsky}{Burkhart et~al.}{2013a}]{burkhart-dendrograms}
Burkhart B.,  Lazarian A.,  Goodman A.,    Rosolowsky E.,  2013a, \apj, 770,
  141

\bibitem[\protect\citeauthoryear{{Burkhart}, {Lazarian}, {Ossenkopf} \&
  {Stutzki}}{{Burkhart} et~al.}{2013b}]{Burkhart2013ApJ...771..123B}
{Burkhart} B.,  {Lazarian} A.,  {Ossenkopf} V.,    {Stutzki} J.,  2013b,
  \mn@doi [\apj] {10.1088/0004-637X/771/2/123}, \href
  {http://adsabs.harvard.edu/abs/2013ApJ...771..123B} {771, 123}

\bibitem[\protect\citeauthoryear{Chepurnov \& Lazarian}{Chepurnov \&
  Lazarian}{2009}]{chepurnov09}
Chepurnov A.,  Lazarian A.,  2009, \apj, 693, 1074

\bibitem[\protect\citeauthoryear{Chepurnov, Gordon, Lazarian \&
  Stanimirovi\'{c}}{Chepurnov et~al.}{2008}]{Chepurnov2008}
Chepurnov A.,  Gordon J.,  Lazarian A.,    Stanimirovi\'{c} S.,  2008, \apj, pp
  1021--1028

\bibitem[\protect\citeauthoryear{Chepurnov, Lazarian, Stanimirovi{\'c}, Heiles
  \& Peek}{Chepurnov et~al.}{2010}]{chepurnov10}
Chepurnov A.,  Lazarian A.,  Stanimirovi{\'c} S.,  Heiles C.,    Peek J. E.~G.,
   2010, \apj, 714, 1398

\bibitem[\protect\citeauthoryear{{Chepurnov}, {Burkhart}, {Lazarian} \&
  {Stanimirovic}}{{Chepurnov} et~al.}{2015}]{Chepurnov2015ApJ...810...33C}
{Chepurnov} A.,  {Burkhart} B.,  {Lazarian} A.,    {Stanimirovic} S.,  2015,
  \mn@doi [\apj] {10.1088/0004-637X/810/1/33}, \href
  {http://adsabs.harvard.edu/abs/2015ApJ...810...33C} {810, 33}

\bibitem[\protect\citeauthoryear{{Collins}, {Xu}, {Norman}, {Li} \&
  {Li}}{{Collins} et~al.}{2010}]{collins10}
{Collins} D.~C.,  {Xu} H.,  {Norman} M.~L.,  {Li} H.,    {Li} S.,  2010,
  \mn@doi [\apjs] {10.1088/0067-0049/186/2/308}, \href
  {http://adsabs.harvard.edu/abs/2010ApJS..186..308C} {186, 308}

\bibitem[\protect\citeauthoryear{{Collins}, {Kritsuk}, {Padoan}, {Li}, {Xu},
  {Ustyugov} \& {Norman}}{{Collins} et~al.}{2012}]{collins-selfgrav}
{Collins} D.~C.,  {Kritsuk} A.~G.,  {Padoan} P.,  {Li} H.,  {Xu} H.,
  {Ustyugov} S.~D.,    {Norman} M.~L.,  2012, \mn@doi [\apj]
  {10.1088/0004-637X/750/1/13}, \href
  {http://adsabs.harvard.edu/abs/2012ApJ...750...13C} {750, 13}

\bibitem[\protect\citeauthoryear{Duch{\^e}ne \& Kraus}{Duch{\^e}ne \&
  Kraus}{2013}]{duchene2013}
Duch{\^e}ne G.,  Kraus A.,  2013, \araa, 51, 269

\bibitem[\protect\citeauthoryear{{Enoch} et~al.,}{{Enoch}
  et~al.}{2006}]{enoch06}
{Enoch} M.~L.,  et~al., 2006, \mn@doi [\apj] {10.1086/498678}, \href
  {http://adsabs.harvard.edu/abs/2006ApJ...638..293E} {638, 293}

\bibitem[\protect\citeauthoryear{Esmaeili \& Shokoohi}{Esmaeili \&
  Shokoohi}{2011}]{pca-econ}
Esmaeili A.,  Shokoohi Z.,  2011, Energy Policy, 39, 1022

\bibitem[\protect\citeauthoryear{Faraway}{Faraway}{2006}]{R_Faraway_2006}
Faraway J.~J.,  2006, Extending Linear Models with R: Generalized Linear, Mixed
  Effects and Nonparametric Regression Models.
Chapman \& Hall/CRC

\bibitem[\protect\citeauthoryear{{Farid} \& {Kosecka}}{{Farid} \&
  {Kosecka}}{2007}]{farid_bispectrum}
{Farid} H.,  {Kosecka} J.,  2007, IEEE Transactions On Image Processing, 8,
  2154

\bibitem[\protect\citeauthoryear{{Federrath}, {Klessen} \&
  {Schmidt}}{{Federrath} et~al.}{2008}]{Federrath08}
{Federrath} C.,  {Klessen} R.~S.,    {Schmidt} W.,  2008, \mn@doi [\apjl]
  {10.1086/595280}, \href {http://adsabs.harvard.edu/abs/2008ApJ...688L..79F}
  {688, L79}

\bibitem[\protect\citeauthoryear{{Federrath}, {Roman-Duval}, {Klessen},
  {Schmidt} \& {Mac Low}}{{Federrath} et~al.}{2010}]{federrath10}
{Federrath} C.,  {Roman-Duval} J.,  {Klessen} R.~S.,  {Schmidt} W.,    {Mac
  Low} M.-M.,  2010, \mn@doi [\aap] {10.1051/0004-6361/200912437}, \href
  {http://adsabs.harvard.edu/abs/2010A%26A...512A..81F} {512, A81}

\bibitem[\protect\citeauthoryear{{Gaches}, {Offner}, {Rosolowsky} \&
  {Bisbas}}{{Gaches} et~al.}{2015}]{Gaches2015ApJ...799..235G}
{Gaches} B.~A.~L.,  {Offner} S.~S.~R.,  {Rosolowsky} E.~W.,    {Bisbas} T.~G.,
  2015, \mn@doi [\apj] {10.1088/0004-637X/799/2/235}, \href
  {http://adsabs.harvard.edu/abs/2015ApJ...799..235G} {799, 235}

\bibitem[\protect\citeauthoryear{Gill \& Henriksen}{Gill \&
  Henriksen}{1990}]{gill-wavelet}
Gill A.~G.,  Henriksen R.~N.,  1990, \apj, 365, L27

\bibitem[\protect\citeauthoryear{Goldreich \& Sridhar}{Goldreich \&
  Sridhar}{1995}]{gs-mhd2}
Goldreich P.,  Sridhar S.,  1995, \apj, 438, 763

\bibitem[\protect\citeauthoryear{Goodman}{Goodman}{2011}]{taste-testing}
Goodman A.~A.,  2011, Computational Star Formation, 270, 511

\bibitem[\protect\citeauthoryear{Goodman, Rosolowsky, Borkin, Foster, Halle,
  Kauffmann \& Pineda}{Goodman et~al.}{2009}]{dendrograms-nature}
Goodman A.~A.,  Rosolowsky E.~W.,  Borkin M.~A.,  Foster J.~B.,  Halle M.,
  Kauffmann J.,    Pineda J.~E.,  2009, Nature, 457, 63

\bibitem[\protect\citeauthoryear{{Habib}, {Heitmann}, {Higdon}, {Nakhleh} \&
  {Williams}}{{Habib} et~al.}{2007}]{Habib07}
{Habib} S.,  {Heitmann} K.,  {Higdon} D.,  {Nakhleh} C.,    {Williams} B.,
  2007, \mn@doi [\prd] {10.1103/PhysRevD.76.083503}, \href
  {http://adsabs.harvard.edu/abs/2007PhRvD..76h3503H} {76, 083503}

\bibitem[\protect\citeauthoryear{Hagihira, Takashina, Mori, Mashimo \&
  Yoshiya}{Hagihira et~al.}{2001}]{Bicoherence-Hagihira}
Hagihira S.,  Takashina M.,  Mori T.,  Mashimo T.,    Yoshiya I.,  2001,
  Anesthesia {\&} Analgesia, 93, 966

\bibitem[\protect\citeauthoryear{{Heitmann}, {Higdon}, {Nakhleh} \&
  {Habib}}{{Heitmann} et~al.}{2006}]{Heitmann06}
{Heitmann} K.,  {Higdon} D.,  {Nakhleh} C.,    {Habib} S.,  2006, \mn@doi
  [\apjl] {10.1086/506448}, \href
  {http://adsabs.harvard.edu/abs/2006ApJ...646L...1H} {646, L1}

\bibitem[\protect\citeauthoryear{{Heitmann}, {Higdon}, {White}, {Habib},
  {Williams}, {Lawrence} \& {Wagner}}{{Heitmann} et~al.}{2009}]{Heitmann09}
{Heitmann} K.,  {Higdon} D.,  {White} M.,  {Habib} S.,  {Williams} B.~J.,
  {Lawrence} E.,    {Wagner} C.,  2009, \mn@doi [\apj]
  {10.1088/0004-637X/705/1/156}, \href
  {http://adsabs.harvard.edu/abs/2009ApJ...705..156H} {705, 156}

\bibitem[\protect\citeauthoryear{{Heitmann}, {White}, {Wagner}, {Habib} \&
  {Higdon}}{{Heitmann} et~al.}{2010}]{Heitmann10}
{Heitmann} K.,  {White} M.,  {Wagner} C.,  {Habib} S.,    {Higdon} D.,  2010,
  \mn@doi [\apj] {10.1088/0004-637X/715/1/104}, \href
  {http://adsabs.harvard.edu/abs/2010ApJ...715..104H} {715, 104}

\bibitem[\protect\citeauthoryear{{Heyer} \& {Brunt}}{{Heyer} \&
  {Brunt}}{2004}]{heyer04}
{Heyer} M.~H.,  {Brunt} C.~M.,  2004, \mn@doi [\apjl] {10.1086/425978}, \href
  {http://adsabs.harvard.edu/abs/2004ApJ...615L..45H} {615, L45}

\bibitem[\protect\citeauthoryear{Heyer \& Schloerb}{Heyer \&
  Schloerb}{1997}]{heyer-pca}
Heyer M.~H.,  Schloerb F.~P.,  1997, \apj, 475, 173

\bibitem[\protect\citeauthoryear{Higdon, Gattiker, Williams \& Rightley}{Higdon
  et~al.}{2012}]{higdon2012computer}
Higdon D.,  Gattiker J.,  Williams B.,    Rightley M.,  2012, Journal of the
  American Statistical Association

\bibitem[\protect\citeauthoryear{Huber, Wiley \& InterScience}{Huber
  et~al.}{1981}]{Huber-RobustStatistics}
Huber P.,  Wiley J.,    InterScience W.,  1981, {Robust statistics}.
Wiley New York

\bibitem[\protect\citeauthoryear{Hunter}{Hunter}{2007}]{matplotlib}
Hunter J.~D.,  2007, \mn@doi [Computing In Science \& Engineering]
  {10.1109/MCSE.2007.55}, 9, 90

\bibitem[\protect\citeauthoryear{{Ivezi{\'c}}, {Connelly}, {VanderPlas} \&
  {Gray}}{{Ivezi{\'c}} et~al.}{2014}]{Ivezic2014sdmm.book.....I}
{Ivezi{\'c}} {\v Z}.,  {Connelly} A.~J.,  {VanderPlas} J.~T.,    {Gray} A.,
  2014, {Statistics, Data Mining, and Machine Learning in Astronomy}

\bibitem[\protect\citeauthoryear{{Kandel}, {Lazarian} \& {Pogosyan}}{{Kandel}
  et~al.}{2016}]{kandel-vca}
{Kandel} D.,  {Lazarian} A.,    {Pogosyan} D.,  2016, \mn@doi [\mnras]
  {10.1093/mnras/stw1296}, \href
  {http://adsabs.harvard.edu/abs/2016MNRAS.461.1227K} {461, 1227}

\bibitem[\protect\citeauthoryear{Kennicutt}{Kennicutt}{1998}]{k98}
Kennicutt R.~C.,  1998, ApJ, 498, 541

\bibitem[\protect\citeauthoryear{Kirk, Johnstone \& Basu}{Kirk
  et~al.}{2009}]{kirk2009}
Kirk H.,  Johnstone D.,    Basu S.,  2009, \apj, 699, 1433

\bibitem[\protect\citeauthoryear{{Kitsionas} et~al.,}{{Kitsionas}
  et~al.}{2009}]{Kitsionas2009A&A...508..541K}
{Kitsionas} S.,  et~al., 2009, \mn@doi [\aap] {10.1051/0004-6361/200811170},
  \href {http://adsabs.harvard.edu/abs/2009A%26A...508..541K} {508, 541}

\bibitem[\protect\citeauthoryear{Kowal, Lazarian \& Beresnyak}{Kowal
  et~al.}{2007}]{Kowal2007}
Kowal G.,  Lazarian A.,    Beresnyak A.,  2007, \apj, 658, 423

\bibitem[\protect\citeauthoryear{Kroupa, Tout \& Gilmore}{Kroupa
  et~al.}{1993}]{kroupa-imf}
Kroupa P.,  Tout C.~A.,    Gilmore G.,  1993, MNRAS, 262, 545

\bibitem[\protect\citeauthoryear{Krumholz, Dekel \& McKee}{Krumholz
  et~al.}{2012a}]{kdm12}
Krumholz M.~R.,  Dekel A.,    McKee C.~F.,  2012a, \apj, 745, 69

\bibitem[\protect\citeauthoryear{{Krumholz}, {Klein} \& {McKee}}{{Krumholz}
  et~al.}{2012b}]{Krumholz2012ApJ...754...71K}
{Krumholz} M.~R.,  {Klein} R.~I.,    {McKee} C.~F.,  2012b, \mn@doi [\apj]
  {10.1088/0004-637X/754/1/71}, \href
  {http://adsabs.harvard.edu/abs/2012ApJ...754...71K} {754, 71}

\bibitem[\protect\citeauthoryear{{Larson}}{{Larson}}{1981}]{larson81}
{Larson} R.~B.,  1981, \mnras, \href
  {http://adsabs.harvard.edu/abs/1981MNRAS.194..809L} {194, 809}

\bibitem[\protect\citeauthoryear{{Lawrence}, {Heitmann}, {White}, {Higdon},
  {Wagner}, {Habib} \& {Williams}}{{Lawrence} et~al.}{2010}]{Lawrence10}
{Lawrence} E.,  {Heitmann} K.,  {White} M.,  {Higdon} D.,  {Wagner} C.,
  {Habib} S.,    {Williams} B.,  2010, \mn@doi [\apj]
  {10.1088/0004-637X/713/2/1322}, \href
  {http://adsabs.harvard.edu/abs/2010ApJ...713.1322L} {713, 1322}

\bibitem[\protect\citeauthoryear{Lazarian \& Esquivel}{Lazarian \&
  Esquivel}{2003}]{Lazarian2003}
Lazarian A.,  Esquivel A.,  2003, \apjl, pp 37--40

\bibitem[\protect\citeauthoryear{Lazarian \& Pogosyan}{Lazarian \&
  Pogosyan}{2000}]{vca}
Lazarian A.,  Pogosyan D.,  2000, \apj, 537, 720

\bibitem[\protect\citeauthoryear{Lazarian \& Pogosyan}{Lazarian \&
  Pogosyan}{2004}]{lp04}
Lazarian A.,  Pogosyan D.,  2004, ApJ, 616, 943

\bibitem[\protect\citeauthoryear{Lazarian \& Pogosyan}{Lazarian \&
  Pogosyan}{2006}]{lp06}
Lazarian A.,  Pogosyan D.,  2006, \apj, 652, 1348

\bibitem[\protect\citeauthoryear{Leroy et~al.,}{Leroy
  et~al.}{2013}]{leroy-kslaw}
Leroy A.~K.,  et~al., 2013, \aj, 146, 19

\bibitem[\protect\citeauthoryear{Lithwick \& Goldreich}{Lithwick \&
  Goldreich}{2001}]{compressible-mhd}
Lithwick Y.,  Goldreich P.,  2001, \apj, 562, 279

\bibitem[\protect\citeauthoryear{{Lombardi}, {Alves} \& {Lada}}{{Lombardi}
  et~al.}{2015}]{Lombardi2015}
{Lombardi} M.,  {Alves} J.,    {Lada} C.~J.,  2015, \mn@doi [\aap]
  {10.1051/0004-6361/201525650}, \href
  {http://adsabs.harvard.edu/abs/2015A%26A...576L...1L} {576, L1}

\bibitem[\protect\citeauthoryear{Mac~Low}{Mac~Low}{1999}]{maclow99}
Mac~Low M.-M.,  1999, \apj, 524, 169

\bibitem[\protect\citeauthoryear{McKee \& Ostriker}{McKee \&
  Ostriker}{2007}]{mo07}
McKee C.~F.,  Ostriker E.~C.,  2007, \araa, 45, 565

\bibitem[\protect\citeauthoryear{{McKee}, {Li} \& {Klein}}{{McKee}
  et~al.}{2010}]{mckee10}
{McKee} C.~F.,  {Li} P.~S.,    {Klein} R.~I.,  2010, \mn@doi [\apj]
  {10.1088/0004-637X/720/2/1612}, \href
  {http://adsabs.harvard.edu/abs/2010ApJ...720.1612M} {720, 1612}

\bibitem[\protect\citeauthoryear{Muggeo}{Muggeo}{2003}]{muggeo2003estimating}
Muggeo V.~M.,  2003, Statistics in medicine, 22, 3055

\bibitem[\protect\citeauthoryear{O'Shea, Bryan, Bordner, Norman, Abel, Harkness
  \& Kritsuk}{O'Shea et~al.}{2004}]{enzo}
O'Shea B.~W.,  Bryan G.,  Bordner J.,  Norman M.~L.,  Abel T.,  Harkness R.,
  Kritsuk A.,  2004, arXiv, p.~3044

\bibitem[\protect\citeauthoryear{Offner, Krumholz, Klein \& McKee}{Offner
  et~al.}{2008}]{offner2008}
Offner S. S.~R.,  Krumholz M.~R.,  Klein R.~I.,    McKee C.~F.,  2008, \aj,
  136, 404

\bibitem[\protect\citeauthoryear{{Offner}, {Clark}, {Hennebelle}, {Bastian},
  {Bate}, {Hopkins}, {Moraux} \& {Whitworth}}{{Offner}
  et~al.}{2014}]{offner-imf}
{Offner} S.~S.~R.,  {Clark} P.~C.,  {Hennebelle} P.,  {Bastian} N.,  {Bate}
  M.~R.,  {Hopkins} P.~F.,  {Moraux} E.,    {Whitworth} A.~P.,  2014, \mn@doi
  [Protostars and Planets VI] {10.2458/azu_uapress_9780816531240-ch003}, \href
  {http://adsabs.harvard.edu/abs/2014prpl.conf...53O} {pp 53--75}

\bibitem[\protect\citeauthoryear{Ossenkopf, Krips \& Stutzki}{Ossenkopf
  et~al.}{2008a}]{oss08II-delvar}
Ossenkopf V.,  Krips M.,    Stutzki J.,  2008a, \aap, 485, 719

\bibitem[\protect\citeauthoryear{Ossenkopf, Krips \& Stutzki}{Ossenkopf
  et~al.}{2008b}]{oss08I-delvar}
Ossenkopf V.,  Krips M.,    Stutzki J.,  2008b, \aap, 485, 917

\bibitem[\protect\citeauthoryear{{Padoan} \& {Nordlund}}{{Padoan} \&
  {Nordlund}}{2002}]{padoan02}
{Padoan} P.,  {Nordlund} {\AA}.,  2002, \mn@doi [\apj] {10.1086/341790}, \href
  {http://adsabs.harvard.edu/abs/2002ApJ...576..870P} {576, 870}

\bibitem[\protect\citeauthoryear{{Padoan}, {Nordlund} \& {Jones}}{{Padoan}
  et~al.}{1997}]{Padoan1997MNRAS.288..145P}
{Padoan} P.,  {Nordlund} A.,    {Jones} B.~J.~T.,  1997, \mn@doi [\mnras]
  {10.1093/mnras/288.1.145}, \href
  {http://adsabs.harvard.edu/abs/1997MNRAS.288..145P} {288, 145}

\bibitem[\protect\citeauthoryear{Padoan, Rosolowsky \& Goodman}{Padoan
  et~al.}{2001}]{padoan2001}
Padoan P.,  Rosolowsky E.~W.,    Goodman A.~A.,  2001, \apj, 547, 862

\bibitem[\protect\citeauthoryear{Padoan, Goodman \& Juvela}{Padoan
  et~al.}{2003}]{padoan-scf}
Padoan P.,  Goodman A.~A.,    Juvela M.,  2003, \apj, 588, 881

\bibitem[\protect\citeauthoryear{{Padoan}, {Haugb{\o}lle} \&
  {Nordlund}}{{Padoan} et~al.}{2012}]{padoan2012}
{Padoan} P.,  {Haugb{\o}lle} T.,    {Nordlund} {\AA}.,  2012, \mn@doi [\apjl]
  {10.1088/2041-8205/759/2/L27}, \href
  {http://adsabs.harvard.edu/abs/2012ApJ...759L..27P} {759, L27}

\bibitem[\protect\citeauthoryear{{Pineda}, {Caselli} \& {Goodman}}{{Pineda}
  et~al.}{2008}]{pineda08}
{Pineda} J.~E.,  {Caselli} P.,    {Goodman} A.~A.,  2008, \mn@doi [\apj]
  {10.1086/586883}, \href {http://adsabs.harvard.edu/abs/2008ApJ...679..481P}
  {679, 481}

\bibitem[\protect\citeauthoryear{Ridge et~al.,}{Ridge
  et~al.}{2006}]{complete-data}
Ridge N.~A.,  et~al., 2006, \aj, 131, 2921

\bibitem[\protect\citeauthoryear{Rosolowsky}{Rosolowsky}{2012}]{rosolowsky-scma}
Rosolowsky E.,  2012, in , Statistical Challenges in Modern Astronomy V.
Springer New York, New York, NY, pp 367--382

\bibitem[\protect\citeauthoryear{Rosolowsky, Goodman, Wilner \&
  Williams}{Rosolowsky et~al.}{1999}]{scf}
Rosolowsky E.~W.,  Goodman A.~A.,  Wilner D.~J.,    Williams J.~P.,  1999,
  \apj, 524, 887

\bibitem[\protect\citeauthoryear{Rosolowsky, Pineda, Kauffmann \&
  Goodman}{Rosolowsky et~al.}{2008}]{dendrograms}
Rosolowsky E.~W.,  Pineda J.~E.,  Kauffmann J.,    Goodman A.~A.,  2008, ApJ,
  679, 1338

\bibitem[\protect\citeauthoryear{Sacks, Welch, Mitchell \& Wynn}{Sacks
  et~al.}{1989}]{sacks1989design}
Sacks J.,  Welch W.~J.,  Mitchell T.~J.,    Wynn H.~P.,  1989, Statistical
  science, pp 409--423

\bibitem[\protect\citeauthoryear{Santner, Williams \& Notz}{Santner
  et~al.}{2013}]{santner2013design}
Santner T.~J.,  Williams B.~J.,    Notz W.~I.,  2013, The design and analysis
  of computer experiments.
Springer Science \& Business Media

\bibitem[\protect\citeauthoryear{{Schneider}, {Knox}, {Habib}, {Heitmann},
  {Higdon} \& {Nakhleh}}{{Schneider} et~al.}{2008}]{Schneider08}
{Schneider} M.~D.,  {Knox} L.,  {Habib} S.,  {Heitmann} K.,  {Higdon} D.,
  {Nakhleh} C.,  2008, \mn@doi [\prd] {10.1103/PhysRevD.78.063529}, \href
  {http://adsabs.harvard.edu/abs/2008PhRvD..78f3529S} {78, 063529}

\bibitem[\protect\citeauthoryear{Sch{\"o}ier, van~der Tak, van Dishoeck \&
  Black}{Sch{\"o}ier et~al.}{2005}]{lamda}
Sch{\"o}ier F.~L.,  van~der Tak F.~F.~S.,  van Dishoeck E.~F.,    Black J.~H.,
  2005, \aap, 432, 369

\bibitem[\protect\citeauthoryear{Seabold \& Perktold}{Seabold \&
  Perktold}{2010}]{seabold2010statsmodels}
Seabold S.,  Perktold J.,  2010, in 9th Python in Science Conference.

\bibitem[\protect\citeauthoryear{Shetty, Glover, Dullemond \& Klessen}{Shetty
  et~al.}{2011}]{shetty-xfac}
Shetty R.,  Glover S.~C.,  Dullemond C.~P.,    Klessen R.~S.,  2011, \mnras,
  412, 1686

\bibitem[\protect\citeauthoryear{Shih}{Shih}{2009}]{mathematical-morphology}
Shih F.~Y.,  2009, {Image Processing and Mathematical Morphology}.
Fundamentals and Applications, CRC PressI Llc

\bibitem[\protect\citeauthoryear{Sridhar \& Goldreich}{Sridhar \&
  Goldreich}{1994}]{gs-mhd1}
Sridhar S.,  Goldreich P.,  1994, \apj, 432, 612

\bibitem[\protect\citeauthoryear{Stanimirovi{\'c} \& Lazarian}{Stanimirovi{\'c}
  \& Lazarian}{2001}]{stan01-sps}
Stanimirovi{\'c} S.,  Lazarian A.,  2001, \apj, 551, L53

\bibitem[\protect\citeauthoryear{Stutzki, Bensch, Heithausen, Ossenkopf \&
  Zielinsky}{Stutzki et~al.}{1998}]{stutzki98}
Stutzki J.,  Bensch F.,  Heithausen A.,  Ossenkopf V.,    Zielinsky M.,  1998,
  A{\&}A, 336, 697

\bibitem[\protect\citeauthoryear{Truelove, Klein, McKee, Holliman, Howell \&
  Greenough}{Truelove et~al.}{1997}]{truelove97}
Truelove J.~K.,  Klein R.~I.,  McKee C.~F.,  Holliman J. H.~I.,  Howell L.~H.,
    Greenough J.~A.,  1997, \apjl, 489, L179

\bibitem[\protect\citeauthoryear{{Wang}, {Li}, {Abel} \& {Nakamura}}{{Wang}
  et~al.}{2010}]{wang10}
{Wang} P.,  {Li} Z.-Y.,  {Abel} T.,    {Nakamura} F.,  2010, \mn@doi [\apj]
  {10.1088/0004-637X/709/1/27}, \href
  {http://adsabs.harvard.edu/abs/2010ApJ...709...27W} {709, 27}

\bibitem[\protect\citeauthoryear{Williams, de Geus \& Blitz}{Williams
  et~al.}{1994}]{clumpfind}
Williams J.~P.,  de Geus E.~J.,    Blitz L.,  1994, ApJ, 428, 693

\bibitem[\protect\citeauthoryear{{Wiseman} \& {Adams}}{{Wiseman} \&
  {Adams}}{1994}]{wiseman94}
{Wiseman} J.~J.,  {Adams} F.~C.,  1994, \mn@doi [\apj] {10.1086/174848}, \href
  {http://adsabs.harvard.edu/abs/1994ApJ...435..708W} {435, 708}

\bibitem[\protect\citeauthoryear{Yang et~al.,}{Yang
  et~al.}{2010}]{pca-genetics}
Yang J.,  et~al., 2010, Nature genetics, 42, 565

\bibitem[\protect\citeauthoryear{Yeremi, Flynn, Offner, Loeppky \&
  Rosolowsky}{Yeremi et~al.}{2014}]{Yer14}
Yeremi M.,  Flynn M.,  Offner S.,  Loeppky J.,    Rosolowsky E.,  2014, \apj,
  783, 93

\makeatother
\end{thebibliography}

\appendix

\section{Method Sensitivities} 
\label{app:method_sensitivities}

We describe the parameter sensitivities of the individual methods modelled in \S\ref{sub:sensitivity_analysis}, combined with the added analyses in \S\ref{sub:sensitivity_to_basic_observables} and Appendices \ref{app:resn}, \ref{app:fiducial_sensitivity_to_temperature} and \ref{app:re_scaling_to_a_common_linewidth_and_intensity}.

\subsection{VCS} 
\label{appsub:vcs_response}

The VCS appears to show sensitivity to all first-order parameter changes. The sensitivities change when split into its large- and small-scale regimes, where the transition point varies between 700 and 1000 m/s across the simulation set. The strongest sensitivity is to changes in \mach. The large-scale region is also sensitive to changes in \virial, along with the second-order \mach:\virial\ interaction.  The large-scale is also sensitive to changes in the driving scale, \drive, though this is set by the location of the break point in the fitting.  Finally, the large-scale regime is sensitive to changes in \plasbeta, perhaps reflecting where the additional energy in the magnetic field leads to a more anisotropic distribution.

Unlike the rest of the methods, our modelling of the VCS makes comparing the noise-added and noiseless cases more difficult. With the inclusion of noise, small spectral scales within the data are lost and the slope flattens to zero, as expected for white noise. Our analysis of the VCS is then limited since the range where the small-scale component is eliminated or severely reduced and the slope in this region is significantly flattened. The large-scale VCS response shown in Figure \ref{fig:noisy_all_results} is then a mix of the two components in Figure \ref{fig:all_results}. Indeed, the responses in the noise-added case are a weak combination of the two components in the noiseless case.

We also find that the VCS response changes with resolution and AMR (Appendix \ref{app:resn}), and changes in the fiducial temperature (Appendix \ref{app:fiducial_sensitivity_to_temperature}).
These results also suggest that caution should be used when directly comparing the VCS spectra between inhomogeneous data sets.


\subsection{SCF} 
\label{appsub:scf_response}

The SCF shows a reliable response to physical changes for each analysis based on the quality testing procedure (\S\ref{sub:determining_quality_of_statistics}). It shows strong responses to changes in \mach\ and \drive, along with their interaction. One of the strongest terms is the interactions \solfrac:\mach. Because we define the distance metric using the two-dimensional correlation surface, the anisotropy difference from changing \solfrac\ is more readily measurable. As noted in \S\ref{sub:sensitivity_to_basic_observables}, the SCF's dependency on \mach\ may be driven by line width changes. However, this sensitivity persists when the analysis is performed on simulated cubes that are regridded to have the same average line width (Appendix \ref{app:re_scaling_to_a_common_linewidth_and_intensity}).  Thus, this sensitivity is not {\em exclusively} a measure of line width differences.

Like the VCS, the SCF is sensitive to a large fraction of the parameter space, showing a strong response to all first order effects except \plasbeta. While this makes disentangling changes in distance more difficult, it allows for a larger range of applications. For example, if reasonable estimates can be attained for some of the physical parameters, such that they are now fixed in the model and not fit for, the SCF may be used to constrain the remaining parameters. This may be of particular use for parameters, like \solfrac, which are more difficult to measure directly from observational data.


\subsection{Delta-Variance} 
\label{appsub:delvar_response}

The delta-variance applied to the centroid field primarily shows responses to changing \drive, particularly for comparisons of the slopes. Since it measures the amount of structure (variance) on different size scales, the primary dependence on \drive\ is expected. Changes in the range of \drive\ affect the size distribution more dramatically than changes in the other parameters.

Since we define two distance metrics for the delta-variance, we can explore how different responses can be obtained from the same statistic. The slope-based method shows no sensitivity to \solfrac, but the normalized difference between the curves themselves does. Changing \solfrac\ changes the shape of the underlying density distribution, particularly on large-scales \citep[e.g.,][]{federrath10}. Though a symmetric kernel is used for the delta-variance, its response will differ slightly when applied to asymmetric regions. The slope-based distance does not account for this change since the larger-scale response deviates from a power-law relation and is excluded from the fitting range (\S\ref{sub:delta_variance}). This may be of limited use when applied to observations, however, since the non-periodic boundaries will be dominated by the spatial coverage of the data.

When applied to the integrated intensity maps, we find no discernable signal in either distance metric. We suggest in \S\ref{sec:limitations_of_this_study} that this is related to the choice of tracer and should be an informative method applied to a column density map, as has been shown in much of the previous work on delta-variance \citep[e.g.,][]{oss08II-delvar}.

\subsection{VCA} 
\label{appsub:vca_response}

Similar to the complementary VCS, the VCA exhibits sensitivity to all first-order effects, except \plasbeta. Unlike the VCS, though, the VCA sensitivity is most strongly affected by changes in \mach, along with its interaction with \drive. Since the VCA is a measure of the spatial power spectrum, with modifications from the velocity component averaged over, it should show several similatites to the SPS. Indeed, the SPS is most strongly affected by changes in \mach. However, as discussed below, the fluctuations between fiducials cause significant scatter in the SPS distances, which are less prominent for the VCA since the information along the spectral dimensions is utilized. Unlike the VCS, whose response we find depends on changes in the resolution and the thermal line width, the VCA shows no sensitivities to these changes. However, increasing the resolution of the simulation to increase the inertial range will provide a more reliable measure for the VCA. Furthermore, the VCA distance can be re-formulated to use the two-dimensional power spectrum, providing a measure of anisotropy \citep{kandel-vca}.

\subsection{Spatial Power Spectrum} 
\label{appsub:sps_response}

The SPS, formulated from the integrated intensity map, is primarily sensitive to changes in \mach\ and \plasbeta. While the quality testing suggests there is discernible signal in the responses, we are limited in our interpretation of the sensitivities due to the measurable fluctuations between the fiducials, which causes a significant amount of scatter (see \S\ref{sec:limitations_of_this_study}). Formally, the SPS is the thick velocity slice limit of the VCA, discussed above. The scatter is more significant here than the VCA since the VCA makes use of the velocity channels, and therefore is based on a larger number of samples.

\subsection{Bispectrum} 
\label{appsub:bispec_response}

The bispectrum encodes the phase information lost when calculating the spatial power spectrum, and thus should show {\it more} sensitivity to parameters than the spatial power spectrum. We define the distance in a simplified manner using the bicoherence (Equation \ref{eq:bispec_distance}), and this simplification appears to capture interesting variations nonetheless. The time-averaged analysis shows significant scatter between the different fiducials, but the free-fall only analysis does not. This suggests an evolution in the phase structure with time. Considering the free-fall sensitivities, we find the that bispectrum shows sensitivity to most of the first order parameters, except \plasbeta. Perhaps more interesting are the very strong higher-order interactions. The most significant terms in the modelling (Figure \ref{fig:freefall_all_results}) are actually {\it third-order} interaction terms, something that no other method examined here shows. This is a strong suggestion that the bispectrum and the related bicoherence may be useful tools in a variety of cases. In future work, we will explore the time evolution of the bispectrum and how this relates to the strong interaction sensitivities. The azimuthal and radial averaging approaches from \citet{Burkhart2016ApJ...827...26B} are another comparison approach that should be examined in the statistical framework we present here.

\subsection{MVC} 
\label{appsub:mvc_response}

The MVC provides a measure of the velocity field index with an added correction term to account for density fluctuations \citep{Lazarian2003}. We find that the MVC response is nearly identical to that of the SPS. However, it also suffers from the turbulence fluctuations between the fiducial runs and that inhibits a closer inspection of the derived sensitivities.

We clarify here a statement from \citet{Boyden2016ApJ...833..233B} which claimed the MVC was sensitive only to the initial random seed, leading to its exclusion from their analysis. This lack of sensitivity was due to a bug in the {\sc TurbuStat} code, which has since been fixed.

\subsection{PCA Eigenvalues} 
\label{appsub:pca_response}

Our formulation of the PCA, comparing the magnitudes of the eigenvalues, is primarily sensitive to changes in \mach\ and \drive. Since this is a decomposition of the covariance matrix, this suggests these parameters play the primary role in shaping the correlations between channels. However, this interpretation is limited by the significant fiducial-fiducial scatter when averaging over all time-steps. When only comparing the simulations at a fixed free-fall time, the scatter is dramatically reduced, suggesting that changes in the covariance structure with time is measurable using this approach. In the free-fall time analysis, the PCA shows sensitivity to \virial, \solfrac\ and \plasbeta\ but is no longer sensitive to changes in \drive. Since the free-fall analysis shows significantly less scatter, the sensitivities in this case are more reliable.

\subsection{Higher Order Statistical Moments} 
\label{appsub:hosa_response}

The skewness and kurtosis are sensitive to all first order terms except \solfrac.  As both statistics consider the distribution of small scale intensity variations, this lack of sensitivity to \solfrac\ is unsurprising since this definition will not take spatial asymmetries into account.  Both measures show similar sensitivities, with the skewness having a stronger measured response since it has less intrinsic scatter than kurtosis. Since both are measures of how the local intensity distribution changes, it makes sense that they are sensitive to how the structure changes with the various parameters, particularly with \drive. In many ways, this is encoding the structure differences that can be seen in Figure \ref{fig:moment0s}.

Since the skewness and kurtosis are measuring small-scale fluctuations, they are sensitive to changes in the resolution, as well as the effect of AMR. In both cases, the maximum intensity variation within the same pixel area is larger, leading to more prominent tails in the skewness and kurtosis distributions.

\subsection{Integrated Intensity PDF} 
\label{appsub:pdf_response}

The connection between \mach\ and \solfrac\ for the underlying three-dimensional density distribution is well-established theoretically \citep{Padoan1997MNRAS.288..145P} and numerically \cite[e.g.,][]{Federrath08}. The turbulent field will approach a log-normal distribution, the width of which depends on \mach\ and \solfrac. Simulations show that the two-dimensional PDF, from the column density or integrated intensity, is also well-modelled by a log-normal form. We find that the width of the integrated intensity PDF strongly depends on changes in \mach\ but also has some dependence on \drive\ and \virial\ due to the degeneracy of these variables in the experimental design (\S\ref{sub:sensitivity_to_basic_observables}). There is no significant sensitivity to \solfrac. This likely results from the scatter between the distances, which results from few pixels having high-intensities for the small map sizes. Previous work at higher pixel resolutions show a clear response to changes in \solfrac\ \citep[or the related forcing parameter, $b$;][]{Federrath08}.

\subsection{Cramer} 
\label{appsub:cramer_response}

The Cramer statistic, as proposed in \citet{Yer14}, primarily shows sensitivity only to changes in \mach. In this way, it shows similarities to the PDF log-normal width, as the definition (Equation \ref{eq:cramer_met}) is a measure of the difference of the scatter within each pairs of channels between two data sets. There are concerns, however, due to its sensitivity to changes in the temperature (Appendix \ref{app:fiducial_sensitivity_to_temperature}). The inter-comparison between velocity channels introduces a dependency on the difference in the line widths of the data. In Appendix \ref{app:re_scaling_to_a_common_linewidth_and_intensity}, we discuss how removing the line width difference changes the Cramer statistics sensitivity dramatically. Rather than showing a dependence to only the \mach, scaling to a common line width makes the Cramer statistic sensitive to changes in all of the first order parameters, except \plasbeta. Because of this line width dependency, however, its usefulness from comparisons with observations is likely limited in the form presented here.

\subsection{Genus} 
\label{appsub:genus_response}

The genus statistic only shows a strong response to changes in \drive, with weaker sensitivities to second order terms with \mach\ and \virial. This response is similar to the delta-variance centroid slopes response, as both provide a measure of the dominant size scale of regions in the data.


\section{Resolution Effects in the Simulations}
\label{app:resn}

Our specific conclusions depend in some ways on the limited root grid resolution of the simulation set used for the analysis.  This may prevent these results from being connected to other studies and may undermine the comparison to observations.  Here we explore the effect of changing the resolution of the fiducial simulations and evaluate how it impacts our results.

\begin{figure}
\includegraphics[width=0.5\textwidth]{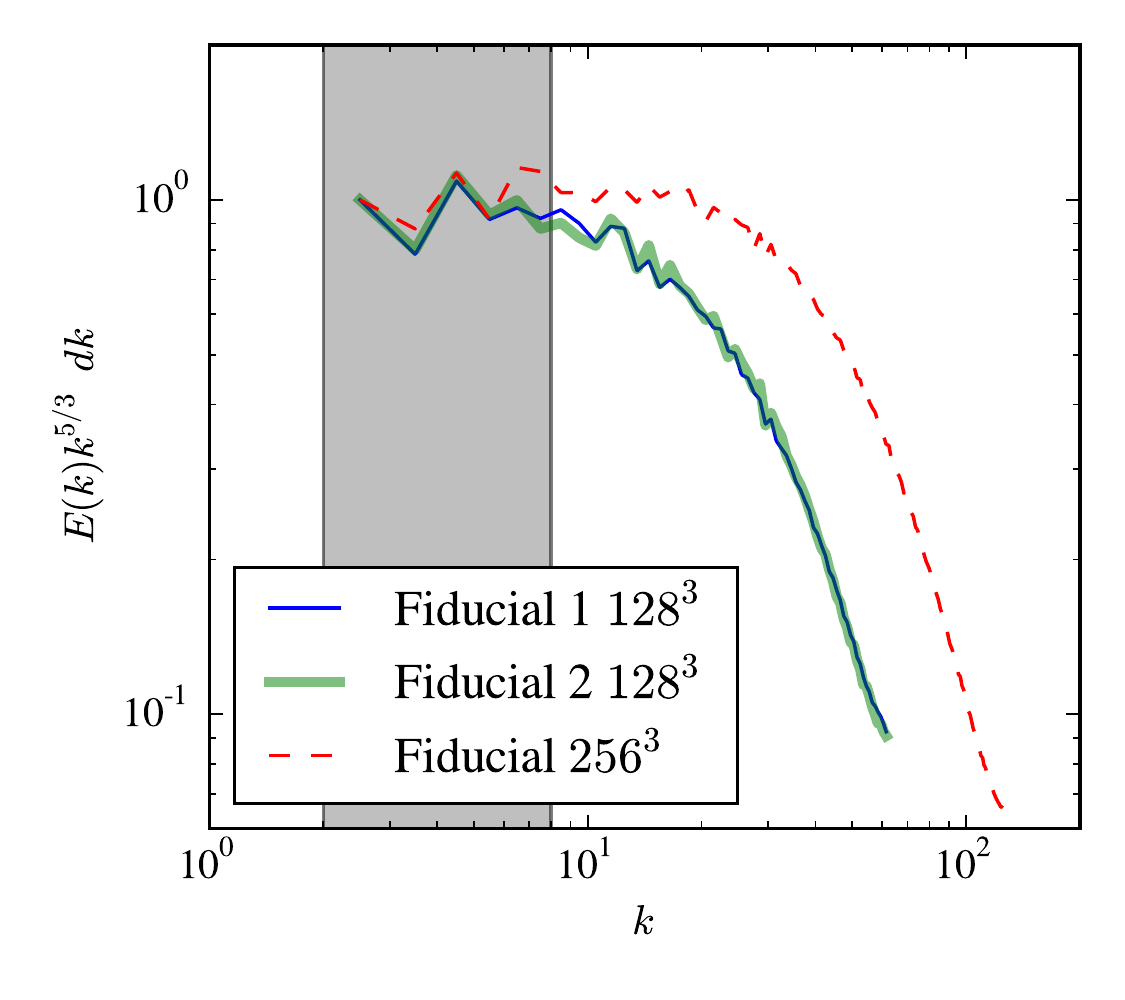}
\caption{\label{fig:powerspec} Normalized, compensated power spectra for two fiducial simulations (same initial conditions, different random seeds, $128^3$) and one higher resolution simulation (same initial conditions, $256^3$).  The figure shows the small inertial ranges of the design simulations.  Higher resolution simulations show a broader inertial range and the limited size of the simulations will affect measures based on the turbulent power spectrum.  The shaded area indicates the driving scales for the simulation.}
\end{figure}

In Figure \ref{fig:powerspec}, we present the power spectra for three different fiducial simulations.  All simulations have the same initial conditions (Table \ref{tab:design}).  The figure presents two of the fiducial simulations at $128^3$ resolution but with different random seeds and one fiducial at $256^3$ resolution.  The figure shows that the simuations have a small inertial range: to $k\sim 16$ in the case of the $128^3$.  The turnover in the power spectrum shows that excessive energy is being lost on small scales when compared to a better-resolved cascade.  The power spectra are shown at $t/t_c = 2.0$, right before self-gravity and adaptive refinement are turned on.

We also process the simulations through our framework to compare the statistics at high and low resolution root grid.  We compute \thco\ emission line data for the $256^3$ simulation using the same approach as for the $128^3$.  We compare these maps to those maps generated from $128^3$ fiducials and we average and regrid the emission from the $256^3$ to a spatial resolution of $128$ pixels along the spatial dimensions.  The regridding experiment mimics the importance of unresolved physics on scales smaller than the resolution that are averaged over in the analysis.  We then calculate the empirical probability that the pseudo-distances from the $256^3$ data are significantly larger than the $128^3$ fiducial distances from each other.  This highlights which statistics are affected by resolution effects.  We calculate the probability $p$ that the distance of the high resolution is consistent with the low-resolution fiducials.  If we adopt $p<0.01$ to indicate significance, we find that a small set of statistics show a significant sensitivity to root grid resolution: Cramer, skewness and kurtosis, and VCS.  This list includes primarily those statistics most connected to the pixel value distributions.  Since the high resolution simulations have more power at the small scales, this behaviour is expected, particularly for the VCS, which is most affected on small velocity scales.

We then carried out a parallel analysis by regridding the emission line data cubes generated from the high resolution simulations to the same resolution as those from the low resolution simulations (i.e., $128\mbox{ pixels}\times 128 \mbox{ pixels}\times 500\mbox{ velocity channels}$).  After this regridding, the statistics that show significant difference ($p<0.01$) from the low-resolution data are similar to the set before: bispectrum, skewness and kurtosis, and VCS.  Of note, the Cramer test no longer registers differences between the different resolutions because the brightest values in the high resolution simulation are smoothed over.  The addition of the bispectrum appears to stem from the averaging process changing the underlying phase correlation structure for which the bispectrum --- and bicoherence --- is sensitive to.  These tests are particularly relevant since they mimic the behaviour of telescopes which average over physical affects playing out at sub-resolution scales.  Most of the statistics are unaffected by the coarse root grid resolution, though statistics that depend on the pixel-to-pixel variations are.

\subsection{The Effect of AMR on the Simulated Data Cubes} 
\label{sub:the_effect_of_amr_on_the_simulated_data_cubes}

We perform a similar analysis to the resolution comparison on the effect of using AMR during the radiative transfer post-processing. We run the post-processing step on a sub-set of the $128^3$ fiducials, allowing RADMC to use the AMR from {\sc Enzo}, and calculate the distance between the AMR and non-AMR fiducial cubes. We then perform the same $p$-value test to determine if the distances are significantly different between the two groups.

First, we examine the visible differences in the fiducials with AMR enabled through the moment arrays. The regions affected by the AMR are the two or three brightest regions within the simulation. These cause minute differences in the line widths in these regions but can dramatically change the peak brightness in the integrated intensity.

We find that AMR has an effect on the bispectrum, the skewness and kurtosis, and the VCS, which is nearly the same set that is sensitive to resolution changes.  Furthermore, the reason driving these changes is related to the skew of the greater integrated intensity in the AMR cubes.  The phase information that the bispectrum is sensitive to changes with the AMR due to the brighter, more compact intensity peaks. Skewness and kurtosis are also sensitive to these larger spatial fluctuations. The VCS, meanwhile, is likely influenced by the larger intensities on small scales. Indeed, we find that the difference is driven mostly by the small-scale regime, where the AMR has a less steep slope consistent with this prediction.

As with the resolution comparison, the difference in these statistics with and without AMR may be more attenuated when using a larger root grid, and this effect should be further compared in the high resolution regime.


\section{Fiducial Sensitivity to Temperature} 
\label{app:fiducial_sensitivity_to_temperature}

We re-ran the five fiducial simulations using a temperature of 40 K, instead of the 10 K used for the simulations set presented in \S\ref{sec:sims}, to test for sensitivities to temperature changes in the statistics.  We also increased the simulation density, energy injection, and magnetic field so that \virial, \mach, and \plasbeta\ were the same for the hot simulation as for the fiducial simulations.  This provides a check on our results since isothermal MHD simulations can be rescaled without changing the underlying physics.  However, the radiative transfer and resulting mock observations should be affected by these changes.  Ideally, our statistical formulations will show no significant response to these changes, since they are intended to track these underlying dimensionless parameters rather that temperature fluctuations.

We calculate the distance between the original fiducials to the hot fiducials between each of the set of five. Essentially, this treats the hot fiducials as the design simulations. From this, we use the fiducial-design p-value test presented in \S\ref{sub:determining_quality_of_statistics} to test whether the two groups of distances are significantly different. We find that only two statistics are sensitive to the temperature change: Cramer and VCS.

The Cramer statistic exhibits a difference between the hot and cool fiducials due to the larger line widths in the hot case. The Cramer statistic is calculated on a two-dimensional data matrix, where each column contains the top 20\% pixel intensities, above a set noise level. Thus, when comparing data where the line widths in one are significantly broadened, near empty channels are compared against channels with significant emission. Given the primary sensitivity of Cramer to \mach\ (\S\ref{sub:sensitivity_analysis}), this makes sense. This suggests that the Cramer statistic does not provide much information beyond measuring differences in the line width distributions of the data cubes.

The dependence of the VCS on temperature is also being driven by the thermal line width change since this removes information about the underlying turbulent fields on larger velocity scales than in the cold fiducial case. When computing the VCS distances, we choose the velocity slice thickness to be the same factor larger than the thermal line width in both cases: $300$ m s$^{-1}$ (or five channels in the original data cubes; \S\ref{sec:sims}) for the cold fiducial, and $600$ m s$^{-1}$ (or 10 channels) for the hot fiducials. At those velocity channel widths, the slopes of the small and large scales components approach each other, but their differences remain statistically significant, and thus they have larger distances than the distances between the cold fiducials. The differences grow for different relative velocity slice thicknesses. A temperature dependence is not unexpected for the VCS. The more sophisticated modelling by \citet{chepurnov10} and \citet{Chepurnov2015ApJ...810...33C} on H{\sc I} data has an exponential dependence on temperature \citep[see Eq. 19 in][]{chepurnov10}.


\section{Re-scaling to a common line width and intensity} 
\label{app:re_scaling_to_a_common_linewidth_and_intensity}

To determine the effects of basic observables (\S\ref{sub:sensitivity_to_basic_observables}) on the method responses, we normalize the intensity of each cube in the simulation set by its 95\% quantile and regrid the velocity channels such that the mean line width is equal across the set.  The spectral regridding size was chosen to match the largest average line width in the simulation set to avoid interpolating to a channel size smaller than the original.  We then repeat the entire analysis described in \S\ref{sec:analysis}.

Most methods exhibit no sensitivity change with the regridded simulated cube set.  Those that do are the Cramer statistic, VCS, and the dendrogram power-law tail.  The Cramer statistic's primary dependence is still the most prominent, but there are now weak dependencies on \drive, \virial, and \solfrac.  This highlights the role of smaller variations that may otherwise be hidden within the fiducial-to-fiducial scatter.  The dendrogram power-law exhibits remarkably different sensitivities; in fact it loses {\it all} sensitivity to first-order changes in the parameters. This is driven by the regridding procedure, so the number of pixels within a structure varies and is truncated wherever the mean line width was initially smaller. Finally, the VCS is difficult to directly compare to our initial results since it depends on the velocity slice thickness, and we have essentially varied this parameter differently across the whole set.  VCS is only sensitive to first-order changes in \mach\ and \plasbeta\ with the regridded set, and the higher order sensitivities are highly reduced. These results suggest that the majority of these methods are insensitive to trivial scalings in the data.

In \S\ref{sub:sensitivity_to_basic_observables}, we note that the SCF's sensitivity to \mach\ may be driven by differences in the line width alone. In this regridded data set, we find no significant change in the SCF's response. This suggests that the \mach\ sensitivity for the SCF is not entirely driven by changes in the line width.


\label{lastpage}

\end{document}